\documentclass[twoside, 12pt, a4paper]{report}

\usepackage{fancyhdr}     
\usepackage{pdfpages}
\usepackage{textcomp}
\usepackage{amssymb}
\usepackage{amsmath}
\usepackage{amsthm}
\usepackage{amscd}
\usepackage{amsfonts}
\usepackage{mathrsfs}
\usepackage{graphicx}
\usepackage[latin1]{inputenc}
\usepackage[T1]{fontenc}
\usepackage[italian,english]{babel}
\usepackage{varioref}
\usepackage{color}
\usepackage[pagebackref]{hyperref}
\usepackage{cite}
\usepackage[Sonny]{./LatexStyFile/fncychap}
\usepackage{braket}
\usepackage{minitoc}
\usepackage{epsfig}

\makeindex
\DeclareGraphicsExtensions{.jpg, .pdf, .png}
\def\varpi{t}

\def\det{\,{\rm det}\, }

\def\sign{{\rm sign}}

\def\T1{T^{(1)}}
\def\T2{T^{(2)}}
\def\T3{T^{(3)}}

\def\Im{\,{\rm Im}\,}
\def\Re{\,{\rm Re}\,}

\def\({\left(}
\def\){\right)}
\def\[{\left[}
\def\]{\right]}
\def\hf{{1\over 2}}

\def\tai#1{{\tilde\alpha}^{[#1]}}
\def\<{\left\langle}
\def\>{\right\rangle}
\def\gl#1{{\rm g}_{#1}}

\renewcommand{\d}{\mathrm{d}}
\newcommand{\de}{\mathrm{d}}
\newcommand{\De}{\mathrm{D}}
\newcommand{\I}{\mathrm{i}}

\newcommand{\cL}{\mathcal{L}}
\newcommand{\cD}{\mathcal{D}}
\newcommand{\ep}{\varepsilon}
\def\vrh{\varrho}

\newcommand{\p}{\partial}

\newcommand{\cV}{\mathcal{V}}
\newcommand{\cC}{\mathcal{C}}

\newcommand{\cK}{\mathcal{K}}
\newcommand{\cM}{\mathcal{M}}
\newcommand{\cW}{\mathcal{W}}
\newcommand{\cN}{\mathcal{N}}
\newcommand{\cE}{\mathcal{E}}
\newcommand{\cX}{\mathcal{X}}

\newcommand{\cR}{\mathcal{R}}
\newcommand{\cT}{\mathcal{T}}
\newcommand{\cJ}{\mathcal{J}}

\newcommand{\unit}{{\mathbf{1}}}

\newcommand{\cY}{\mathcal{Y}}

\newcommand{\expe}[1]{{\bf E}\!\left( #1\right)}

\newcommand{\cZ}{\mathcal{Z}}
\newcommand{\cI}{\mathcal{I}}
\newcommand{\cO}{\mathcal{O}}

\newcommand{\cU}{\mathcal{U}}
\newcommand{\cA}{\mathcal{A}}
\newcommand{\cB}{\mathcal{B}}
\newcommand{\cQ}{\mathcal{Q}}

\newcommand{\pa}{\partial}
\newcommand{\nn}{\nonumber}

\newcommand{\eps}{\epsilon}
\newcommand{\IR}{\mathbb{R}}
\newcommand{\IC}{\mathbb{C}}
\newcommand{\IZ}{\mathbb{Z}}

\newcommand{\tzeta}{\tilde\zeta}

\newcommand{\txi}{\tilde\xi}

\newcommand{\CP}{\IC P^1}

\def\bea{\begin{eqnarray}}
\def\eea{\end{eqnarray}}
\def\be{\begin{equation}}
\def\ee{\end{equation}}
\def\ba{\begin{align}}
\def\ea{\end{align}}
\def\bse{\begin{subequations}}
\def\ese{\end{subequations}}

\def\bF{\bar F}
\def\bY{\bar Y}

\def\bV{ \bar V }
\def\bW{ \bar W}

\def\bZ{\bar Z}

\def\bl{\bar \lambda }

\def\ba{\bar a}

\def\bl{\bar l}

\def\bn{\bar n}

\def\bz{\bar z}

\def\cl0{\tilde c_0}

\def\ci#1{c^{[#1]}}
\def\cij#1{c^{[#1]}}

\def\txii#1{{\tilde\xi}^{[#1]}}
\def\ai#1{{\alpha}^{[#1]}}

\def\xii#1{\xi_{[#1]}}
\def\alpi#1{\alpha^{[#1]}}

\def\rhoi#1{\rho_{[#1]}}

\def\gi#1{g^{[#1]}}

\def\Hij#1{H^{[#1]}}

\def\hHij#1{\hat H^{[#1]}}

\def\Xigi{\Xi_{\gamma}}

\newcommand{\Li}{{\rm Li}}

\def\ellg#1{\ell_{#1}}
\def\hUks{{\bf U}}
\def\hgam{\hat \gamma}
\def\hkp{\hHij{\hgam}_{k,p}}
\def\tS{\tilde S}

\def\Zg{Z_{\gamma}}
\def\bZg{\bar Z_{\gamma}}

\def\hng{\Omega(\gamma)}
\def\Om#1{\Omega({#1})}
\def\bOm#1{\bar\Omega({#1})}

\def\htt{{\mathtt t}}
\def\htp{\htt_+}
\def\htm{\htt_-}
\def\htpm{\htt_\pm}

\def\rhoi#1{\rho_{[#1]}}

\def\tinv{s}
\def\tinvp{\tinv_+}
\def\tinvm{\tinv_-}
\def\tinvpm{\tinv_\pm}

\def\Cf{\cC^{c,d}}
\def\Cfp{\Cf_+}
\def\Cfm{\Cf_-}
\def\Cfpm{\Cf_\pm}

\def\kk{{\bf k}}

\def\Fi{\mathscr R}

\def\gamD#1{\tilde\gamma}
\def\GamD#1{\Gamma^{(#1)}}
\def\Nq{N}

\def\Ilg{\cJ^{(1)}}
\def\Ilog#1{\cJ^{(1,#1)}}
\def\Irt{\cJ^{(2)}}
\def\Irat#1{\cJ^{(2,#1)}}

\def\Igp{\Ilog{+}_{\gamma}}
\def\Igm{\Ilog{-}_{\gamma}}
\def\Igpm{\Ilog{\pm}_{\gamma}}
\def\Igamp#1{\Ilog{+}_{#1}}

\def\Igg{\Ilg_{\gamma}}
\def\Iggp{\Ilg_{\gamma'}}

\def\Igam#1{\Ilg_{#1}}
\def\rIgam#1{\Irt_{#1}}
\def\rIgamp#1{\Irat{+}_{#1}}
\def\rIgamm#1{\Irat{-}_{#1}}
\def\rIg{\Irt_{\gamma}}
\def\rIgp{\Irat{+}_{\gamma}}
\def\rIgm{\Irat{-}_{\gamma}}
\def\rIgpm{\Irat{\pm}_{\gamma}}

\def\XXint#1#2#3{{\setbox0=\hbox{$#1{#2#3}{\int}$}
\vcenter{\hbox{$#2#3$}}\kern-.5\wd0}}

\newcommand{\cwarrow}{\text{\Large$\curvearrowright$}}
\newcommand{\ccwarrow}{\text{\Large$\curvearrowleft$}}

\def\cij#1{c}
\def\ci#1{c}

\def\vv{v}

\def\vl{v}
\def\bvl{\bar \vl}

\def\Min{M}

\def\Uk{U}
\def\Uin{\mathbf{U}}

\def\cCf{\cC}

\def\tleta{\tilde\eta}
\def\CY{\mathfrak{Y}}
\def\CYm{\mathfrak{\hat Y}}

\def\hH{h}
\def\hHij#1{h^{[#1]}}

\def\Fcl{F^{\rm cl}}

\def\Tmn#1{T^{#1}}

\def\qr{\sigma_D(\gamma)}
\def\qrp{\sigma_D(\gamma')}

\begin{document}

\thispagestyle{empty}
\includepdf{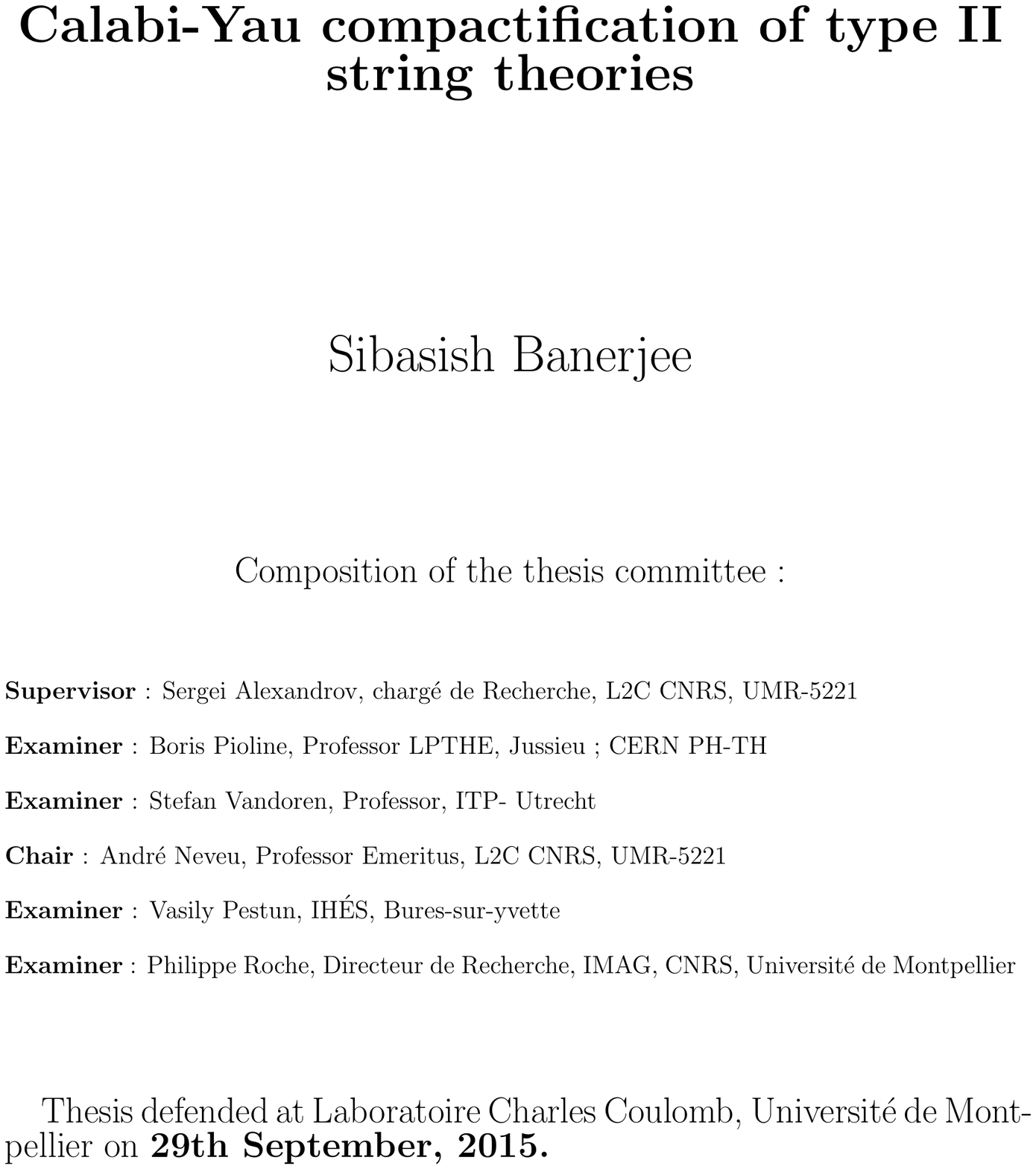}

\pagenumbering{roman}
\dominitoc
\tableofcontents
\clearpage


\newpage

\noindent
{\huge {\bfseries Abstract}}
\vspace{1cm}

\noindent

Superstring theories are the most promising theories for unified description of all fundamental interactions including gravity. However, these theories
are formulated consistently only in 10 spacetime dimensions. Therefore, to connect to the observable world, it is required to compactify 6 out of those 10 dimensions 
in a suitable fashion. 

In this thesis, we mainly consider compactifications of type II string theories  on Calabi-Yau 
threefolds. As a consequence, the resulting four dimensional theories preserve  $\cN=2$ supersymmetry.
 In these cases the metrics on the moduli spaces of the matter multiplets, vector and hypermultiplets, completely determine the  low energy theories. Whereas the 
former are very well understood by now, the complete description of hypermultiplets is more complicated. In fact, hypermultiplets receive 
both perturbative and non-perturbative corrections. The thesis mainly pertains to the understanding of the non-perturbative corrections. Our findings for the hypermultiplets rely on the so called twistorial construction. We discuss this technique
in details and use it throughout 
the course of this thesis.  
 In particular, armed with this, and exploiting 
various symmetries, especially S-duality and mirror symmetry, we discuss the procedure to derive the D-brane and NS5-brane instanton corrections to the hypermultiplet moduli space.

\newpage

\noindent
{\huge {\bfseries Introduction}}
\vspace{1cm}

\noindent

Recent years have witnessed an increasing role of non-perturbative effects in modern day theoretical physics.
For example, in the context of non-abelian gauge theories which describe elementary particle physics in a wide energy scale,
the perturbative expansion is sufficient for high energy scales, but for the low energy limit where the gauge couplings grow strong
and non-abelian symmetries are unbroken the non-perturbative effects become extremely important, in particular to explain the phenomenon of confinement.
Non-perturbative effects are important in the context of string theories as well. For instance, after the compactification string theories receive two kinds of corrections : $\alpha'$-corrections and corrections
from string coupling constant ($g_s$). Both of these corrections can be perturbative and non-perturbative. These non-perturbative contributions are necessary ingredients for consistency with beautiful as well as mathematically intriguing dualities between different theories. In the case of superstring theories they have a nice interpretation in terms of branes wrapping different supersymmetric cycles in the compactification manifold. The content pertaining to the thesis is primarily to understand the dynamics of such non-perturbative objects. 

In the course of pursuing the Ph.D., I have been working on such non-perturbative effects in the case of ten dimensional type II A/B string
theories compactified on Calabi-Yau threefolds, with Sergei Alexandrov. After the compactification, apart from the gravitational multiplet, there are two kinds 
of matter multiplets called vector multiplets (VM) and hypermultiplets (HM). The low energy effective action is completely determined by the geometry of the moduli spaces
of VM and HM. Since the vector multiplets are well-understood by now, the main goal of our work  has been to find  the HM moduli
space including all instanton corrections. For achieving this one needs to invoke various dualities and require consistency with the symmetries of type IIA/B string theories.

But performing this task is difficult compared to the vector multiplet case due to one crucial feature of the hypermultiplets. Whereas the supersymmetry requires the vector multiplet
moduli space to be a special K\"ahler manifold, the hypermultiplet moduli space is  a Quaternion-K\"ahler manifold which is a difficult
geometrical object to study. The most convenient way to work with such manifolds is to pass to their twistor spaces which are  $\mathbb{C}P^1$ bundles on them. The advantage is that on the twistor space there are certain  so-called holomorphic
transition functions that encode the entire geometry of the base space in nice fashion. Therefore, the problem of incorporating instanton corrections to the HM moduli space reduces to
the problem of obtaining transition functions that generate these corrections.
Due to this it is important to understand how different symmetries of the HM moduli space are realized at their twistor spaces. 
It is known that  all of them can be lifted to holomorphic isometries at the twistor space, however one still needs to find their explicit action at that level.

 In our first project (\cite{Alexandrov:2013mha}) we provided an explicit twistorial construction of quaternion-K\"ahler manifolds (obtained by deforming the so-called c-map spaces) carrying an isometric action of the modular group SL(2,$\mathbb{Z}$) which is a manifest symmetry of type IIB string theories. We found how this symmetry is lifted to the twistor space 
and a certain non-linear constraint which must be satisfied by the transition functions to be consistent with isometric action of SL(2,$\mathbb{Z}$).
Besides, we also established the instanton corrected mirror-map between the physical fields of type IIA and type IIB string theories.
 As a byproduct of our analysis, we found a modular invariant function that encodes all deformations
of the quaternion-K\"ahler space in a non-trivial way. Its existence is not evident {\it{a priori}} and it might have an interpretation as 
an S-duality invariant partition function. This work provided a general framework for incorporating
NS5-instantons in string compactifications with N=2 supersymmetry.

In our next work (\cite{Alexandrov:2014mfa}), we applied the above construction to the concrete issue of incorporating NS5 instantons in the type IIB framework. The starting 
point was the construction of D-instantons in type IIA framework as established in Alexandrov,
Pioline, Saueressig and Vandoren in 2008, \cite{Alexandrov:2008gh} and later to {\it all orders} in instanton expansion by Alexandrov, \cite{Alexandrov:2009zh}. Then using mirror symmetry 
and exploiting S-duality symmetry of type IIB string theory, in particular the fact that under SL(2,$\mathbb{Z}$) symmetry D5 and NS5
instantons transform as a doublet, we derived the transition functions for NS5 instantons. Actually finding the transition functions directly is a difficult exercise due to the fact that the constraint imposed on them by SL(2,$\mathbb{Z}$) symmetry is extremely non-linear and  had been obtained in our previous work.
The key idea to overcome this problem was to introduce a new parametrization of the quaternion-K\"ahler manifolds.
Instead of the transition functions it relies on the so-called ``contact hamiltonians" in terms of which the above-mentioned constraint takes a simple linear form.
 From these contact Hamiltonians it is then straightforward to compute
the transition functions which is the main result of this work.
Then we proceeded further and incorporated also D(-1)-D1 instanton  contributions (\cite{Alexandrov:2014rca}) alongwith fivebrane instantons in another project.
There we derived the transition functions in that more general case.
We also provided a thorough study of discrete isometry groups.
We found that the standard representations of these groups obtained by discretization of the classical continuous symmetries is inconsistent with the closure of 
the group action. We showed how to modify the representations to make them consistent.
 It turns out that these modifications have origins in the so-called quadratic refinement, leading to a natural question which we would 
like to understand better. In particular, we expect that this modificaton arises due to subtleties in the definition of the one-loop determinant around D-instanton background,
so they have an origin similar
to the Freed-Witten anomaly (Freed,Witten 1999, \cite{Freed:1999vc}).

One of the main reasons why we work at the level of the twistor space and the transition functions on it is because it is hopelessly complicated to compute the explicit expression for the metric on the 
quaternion-K\"ahler space in the general case. However if one considers only electrically charged instantons it is still possible to evaluate the metric explicitly.
In chapter\ref{chapter 4}, we give the expression for the metric in such cases following the work in \cite{Alexandrov:2014sya}. 
Moreover, we also investigated the fate of the curvature singularity appearing at the perturbative level. This singularity must be resolved by instantons,
but our results show that inclusion of D(-1) instantons is not sufficient for this purpose, indicating that probably one has to include all the instantons, in particular the NS5-instantons.
One of the important cross-checks to our construction is that, in the four dimensional case corresponding to the universal hypermultiplet, our metric fits the Tod ansatz
and provides an exact solution of the continuous Toda equation, for the case considered. We provided some details of this fact in our paper.

The research that I have been pursuing throughout my Ph.D., i.e. to understand instanton-corrected HM moduli space, requires further
investigations on which I would like to continue working in future. 
Despite what we have achieved, there remain several very important issues to be addressed. 
A summary and if known, possible ways to tackle the unsolved issues will be given in the final chapter containing discussions and conclusions. 

Although, this thesis will be mainly concerned with string compactifications, there are several other closely connected and interesting topics to the works. 
In particular, for the four dimensional N=2 gauge theories compactified on a circle, the instanton-corrected moduli space 
constructed by Gaiotto, Moore and Neitzke is dual to the D-instanton corrected HM moduli space in our context in a precise mathematical sense
which goes by the name of ``QK/HK-correspondence".
Another of such examples is that there could be a relation between NS5-instanton
corrections and the quantization of cluster varieties. To state even one more, there also exists a possibility to relate NS5-instanton corrections with amplitudes appearing in the context of topological strings,
as already advocated in (Alexandrov,Persson,Pioline 2010, \cite{Alexandrov:2010ca}). Such dualities and deep interconnections that occur throughout theoretical physics deserve further investigations. 

This thesis is organized as follows :

\begin{itemize}

\item In chapter \ref{chapter 1}, we will start with the recapitulation of some basic facts in string theory,
before proceeding to briefly discuss about the problems that we are going to encounter during the thesis.

\item In the chapter \ref{chapter2} we will discuss about the geometric nature of the
Calabi-Yau manifolds in some details, after which we will move on to describe the
generic structure of the hypermultiplet moduli space and its symmetries at the
classical level. We will also comment on  how incorporation of instanton corrections
break those classical symmetries to their discrete subgroups.

\item In chapter \ref{chapter3} we will discuss the mathematical construction for finding an
efficient description of the geometry of the hypermultiplet moduli space.
This geometry is quaternion-K\"ahler, as we shall see in the next chapter.
The mathematical technique which we will describe is called twistor formulation which is an efficient way to
deal with difficult geometric objects, the quaternion-K\"ahler spaces.

\item In chapter \ref{chapter 4} we will pass to effects of the D-instantons to hypemultiplet
moduli space, after briefly reviewing its perturbative loop corrected metric. 
We will exploit the mathematical theory of twistor spaces
spelled out in the chapter \ref{chapter3}. In the end we will find out the D-instanton corrected
metric explicitly in a particular case and comment on the topology associated with spaces
with such metrics.

\item In chapter \ref{chapter5}, we will talk about the mirror symmetry at the quantum level.
We will start with D-branes and their classifications. Then we will delve into describing realization of
S-duality at the level of the twistor space and give its mathematical construct. We will derive 
the mirror map between type IIA and IIB fields in presence of all possible instanton corrections.

\item In chapter \ref{ch6}, we will discuss about the symmetries in the quantum hypermultiplet moduli space. 
We shall see how to realization a consistent action for all the groups. It will lead us to the consideration of
a certain type of half-integral anomalies in the transformation of the fields under monodromy. 

\item Then chapter \ref{ch7} contains one of the highlights of the works. We give the form of 
NS5-brane instanton contributions. We show how invoking constraints imposed by the S-duality symmetry,
it is possible to derive them. As an aside, we shall also find that the actions of discrete symmetry groups 
discussed in chapter \ref{ch6} render $\cM_{\rm HM}$ invariant in presence of NS5-instantons.

\item Finally, we will summarize all the results and some of the persisting problems.

\end{itemize}

This thesis is primarily based on the work of the following papers (in chronological order), which are attached in the end. 

\begin{itemize}

\item S.~Alexandrov and S.~Banerjee, ``{Modularity, Quaternion-Kahler spaces and
  Mirror Symmetry},'' {\em J. Math. Phys. 54,} {\bf 102301} (2013).

\item S.~Alexandrov and S.~Banerjee, ``{Fivebrane instantons in Calabi-Yau
compactifications},'' {\em Phys.Rev.} {\bf D90} (2014) 041902.

\item S.~Alexandrov and S.~Banerjee, ``{Dualities and fivebrane instantons},'' {\em
  JHEP} {\bf 1411} (2014) 040.

\item S.~Alexandrov and S.~Banerjee ``{Hypermultiplet metric and D-instantons}",
{\em JHEP}, 1502, (2015) 176. 

\end{itemize}


\huge{ \emph {Acknowledgments}}

\vspace{1cm}

\normalsize

This work was done in the Laboratoire Charles Coulomb, in the Universit\'e Montpellier, France.
I would like to take this opportunity to express my gratitude to the people of the laboratory for providing me the facilities, which were instrumental
to accomplish my work. I also express my sincerest gratitude to the university for financially supporting me for the entire duration of 
my Ph.D. I am grateful to all the people in the department for a stimulating and vibrant atmosphere suitable for work.

The biggest part of my gratitude goes to my supervisor Sergei Alexandrov whose continuous guidance paved the 
way to successful completion of my Ph.D. thesis. His creativity and knowledge were indispensable for all the
works we have done together. My interaction with him did not remain confined within academics only. In life and
other matters his helps have been something for which I am immensely grateful to him. 
If I were to list all the reasons for it, I would probably have to write another thesis on it.

I am indebted to all who shared their scientific knowledge and ideas with me. I would like
express my sincerest gratitude to   Andrea Brini, Michele Cirafici, Vicente Cortes,  Samuel Grushevsky, Jeff Harvey, Jan Louis, Jan Manschot,  Thomas Mohaupt,
Greg Moore, Andy Neitzke, Daniel Persson,  Boris Pioline,  Soo-Jong Rey, Philippe Roche,  Bernd Seibert, J\"org Teschner, Stefan Vandoren,  Owen Vaughn,  Erik Verlinde,  Alexander Westphal,
Michele Del Zotto.
In different schools and conferences I have had the opportunity to meet quite a 
few other Ph.D. students with whom I did spend quite enjoyable times both academically and otherwise. Among them,
a special thanks go to Nava Gaddam, Irfan Ilgin, Du Pei, Henrik Gustafsson, Pitro Longhi, Caner Nazaroglu, Maximilian Kelm and Malte Dyckmanns. In our department,
I had a great time while discussing about science and life with Pronob Mitter. A cordial thanks is due for him. I also 
had a very enjoyable relationship with Andr\'e Neveu, Philippe Roche, Nicolas Cramp\'e, Gilbert Moultaka, Vladimir Lorman, Vladimir Fateev,
Maxim Clusel (whose premature demise in October, 2015 was a shock to me from which I am yet to completely recover), Michele Frigario, Jerom\'e Dorignac, Luca Ciandrini, Andrea Parmagianni and especially Dominique Caron of our department. With some of them it has
been more about science and with others mostly it was about having a stimulating conversation. A special ``thank you" goes to Dominique, who not only
fixed my computers a lot of times, but also engaged in occasional friendly banters which I will definitely miss when I go away. In my office, a pleasant atmosphere
was created by Pablo Montalvo, Nicolas Bizot, Stefano Magni and across the corridor I had the pleasure to have the company of Renata Garces,
Jean Charles Walter, Samuel Belliard and the relatively newcomers, Yash Chopra and Gaurav Dhar. I will miss the company of Stefano especially
when my Ph.D. finishes. I will miss the good dinners, sometimes heated and in other times fun discussions that we had together. I will try to 
remember the ``advices" on life that he gave me from time to time. Also I am indebted to him for his unreserved willingness to reach out 
to help me when I needed it most. Coffee-times with him were one of things I used to look forward to when I went to the university.

Thanks are also due for my ex-compatriots with whom I regularly keep in touch. Political and scientific discussions with Amit, Dibya, Debashis, Antareep , Debanjan, Harsh and Mayukh
at odd hours over Skype happen quite frequently and I hope that it will not stop in future. With Antareep I also discuss sometimes mathematics. From him I knew about
several literatures dealing with techniques of modular forms which play a crucial role in this thesis. 

I would like to thank Philippe Roche, Andr\'e Neveu and Vasily Pestun for accepting to be members of the jury. I am also
thankful to Boris Pioline and Stefan Vandoren for agreeing to be referees of my thesis.

Finally, I would like to mention that my parents remained a source of inspiration for me throughout the course of the thesis.
The same goes for my sister, Mihir and the sweet little kid that they have. Without their continuous support and encouragement, 
I could not have completed the thesis, nor could I have kept my sanity intact.

\pagenumbering{arabic}

\newpage

\chapter{Some basic facts about string theory}

\label{chapter 1}

 One of the most challenging frontiers in physics is finding a consistent theory 
of quantum gravity. There are several reasons why this problem is very important.
First of all, from purely aesthetic point of view it is desirable to formulate a
single theory that takes into account all of the fundamental forces together.
{\it A priori} there is no reason why gravity should be treated differently compared to
the other forces : electromagnetic, strong and weak, where the latter forces are consistently described in the framework of {\it quantum} field theory.
However, finding a formulation that includes gravity as well turns out to be complicated. 
The reason actually stems from the fact that gravity behaves in a significantly different fashion, as it is a force that operates at large distance scales. 
The second most important motivation becomes clear when we try to understand the physics at the time of the big-bang when Planck scale effects play
the dominant role. For being able to understand such physics to consider {\it quantum gravity} is of paramount importance. 
On one hand {\it unification} of all fundamental forces and on the other hand the urge to have a firmer grasp on the such fundamental physics, i.e. to
understand the phenomenon of the big-bang, among others are reasons for which such a tremendous amount of intellectual endeavor is poured into
the attempt of quantizing gravity by physicists all around the world.  
 
Among all the theories of quantum gravity proposed till this date,
it appears to be the case that string theory is the most promising candidate.
First, string theory seems to be able to provide the 
correct platform for unification of all fundamental forces. Furthermore, presumably it also provides an answer to the 
question of what happens to the laws of physics at the Planck scale. However despite all of its successes,
it suffers from a serious drawback. One still needs to connect string theory to the experimentally viable realm for making falsifiable
predictions. Although the recent years have witnessed significant progress in this regard, the final goal unfortunately still remains
illusive.

.

\section{A brief history of string theory}

Although string theory from today's standpoint seems to be the most promising theory for explaining all fundamental
forces, it actually has a rather curious history. The fashion in which string theory evolved, sometimes going through
a very barren period on the verge of being rejected and sometimes with a flurry of new ideas with remarkable breakthroughs,
has been nothing short of a engrossing thriller. In this section, we start by reviewing this extremely interesting chain of events.

String theory came into being as an attempt to understand strong interaction. Before QCD was discovered the
tower of particles in the processes involving strong interaction required to be explained. For this Veneziano started considering
the so called dual models. He required the scattering amplitudes in the $s$ and $t$-channels should coincide. 

This requirement along with unitarity and locality were strong enough to fix the amplitudes. Thus it was possible to find them completely
 in some simple cases and in general cases their asymptotic forms could be derived. In particular, it was shown that they are much
softer compared to the field theoretic amplitudes. Moreover, the amplitudes coincided to that of the strings and the duality $s$ and $t$-channel
amplitudes become evident - they are two degenerate limits of the same string configuration. Also the absence of ultraviolet divergences
got a natural interpretation. In field theory they appear due to the local nature of the interaction related to the fact that interacting objects are 
thought to be pointlike. When such pointlike objects (particles) are replaced by strings the singularity gets smoothened out over the string worldsheet.

That was the first time when any theory which has some resemblance to the string theories of the present day was considered.
However, the advent of QCD and interpretation of particles that participate in strong interaction through their constituent
quarks led to the seeming demise of the interest in strings momentarily. Furthermore, the amplitudes computed from dual models did suffer
from the exponential fall off as well. Together it meant for revival strings had to await for some other source of interest.

Luckily, the wait was not for long. It was soon realized that each string possesses a spectrum of excitations.
Considering closed strings, a massless mode with spin 2 was discovered, which does have the same characteristics
as graviton, the particle responsible for gravitational interaction. 
Moreover, the open string spectrum contains a massless vector boson so that they can describe gauge theories, leading to the unification. 
These facts provided yet another impetus towards the investigation of strings which
then started to be considered as theories relevant for quantum gravity.

This completely changed the perspectives from which string theory was looked at. Since then string theory has been developed as
a theory that attempts to unify all the fundamental forces including gravity. Let us have a look at this development in the following in a bit
more systematic fashion.

\section{Bosonic strings and Polyakov action}
\label{bosstring}

The content here closely follows that of Polchinski in \cite{Polchinski:1998cam1, Polchinski:1998cam}.
To start, let us first recapitulate the Nambu-Goto action for a string which is given by the area of world sheet, so
that classical trajectories for strings are its world sheets with minimal area
\bea
S_{\text{NG}}[X] = -\frac{1}{2\pi\alpha'} \int_{\Sigma} \de\sigma\de\tau \sqrt{-h(X)},
\qquad
\det{h_{ab}} = h,
\eea
where $\alpha'$ is a constant of dimension length squared, the metric induced on the world sheet is
\bea
h_{ab} = G_{\mu\nu} \p_a X^\mu \p_b X^\nu,
\label{wsptarsp}
\eea
where $X^\mu(\tau,\sigma)$ describe the embedding of the worldsheet into the target space
and $G_{\mu\nu}$ is the target space metric.

This complicated non-linear action can be classically rewritten as the so called Polyakov action 

\bea
S_P [h,X] = -\frac{1}{4\pi\alpha'} \int_{\Sigma} \de\sigma \de\tau \sqrt{-h}\, G_{\mu\nu}h^{ab} \p_a X^\mu \p_b X^\nu,
\eea
where the worldsheet metric $h_{ab}$ is now considered as dynamical variable and \eqref{wsptarsp} appears as one of the equations of motion. 
This can be further extended by adding the standard Einstein term which is a total derivative in two
dimensions,
\bea
\frac{1}{4\pi} \int_{\Sigma} \de\sigma \de\tau \sqrt{-h} R = \chi,
\eea
which is a total derivative in two dimensions given by the Euler characteristic of the surface. For a Riemann surface with $g$ handles and $b$ boundaries
it is given by,
\bea
\chi = 2 -2g -b.
\eea
Taking into account this contribution, the Polyakov action becomes,
\bea
S_P = -\frac{1}{4\pi\alpha'} \int_{\Sigma} \de\sigma \de\tau \sqrt{-h}\,\[ G_{\mu\nu}h^{ab} \p_a X^\mu \p_b X^\nu +\alpha'\nu R\].
\label{genPol}
\eea

For completeness, one has to also put some boundary conditions in the case of open strings,
for which Neumann condition is imposed on the boundary of the world-sheet $\p\Sigma$.
Moreover, one identifies closed strings as having periodic boundary conditions on the world-sheet.

\subsection{World sheet point of view}

Since we consider the world sheet metric as a dynamical variable, the Polyakov action can be viewed
as a theory of two dimensional gravity coupled to the matter fields $X^\mu$. 
We would like to quantize the world-sheet theory that we found in the previous section. 

We will do this using the framework of path integral quantization. To define the path integral, we perform the Wick rotation to the
Euclidean signature on the worldsheet
\bea
Z= \sum_{\text{surfaces} \,\Sigma} \int \De X_\mu e^{-S_P (X,\Sigma)}.
\label{bospartfn}
\eea
The sum over the surfaces should be understood as a sum over all possible topologies which is
equivalent to the sum over the genus of the Riemann surface, supplemented by the integral over diffeomorphism
invariant classes of the metric,
\bea
 \sum_{\text{surfaces}\, \Sigma} \mapsto \sum_g \int \De \rho(h_{ab}).
\eea
Now to perform the integral one uses the standard trick of Faddeev-Popov gauge fixing where the symmetry
is the world sheet diffeomorphisms. However, there is another symmetry that one needs to consider, the Weyl invariance.

\subsection{Weyl invariance}

Apart from Poincare and diffeomorphism symmetries, the Polyakov action does
possess another symmetry at the classical level, the rescaling symmetry of the world sheet metric or the Weyl
invariance,
\bea
h_{ab} \mapsto e^\phi h_{ab}.
\eea
This symmetry together with the diffeomorphisms is responsible for equivalence of the Polyakov and
Nambu-Goto actions.

However, at the quantum level the above symmetry (Weyl invariance) can be broken, as the integration
measure of the path integral over the metrics has to be regularized and in general there is no regularization
scheme that keeps all the symmetries intact.

This anomaly can be best expressed in terms of the stress-energy tensor which is
\bea
T^{ab}
= -\frac{2\pi}{\sqrt{-h}}\, \frac{\delta S_P}{\delta h_{ab}}
= -\frac{2}{\alpha'}\,\(\p^a X_\mu \p^b X^\mu - \frac12\, h^{ab} \p_c X_\mu \p^c X^\mu \).
\eea
The invariance under infinitesimal Weyl transformation implies that classically
\bea
h_{ab} \,\frac{\delta S_P}{\delta h_{ab}} = 0
\,\, \implies T^a_a = 0.
\eea

On the other hand, at the quantum level the vacuum expectation value of $T^a_a$ becomes non-vanishing 
and can be shown to be  proportional to the curvature of the target space metric,
\bea
\langle T^a_a\rangle = -\frac{c}{12} \cR .
\label{Confanomaly}
\eea
To understand the meaning of the proportionality coefficient $c$, let us consider the Polyakov action with the worldsheet
metric fixed to the flat one
\bea
S_P = \nu \chi + \frac{1}{4\pi\alpha'} \int_\Sigma\de \sigma \de \tau \delta^{ab} \p_a X_\mu \p_b X^\mu.
\label{gaugefixedPol}
\eea
The above action still possesses the conformal symmetry. In two dimensions, it is realized in terms of the Virasoro algebra,
which at the quantum level acquires a central extension. The coefficient $c$ is actually nothing but the central charge of this algebra. 

It can be shown (see for example \cite{Polchinski:1998cam1}) that the ghosts that appear due to gauge fixing
the diffeomorphism symmetry give rise to negative contribution to the central charge of value $-26$.
Since this two dimensional theory is still a conformal theory, to conclude about critical dimensions, it is sufficient to compute the central charge. 
For each bosonic degree of freedom the contribution to the central charge is $+1$.
Taking all contributions into account, then for having an anomaly free theory one must formulate the theory in $26$ dimensions as $c= D-26$. Hence we get
to the well known statement that bosonic strings live in $26$ dimensions.

This gives an exact result for the Weyl symmetry. Thus, one of the gauge symmetries of the classical theory turns out to be broken.
This effect can be seen from other important approaches of string quantization. For example, in the framework of the canonical quantization,
in the flat gauge one finds breakdown of unitarity. Similarly, in the light cone quantization, one finds that the global Lorentz symmtery of the target space 
is broken. This exemplifies why the existence of Weyl symmetry is so important for existence of a viable theory of strings.  

From the gauge fixed action \eqref{gaugefixedPol} one can draw also another important conclusion. 
As we will see later in this chapter, the string theories are defined as asymptotic expansions in the string coupling constant.
This expansion is nothing but a sum over the genera of the string world-sheet . It can be associated with string loop expansion,
because adding a handle or a strip for a closed or an open string cases respectively, can be interpreted as emission and reabsorption of virtual string.

Using the expression for gauge-fixed Polyakov action \eqref{gaugefixedPol}, we see that the partition function \eqref{bospartfn} 
contains terms each of which is weighted by a factor $e^{-\nu\chi}$ depending only on the topology.   
Due to this one needs to associate a factor of $e^{2\nu}$ for attaching a handle to a closed string and $e^{\nu}$ for 
attaching a strip to the open string. On the other hand each interaction is determined by the corresponding coupling constants.
Thus, one arrives at the following relation 
\bea
g_{\rm{ closed}} \sim e^\nu, \qquad g_{\rm{ open}} \sim e^{\nu/2}.
\label{coupling}
\eea

\section{Low energy limits primer}

In this section we are going to deal with string theories in curved spacetime. In the next section we will
incorporate the fermions as well, as they give rise to string theories which are comparatively more well-behaved
than string theories which contain only bosons. However, fixing our attention just to the bosonic part suffices
here as one can always recover the fermionic terms using supersymmetric transformations.

In fact, we have already discussed an action for strings moving in general spacetime. 
It is given by the $\sigma$-model  \eqref{genPol}  with arbitrary $G_{\mu\nu}(X)$.
Actually one can consider the general spacetime
metric as a coherent state of gravitons appearing in the closed string spectrum. Hence, any generic
$G_{\mu\nu}$ roughly can be interpreted as summing of excitations of this mode.

Apart from gravitons there are other two massless modes in the string spectrum, the antisymmetric
tensor $B_{\mu\nu}$ and the dilaton $\phi$. Thus a natural generalization of \eqref{genPol} is given
by the following,
\bea
\!\!\!S_{\sigma} = \frac{1}{4\pi\alpha'}\, \int_{\Sigma} \, \de^2 \sigma \sqrt{h}\,\[
\(h^{ab} G_{\mu\nu}(X) + \I\epsilon^{ab} B_{\mu\nu}(X)\) \p_a X^\mu\p_b X^\nu
+\alpha' R\phi(X)\].
\label{actionsig}
\eea
Notice that the dilaton is actually a generalization of $\nu$. From the identification \eqref{coupling},
the string coupling constant can vary in spacetime, being exponential of the dilaton,
\bea
g_{\rm closed} \sim e^\phi.
\eea 
In contrast to the Polyakov action in flat spacetime the action \eqref{actionsig} is non-linear and represents an interacting theory.
The couplings of this theory are coefficients of $G_{\mu\nu}$, $B_{\mu\nu}$ and $\Phi$ of their expansions in $X^\mu$. 

The effective theory, which appears in the low-energy limit, should be a theory of fields in the
target space. On the other hand, from the world sheet point of view, these fields represent
an infinite set of couplings of a two-dimensional quantum field theory. Therefore, equations
of the effective theory should be some constraints on the couplings. The only constraint which is not put by 
hand is that the \eqref{actionsig} should define a non-linear $\sigma$-model. It means that the resulting quantum field theory must retain Weyl invariance.
It is this condition which gives necessary equations on the target space fields. 

The $\beta$-functions for all the target space fields which play the roles of couplings in the theory on the worldsheet should vanish.
A very non-trivial fact which, on the other hand, can be considered as a sign of consistency
of the approach, is that the equations for $\beta$-functions can be derived from the following spacetime action,
\bea
S_{\rm eff} = \frac12\, \int \de^D x \sqrt{-G} e^{-2\phi} \[ -\frac{D-26}{3\alpha'} + \cR + 4(\nabla\phi)^2 - \frac{1}{12} H_{\mu\nu\lambda} H^{\mu\nu\lambda} \],
\eea
where $H$ is the field strength for the two form field $B$. All terms in the action are very natural representing the simplest
Lagrangian for a scalar, antisymmetric spin-2 fields. Although the first term is very large because of the $\alpha'$ in the denominator, it exactly vanishes for 
$D=26$. The factor of $e^{-2\phi}$ indicates that the action is written in the string frame. Rescaling the metric appropriately, one can recover the
standard Einstein term. Thus it is clear that at the low energy approximation, one gets back the usual Einstein gravity.

\section{Superstring theories}

In section \ref{bosstring} we discussed how to define a consistent string theory in $26$ dimensions.
It is obvious that in order to connect to real world one has to compactify\footnote{Compactification is a
procedure of getting down to the relevant dimension from higher dimensional theories. It was actually
first considered long ago around 1920's by Kaluza and Klein. They tried to construct a unified theory of gravity and
electromagnetism. To get rid of extra dimensions, one considers them to be ``smaller" than the visible
scales and the manifold that is made out of them to be compact.}
 such theories on manifolds of
dimension $22$. The final theory should of course depend on the manifold chosen for conpactification.
However, since there is no general principle which determines this manifold to compactify on, it is impossible to find the {\it right} compactification. 
The situation becomes even worse once it is realized that there actually
exist string modes for which mass squared is negative in this case, called ``tachyons" violating unitarity of the theory. This
pathology is the main culprit which makes it a bad theory to consider.

It turns out that this situation can be cured if one adds fermions, what leads to a theory of superstrings.
The prefix ``super'' comes from the fact that these theories have some amount of supersymmtry. Below we briefly
describe both target space and world-sheet formulation of these theories, even though for computational purposes the latter
is preferred.

The target space description for the superstring theories is called Green-Schwarz formalism. When one considers one or two sets of Majorana-Weyl spinors
with respect to the global Lorentz symmetry of the target space (but scalars in the world-sheet), they determine the amount of supersymmetry
in the final theory, which can be either $\cN =1$ or $\cN=2$. This automatically implies that the dimensions of the theory can not be
arbitrary. It can be either $D=3,4,6$ or $10$. Furthermore for an anomaly free quantum theory it turns out that the number of dimensions
can be only $10$ which we call as the ``critical" dimensions of the superstring theories. In contrast to bosonic string theory,
the target space formulation of superstrings inherently
ensure the absence of the pathological tachyonic modes as the spectrum starts from massless modes.

In the second type of formulation, superstring theories with world-sheet supersymmetry are considered (Ramond-Neveu-Schwarz
formalism). The world-sheet fermions $\psi^\mu$
transform as vectors under the global Lorentz symmetry of the target space. Furthermore since the supersymmetry
is considered on the world-sheet, the conformal symmetry of the bosonic strings is promoted to a superconformal
symmetry.

For such theories again the formula for the anomaly \eqref{Confanomaly} holds, as it is still a conformal theory.
We should again analyze when the conformal anomaly vanishes. In this case one has to take into account that
there are extra world-sheet fermions and ghosts that arise from gauge fixing local fermionic symmetry. It turns out that
the contribution to the central charge of the superconformal ghosts is $11$ and contribution of each fermion is $\frac12$. Hence,
\bea
c= D-26 + \frac12 D + 11 = \frac32 (D-10),
\eea
indicating that $D=10$ for an anomaly free quantum theory, in agreement with the result obtained in the Green-Schwarz formalism.  

To analyze spectrum of this formulation, one should now impose boundary conditions on the worldsheet fermions $\psi^\mu$. 
But now the number of posiibilities is doubled with respect to the bosonic case. For example, in the case of the closed strings, one can not only choose 
periodic boundary conditions on $\psi^\mu$, but also anti-periodic condition. These two sectors are called the the Ramond (R) and and the Neveu-Schwarz (NS) sectors. 
This RNS formulation represents an extension of the conformal field theory \eqref{gaugefixedPol} to the superconformal field theory. 
The additional degrees of freedom in this case precisely correspond to $\psi^\mu$ and the R and NS sectors are related by the so called
{\it spectral flow} symmetry of the superconformal field theory. In each of the sectors R or NS the superstring has different spectra of modes.   
From the worldsheet point of view the R-sector ddescribes fermions and the NS-sector describes the bosons. However, now we again encounter the
all too familiar problem : there exist tachyonic modes for the NS-sector. 

We started to build up a theory on the world-sheet which should amount to the same as the target space theory of Green-Schwarz.
These tachyons mentioned above could potentially spoil the construction. Fortunately, there exists a mechanism going by the name of GSO (Gliozzi-Scherk-Olive) projection
which removes tachyons and several other modes out of the spectrum, rendering the final theory to be well-defined.
Finally, it has also been proven by exploiting intricate relations among symmetries of superstring theories in $10$ dimensions
that both Green-Schwarz and Ramond-Neveu-Schwarz formalisms are equivalent providing the final piece of construction.

Now we have a consuistent formalism for the superstring theory. But one question arises : how many superstring theories exist ? Is it unique ?
At the classical level, it is certainly not unique. There are several possibilities : open/closed, oriented/unoriented, $\cN=1/\cN=2$ string theories. 
Besides, for open strings, one can introduce Yang-Mills gauge symmetries adding charges to the ends of the strings. Also, considering $\cN=1$ closed strings,
one can construct the so called heterotic string theories where one can also introduce gauge groups. 

Generally quantum theories suffer from anomalies which we need to cancel. This fact restricts the gauge group of open strings to
$SO(32)$ and gauge groups for heterotic strings to $SO(32)$ or $E_8 \times E_8$. Eventually, we have five
consistent superstring theories :

\begin{itemize}

\item Type IIA : $\cN=2$ non-chiral closed strings,

\item Type IIB : $\cN=2$ chiral closed strings,

\item Type I    :  $\cN=1$ uoriented open strings with gauge group $SO(32)$ 
                           and unoriented closed strings,

\item   heterotic strings with gauge group $SO(32)$,

\item heterotic strings with gauge group $E_8\times E_8$.

\end{itemize}

\subsection{M-theory and dualities}

Since there are five consistent superstring theories, the situation is not completely satisfactory. It is natural to look for further unification.
In \cite{Witten:1995ex}, Witten showed that all of the string theories can actually be realized as different limits of a single theory, which is
one dimensional higher, the so called M-theory and all of the string theories can be related by various dualities to each other. 

The picture that was found for M-theory, tells that different superstring theories are manifestations of different vacua of 
the M-theory. A generic point in the moduli space corresponds to an 11-dimensional vacuum, as the M-theory is itself 11-dimensional.
Furthermore, it has got a special 11-dimensional Lorentz invariant vacuum which is described by the flat spacetime. 
 
To understand how different superstring theories can be combined into M-theory, one needs
to compactify this very special vacuum in different fashions. 
If one compactifies it on a torus,  $\cN=2$ theories are obtained, whereas
compactification on a cylinder leads to $\cN=1$ theories.
The known superstring theories are then produced as degenerate limits of the torus or the cylinder.
For example, to get the type IIA one considers a torus with one of the radii is much bigger than the other.
Then one effectively performs the compactification on a circle and the string coupling constant in this case
is determined by the smaller radius. For the type IIB one actually considers the situation when both of the radii of the torus
vanish but the ratio is finite which is related to the string coupling in this case.
 For heterotic or type I string theories
one has to consider different radii or lengths of the cylinders to realize them as different limits of the M-theory. All these theories are related through
T or S-duality. This T-duality is the symmetry relating two theories by inverse compactification radii, correspondingly inverting the coupling and therefore,
interchanging the momentum and winding modes \footnote{T-duality on the world-sheet changes the sign of the right movers.}. S-duality on the other hand allows to
go from weak coupling to strong coupling limit. We will see detailed applications of them later.

There was a problem before, namely there was no way through which one could assign a value for $\nu$ in \eqref{coupling}.
In light of the M-theory, this inadequacy is also resolved. 
Now the string coupling constant depends on the background on which the theory is considered. 
Thus, it can be a moduli of the underlying M-theory so that it is not a free parameter any longer.

The two main ingredients that propelled second superstring revolution are the discovery of the web of dualities relating various superstring
theories and the possibility to realize them as various limits of M-theory, which we have just described, as well as the discovery of D-branes which play a crucial role in establishing the dualities.
These D-branes have dual interpretations. On one hand they are solitonic solutions to supergravity equations and on the other they can be interpreted as objects
on which open strings end. For open strings ending on D-branes, Dirichlet boundary conditions are imposed on fields moving on open string world-sheet. From the point
of view of T-duality, the presence of D-branes is necessary because T-duality interchanges Neumann and Dirichlet boundary conditions.
Apart from D-branes, one should also consider another kind of non-perturbative object, the
so called Neveu-Schwarz (NS) five-branes which are dual to the fundamental strings carrying the $B$-field charges.

\section{String compactifications}

We have already seen that in order to be able to connect to the four dimensional real world, it is necessary to compactify
higher dimensional string theories on some compact manifold. The extra dimensions are compact and small enough for being invisible 
at the usual scales. According to our previous discussion, bosonic string theory is not suitable to study because there is no criterion
that dictates what kind of manifolds one should choose to perform the compactification, as there is no particular preference {\it a priori}.
Since the theory obtained finally crucially depends on the kind of manifold chosen, the inability to narrow down the possibilities, makes
it an inviable theory to consider.

Throughout this thesis, we will focus on the compactification of type II superstring theories on Calabi-Yau threefolds. Although,
we will discuss such theories in details in the next chapter, here we want to make some cursory remarks about them, so that
we can connect to the next chapter seamlessly. Upon choosing the compactification manifold to be Calabi-Yau, one finds that
it is possible to retain $\cN=2$ supersymmetry in the low energy effective action, where typically one finds a supegravity theory
coupled to matter fields. We work with $\cN=2$ supersymmetry because, it is far less rigid than a theory with $\cN=4$ supersymmetry,
and on the other hand it is still amenable to analytical tools in contrast to $\cN=1$ theories, which although are phenomenologically
more relevant, often pose rather technically extremely difficult situations, if not intractable. 

An implication of $\cN=2$ supersymmetry is that the low energy theory is completely determined by the metric on the moduli space,
which can be parametrized by the vacuum expectation values of the four dimensional scalars. Moreover, the moduli space gets 
factorized into two disconnected components, known as the vector multiplet (VM) sector and the hypermultiplet sector (HM). A more thorough
exposition to them is delegated in the next chapter. 

We shall find that the vector multiplets are classically exact and well understood, whereas understanding the hypermultiplet
sector is more challenging. During the course of the thesis, our emphasis will primarily lie upon finding a quantum corrected description
of the HM moduli space. With the input data being some topological characteristics of the Calabi-Yau threefolds, our main purpose will
be to go to the deep quantum or non-perturbative regime. Even though the final objective has not been achieved yet, some recent progresses,
as we are going to find in the thesis, certainly give us encouragements that we are actually not very far away.

\newpage

\chapter{Compactification of type II string theories on Calabi-Yau threefolds and their moduli spaces}
\label{chapter2}


As mentioned in the previous chapter, all of the five consistent superstring theories are formulated in ten dimensions.
To connect them to the realm of viable experiments for making falsifiable predictions, one has to compactify
them on six dimensional manifolds, so that in the end we obtain theories in four dimensions.  
At the low energies this procedure typically gives rise to supergravity coupled to matters. For these effective theories, we will restrict ourselves to
the lowest order of $\alpha'$-corrections, necessarily leading to the fact that the low energy effective action (LEEA) is restricted to the two derivative terms only.

Of course the precise form of these effective actions depends on what kind of manifold is chosen for compactification. It turns out that
for retaining eight unbroken supercharges, one has to choose the compactification manifolds to be Calabi-Yau threefolds (where we
neglect the possibility of presence of fluxes and consider only torsionless Calabi-Yau).
In this thesis, we will mainly focus on four dimensional theories preserving $N=2$ supersymmetry. 
 In the rest of this chapter, we will describe the general structures of the moduli spaces obtained after performing
such compactifications of type II A/B string theories. We will conclude by mentioning, what features of classical low energy effective actions are expected
to be retained when quantum corrections, both perturbative and non-perturbative are included, and justify the reasons for such expectations.

\section{Calabi-Yau manifolds and their moduli spaces}

 A compact Calabi-Yau threefold ($\CY$) is a Ricci-flat K\"ahler manifold, which necessarily means the first Chern class for them vanishes.
They are characterized by the existence of a unique globally well-defined covariantly constant holomorphic $(3,0)$ form which we will call by $\Omega$
and a canonically defined complex structure.

Calabi-Yau threefolds do not admit any non-trivial 1 and 5-cycle which is implied by the condition $h^{0,1}(\CY) = h^{1,0}(\CY) = 0$. Moreover
uniqueness of the non-degenerate form $\Omega$ is expressed by the condition $h^{3,0} (\CY) = 1 $.

The space of deformations of a Calabi-Yau manifold comprises of two sectors : the complex structure deformations and the K\"ahler class deformations.  
Clearly, $\Omega$ determines the complex structure uniquely on $\CY$. On the other hand the information about the
K\"ahler structure is contained in the $(1,1)$ K\"ahler form $J$. Locally the moduli space of complex structure deformation
is then given by $H^{2,1}(\CY,\IC)$ and the K\"ahler class deformation locally coincides with $H^{1,1}(\CY,\IR)$.
In fact, one should consider the complexification of the K\"ahler form $J$ as it always appears with the two form $B$-field as $B+\I J$.
Then the moduli space that we need to analyze is the product of the above two moduli spaces 
\bea
\cM_{\CY} = \cM_{C} (\CY) \times \cM_K(\CY).
\eea

Without going into the details of intrinsic definition, let us remind the reader that a special
K\"ahler geometry can be defined as a K\"ahler geometry with the special structure, that all
geometric quantities can be determined in terms of a single holomorphic function called the prepotential.
Both the above sectors are equipped with such geometries, more precisely, {\it local} special K\"ahler geometries. 
For this case, the special K\"ahler manifolds are completely determined by a degree two homogeneous function $F(X)$,
where $X^\Lambda$ are the coordinates with indices $\Lambda = 0,...,n$ and $n$ is the complex dimension of the manifold.  
Then the K\"ahler potential is given by
\bea
\cK = -\log \[ \I (X^\Lambda \bF_\Lambda (X) - \bar{X}^\Lambda F_\Lambda (X) )\].
\label{locKpot}
\eea
Using the projective coorinates $X^\Lambda/X^0 = (1,z^a)$ one can find the metric from the K\"ahler potential by
\bea
g_{a\bar{b}} = \p_{a} \bar{\p}_{\bar{b}} \,\cK(z,\bz).
\label{metSK}
\eea
The holomorphic three form $\Omega$ and the complexified K\"ahler $(1,1)$-form provide natural candidates for the metric,
the above defined coordinates and the prepotential. We discuss these above two cases separately and introduce a few formulas on the way.

\subsection{Complex structure moduli}
\label{cxstrmod}
The holomorphic three form $\Omega$ gives rise to the following K\"ahler potential on $\cM_C(\CY)$ given by,
\bea
\cK = -\log \I \int_{\CY} \Omega \wedge \bar{\Omega}.
\label{cxKahler}
\eea
We now want to introduce holomorphic Darboux coordinates on $\cM_C(\CY)$. For doing so,
let us consider the full cohomology group 
\bea
H^3 = H^{3,0} \oplus H^{2,1} \oplus H^{1,2} \oplus H^{0,3}.
\eea
This is dual to $H_3(\CY)$ where we a choose a symplectic marking $\cA^\Lambda,\cB_\Lambda$ such that $\cA^\Lambda \# \cB_\Sigma = \delta^\Lambda_\Sigma$.
Given such a basis, let us define
\bea
X^\Lambda = \int_{\cA^\Lambda} \Omega,
\qquad
F_\Lambda = \int_{\cB_\Lambda} \Omega.
\eea
Clearly this is a highly redundant coordinate system ($=2(h^{2,1}+1$)), as the space of complex structure deformations is only
$h^{2,1}$ dimensional. As a result, one can always choose a basis of cycles for which $F_\Lambda$ becomes the derivative of $F(X)$  with respect to $X^\Lambda$.
Let us show it explicitly. For that we need to use the Riemann bilinear identity,
\bea
\int_{\CY} \psi \wedge \chi = {\sum_\Lambda} \(\int_{\cA^\Lambda} \psi \int_{\cB_\Lambda} \chi - \int_{\cA^\Lambda} \chi \int_{\cB_\Lambda} \psi\)
\eea
Since the deformation of $\Omega$ can be at most a $(2,1)$-form 
\bea
\begin{split}
0 &= \int_{\CY} \Omega \wedge \p_\Lambda\Omega 
\\
& = \int_{\CY} \(X^\Sigma \cB_\Sigma + F_\Sigma \cA^\Sigma\) \wedge \(\cB_\Lambda + \p_\Lambda F_\Sigma \cA^\Sigma\).
\end{split}
\eea
Then using the bilinear identity one immediately finds
\bea
0 = F_\Lambda - X^\Sigma \p_\Lambda F_\Sigma =2\[ F_\Lambda - \frac12\, \p_\Lambda \(X^\Sigma F_\Sigma\)\].
\eea
Integrating, we find that prepotential is a second degree homogeneous function as $F=\frac12\, X^\Lambda F_\Lambda$. 
This defines the prepotential for the complex structure moduli and in particular, the K\"ahler potential \eqref{cxKahler} coincides with the
one following from \eqref{locKpot}.

\subsection{K\"ahler moduli}

In this case, one should consider instead the even cohomology group. 
\bea
H^{\rm even} = H^0 \oplus H^2 \oplus H^4 \oplus H^6.
\eea
Let us choose a basis in this space parametrized by $\omega_I = (1,\omega_i)$ and $\omega^I =(\omega_{\CY},\omega^i)$,
where $\omega_{\CY}$ is the volume form, such that
\bea
\omega_i \wedge \omega^j = \delta^j_i \omega_{\CY},
\qquad
\omega_i \wedge \omega_j = \kappa_{ijk} \omega^k
\eea
and the indices $I =(0,i) = 0, 1,...,h^{1,1}(\CY)$. The second equation defines the triple intersection numbers.
Choosing the dual basis of two cycles $\gamma^i$ and a basis of four cycles $\gamma_i$, the triple 
intersection number can be found as 
\bea
\kappa_{ijk} = \int_{\CY} \omega_i \wedge \omega_j \wedge \omega_k = \langle \gamma_i,\gamma_j,\gamma_k\rangle .
\eea
Then the natural metric on $\cM_K(\CY)$ is given by
\bea
g_{i\bar{j}} = \frac{1}{4V} \int_{\CY} \omega_i \wedge \ast \omega_j = \p_i \p_{\bar{j}} (-\log 8 \cV),
\label{Kmet}
\eea
where $V = \frac16\,\int_{\CY} J\wedge J\wedge J$ is the Calabi-Yau volume. 

Then the holomorphic coordinates on $\cM_K(\CY)$ are 
\bea
v^i = b^i + \I t^i = \int_{\CY} (B+\I J).
\eea
Let us introduce the homogeneous coordinates $X^I = (1,v^i)$. In terms of $X^I$ the prepotential for the
moduli space of complexified  K\"ahler class deformations takes the form,
\bea
F(X) = -\frac16\, \kappa_{ijk} \frac{X^i X^j X^k} {X^0},
\label{Kprepot}
\eea
where \eqref{locKpot} indeed leads to the K\"ahler potential from \eqref{Kmet}. We will see in the following
that this cubic prepotential is only an approximation in the large volume limit $t^i \to \infty$ of the prepotential
that appears in string compactification.

\section{Field content of type II supergravities in four dimensions}

In order to obtain the low energy effective action in four dimensions, one starts
from ten dimensional supergravity theories. We will focus particularly on type IIA and type IIB supergravity theories.
We will be interested in the bosonic sectors only, because the fermionic sectors
can in principle be restored by supersymmetic transformations.

The bosonic sectors in 10 dimensions are subdivided into two parts, the NS and the RR sectors. The former comprises the ten-dimensional metric $\hat{g}_{XY}$, the two form
field $\hat{B}_2$ and the ten dimensional dilaton $\hat{\phi}$, whereas the latter sector contains p-form potentials $\hat{A}_p$. For the type IIA
theory $p=1,3$ and for type IIB, $p=0,2,4$, with an additional constraint on $\hat{A}_4$, such that its field strength is self-dual, $F_5 = \ast F_5$.

To perform the compactification these differential forms should be expanded in the basis of harmonic forms. For type IIB side such forms have already been introduced
in the previous subsection. For type IIA one has to remember that there is no nontrivial 1 or 5-cycles so that one has to consider only 3-forms. One can choose dual forms to
$\cA^\Lambda$ and $\cB_\Lambda$ cycles introduced in section \ref{cxstrmod} to be $\alpha_\Lambda$ and $\beta^\Lambda$ respectively, such that
\bea
\int_{A^\Lambda} \alpha^\Sigma = \delta_\Lambda^\Sigma,\,\,\,\,\,  \int_{A^\Lambda} \beta_\Sigma = 0,\,\,\,\, \int_{B_\Lambda} \alpha^\Sigma  = 0,\,\,\,\,
\int_{B_\Lambda} \beta_\Sigma = - \delta^\Lambda_\Sigma. 
\eea
The procedure of compactification to four dimensions on Calabi-Yau 
is described in details in \cite{Polchinski:1998cam} and a relevant review is present in \cite{Alexandrov:2011va}. In the following
we will only discuss the field contents in the two cases of  in four dimensions. 

\subsection{Type IIA}

The bosonic sector of type IIA supergravity in $10$-dimensions is described by the following action \cite{Polchinski:1998cam} :
\bea
\begin{split}
S_{\rm IIA} &= \frac12\, \int\[e^{-2\hat{\phi}}\(\hat{R}\ast 1 + 4 \de \hat{\phi} \wedge \ast\hat{\phi}-\frac12\, \hat{H_3}\wedge \ast \hat{H_3}\)
- \(\hat{F_2} \wedge \ast\hat{F_2} + \hat{F_4} \wedge \ast\hat{F_4}\)
\right.
\\ & \left. \qquad \qquad
- \hat{B_2}\wedge\de \hat{A_3}\wedge \de \hat{A_3} \],
\end{split}
\eea
where hats stand for fields in $10$-dimensions and
\bea
\hat{H_3} = \de \hat{B_2}, \qquad \hat{F_2} = \de \hat{A_1}, \qquad \hat{F_4} = \de \hat{A_3} - \hat{A_1} \wedge \hat{H_3}.
\eea
Let us now analyze what happens due to compactification.

\begin{itemize}

\item First let us try to understand the NS-NS sector to which
$\hat{g}_{XY},\hat{B}_2, \hat{\phi}$ belong.
Then 
\bea
\begin{split}
& (\hat{g}_{XY})_{\rm 10d} \mapsto \(g_{\mu\nu}, t^i,z^a\)_{\rm 4d} \mapsto \(g_{\mu\nu},t^i,z^a\)_{\rm 4d-dual},
\\ &
(\hat{B_2})_{\rm 10d} \mapsto \(B_2 + b^i\omega_i\)_{\rm 4d} \mapsto \(\sigma, b^i\)_{\rm 4d-dual},
\\ &
\hat{\phi}_{\rm 10d} \mapsto \phi_{\rm 4d} \mapsto \phi_{\rm 4d-dual},
\end{split}
\eea
where the first arrow indicates dimensional reduction from 10 to 4 dimensions and the second arrow indicates 
dualization in 4 dimensions. 

\item Then for the RR sector,
\bea
\begin{split}
& (\hat{A_1})_{\rm 10d} \mapsto (A^0_1)_{\rm 4d} \mapsto (A^0_1)_{\rm 4d-dual},
\\ &
(\hat{A_3})_{\rm 10d} \mapsto \(A_3 + A^i_1\omega_i + \zeta^\Lambda \alpha_\Lambda + \tzeta_\Lambda \beta^\Lambda\)_{\rm 4d}
\mapsto \({\rm constant}, A^i_1,\zeta^\Lambda,\tzeta_\Lambda\)_{\rm 4d-dual}. 
\end{split}
\eea
\end{itemize}

Thus, in four dimensions the fields organize in the following multiplets :
\begin{itemize}

\item gravitational multiplet                $(g_{\mu\nu},A^0_1)$,
\item tensor multiplet                          $(B_2,\phi,\zeta^0,\tzeta_0)$.,
\item $h^{2,1}$ hypermultiplets        $(z^a,\zeta^a,\tzeta_a)$ ,
\item $h^{1,1}$ vector multiplets      $(A^i_1,v^i = b^i + \I t^i)$.
\end{itemize}

$A_3$ is not included because in four dimensions it can be dualized to a constant which we put to zero as it does not carry any degree of freedom.
However in \cite{Louis:2002ny} it was shown that it can play an important role by inducing a gauge charge for the NS-axion $\sigma$,
where $\sigma$ appears due to dualization of the field $B_2$.
After this dualization the tensor multiplet gets converted into another hypermultiplet and
there are now $h^{2,1}+1$ hypermultiplets in total. The hypermultiplet that comes from the dualization of tensor multiplet is always present
and is called the universal hypermultiplet. When the compactification manifold is chosen to be a {\it{rigid}} Calabi-Yau threefold, by which one means that
there is no complex structure moduli i.e. $h^{2,1} = 0$, this is the only hypermultiplet that remains in the spectrum.

\subsection{Type IIB}

The bosonic sector of type IIB supergravity in $10$-dimensions is described by the following action \cite{Polchinski:1998cam} :
\bea
\begin{split}
S_{\rm IIA} &= \frac12\, \int\[e^{-2\hat{\phi}}\(\hat{R}\ast 1 + 4 \de \hat{\phi} \wedge \ast\hat{\phi}-\frac12\, \hat{H_3}\wedge \ast \hat{H_3}\)
\right.
\\ & \left. \qquad
- \(\hat{F_1} \wedge \ast\hat{F_1} + \hat{F_3} \wedge \ast\hat{F_3} + \frac12\, \hat{F_3} \wedge \ast\hat{F_3}\)
- \hat{A_4}\wedge\de \hat{H_3}\wedge \de \hat{A_2} \],
\end{split}
\eea
and
\bea
\begin{split}
& \hat{H_3} = \de \hat{B_2}, \qquad \hat{F_1} = \de \hat{A_0},
\\
& \hat{F_3} = \de \hat{A_2} - \hat{A_0} \wedge \hat{H_3}
\qquad \hat{F_5} = \de \hat{A_4} - \frac12\, \hat{A_2}\wedge \hat{H_3} + \frac12\, \hat{B_2} \wedge \de\hat{A_2}.
\end{split}
\eea
Let us find out the situation after compactification.
\begin{itemize}

\item
In the NS-NS sector
\bea
\begin{split}
& (\hat{g}_{XY})_{\rm 10d} \mapsto \(g_{\mu\nu}, t^i,z^a\)_{\rm 4d} \mapsto \(g_{\mu\nu},t^i,z^a\)_{\rm 4d-dual},
\\ &
(\hat{B_2})_{\rm 10d} \mapsto \(B_2 + b^i\omega_i\)_{\rm 4d} \mapsto \(\psi, b^i\)_{\rm 4d-dual},
\\ &
\hat{\phi}_{\rm 10d} \mapsto \phi_{\rm 4d} \mapsto \phi_{\rm 4d-dual}.
\end{split}
\eea

\item Then for the RR sector,
\bea
\begin{split}
& (\hat{A_0})_{\rm 10d} \mapsto (c^0)_{\rm 4d} \mapsto (c^0)_{\rm 4d-dual},
\\ &
(\hat{A_2})_{\rm 10d} \mapsto \(A_2 + c^i\omega_i\)_{\rm 4d} \mapsto \(c_0,c^i\)_{\rm 4d-dual}
\\ &
(\hat{A_4})_{\rm 10d} \mapsto \(D^i_2 \omega_i + \tilde{D}_i \omega^i + A^\lambda_1 \alpha_\Lambda + \tilde{A}_{1,\Lambda}\beta^\Lambda \)_{\rm 4d}
\mapsto \(c_i,A^0_1,A^a_1\)_{\rm 4d-dual} 
\end{split}
\eea
\end{itemize}
Due to the self-duality condition on $F_5$,  the scalars $\tilde{D}_i$ and vectors $\tilde{A}_{1,\Lambda}$ are dual to the two forms $D^i_2$ and 
the vectors $A^\Lambda_a$. Thus they are redundant fields not contributing to the spectrum. 
The fields arrange as in multiplets below :
\begin{itemize}

\item gravitational multiplet                $(g_{\mu\nu},A^0_1)$,
\item double tensor multiplet              $(B_2,A_2,\phi,c^0)$,
\item $h^{1,1}$ hypermultiplets        $(t^i,b^i,c^i,D^i_2)$ ,
\item $h^{2,1}$ vector multiplets      $(A_1^a, z^a)$.
\end{itemize}

As above, the double tensor multiplet and $h^{1,1}$ tensor multiplets can be dualized to $h^{1,1} + 1$ hypermultiplets and we have the final form of the spectrum.

\vspace{2 mm}

Here, we would like to emphasize that the above conditions on the number of vector multiplets (VM) and hypermultiplets (HM) in type IIA and type IIB
theories are in the perfect agreement with the statement of mirror symmetry, extension of which when quantum corrections are present is one of the principal
focus of the entire work. Given a mirror pair of Calabi-Yau threefolds $(\CY,\CYm)$ the Hodge numbers are related as the following,
\bea
h^{1,1}(\CY) = h^{2,1} (\CYm), \,\,\,\, h^{2,1} (\CY) = h^{1,1} (\CYm).
\eea
This means that under mirror symmetry the numbers of the vector multiplets and the hypermultiplets of type IIA and type IIB theories are exchanged to each other
if the compactification manifolds are chosen to be mirror of one another.

\section{Low energy effective actions}

When one restricts to the lowest order of $\alpha'$-corrections, the relevant terms in the LEEA should be considered upto
two derivatives terms only. In this approximation the VM and HM stay decoupled from each other and supersymmetry puts severe restrictions
on the form of the action. It becomes
\bea
S = \frac12\,\int\, R \ast 1 + S_{\text{VM}} + S_{\text{HM}}.
\eea
The form of $S_{\text{VM}}$ is well known, see for example \cite{Cecotti:1989qn}. It is as the following,
\bea
S_{\text{VM}} = \int \[-\frac12\, \Im \cN_{IJ} F^I \wedge \ast F^J + \frac12\, \Re \cN_{IJ} F^I \wedge F^J + \cK_{i\bar{j}} (v,\bar{v}) \de v^i \wedge \de \bar{v}^{\bar{j}} \].
\eea
The graviphoton field strength $F^0$ has been combined with the other gauge field strengths to $F^I$'s. $N=2$ supersymmetry dictates the forms of the matrices
$\cK_{i\bar{j}}$ and $\cN_{IJ}$. Both of them are determined by a holomorphic prepotential on the special K\"ahler manifold spanned by the moduli $v^i$. 
The former is just the metric on this space and is given by the derivatives of the K\"ahler potential. The latter is defined as
\bea
\cN_{IJ} (v,\bar{v}) = \bF_{IJ} - \I \frac{N_{IK} N_{JL} v^K v^L}{N_{XY} v^X v^Y} ,
\label{periodrel}
\eea
where $N_{IJ} = \I (F_{IJ} - \bF_{IJ}),\, v^I =(1,v^i)$. The matrix $\cN_{IJ}$ has a physical interpretation as a matrix of gauge couplings.
Moreover unlike $N_{IJ}$, which has a Lorentzian signature, the matrix $\cN_{IJ}$ is negative definite and hence the kinetic term is positive definite,
as it must be.

Next we turn to the HM part of the action. It takes the form of a non-linear $\sigma$-model,
\bea
S_{\text{HM}} = \int \de^4 x g_{\alpha\beta} (q) \p_\mu q^\alpha \p^\mu q^\beta.
\eea
Here $\alpha = 1,...,4n_H$, where $n_H$ is the number of HM's and $q^\alpha$ denotes all of the scalars.
Again $g_{\alpha\beta}(q)$ can not be arbitrary, $N=2$ supersymmetry restricts it to be quaternion K\"ahler (QK) \cite{Bag:1983np}.
We will discuss about these types of geometries in the next chapter.

This shows that the LEEA of Calabi-Yau compactifications of type II string theories is completely determined by the geometry
of the VM and HM moduli spaces. Moreover at the two derivative level the full moduli space factorizes ,
\bea
\cM_{4\text{d}} = \cM_{\text{VM}} \times \cM_{\text{HM}},
\eea
where $\cM_{\rm VM}$ and $\cM_{\rm HM}$ are vector multiplet and hypermultiplet moduli spaces, respectively.
Since the above statements are consequences of $N=2$ supersymmetry, they should continue
to hold even in presence of all quantum corrections. As mentioned earlier, the former is a special K\"ahler (SK) manifold and
the latter is a QK manifold.

The holomorphic prepotential describing the vector multiplet moduli space is tree-level exact, more precisely, it does not receive 
any correction in $g_s$ and can be explicitly computed. In particular, for the type IIB theory, where the vector multiplet moduli space
coincides with the space of complex structure deformations $\cM_{\rm VM} = \cM_C(\CYm)$, the prepotential
is also independent of the $\alpha'$-corrections. Therefore in this case the VM's obtained by Kaluza-Klein reduction is exact and hence is determined by \eqref{cxKahler}.

On the other hand, the $\cM_{\rm VM}$ in the type IIA side coincides with the moduli space of complexified K\"ahler class deformations $\cM_{\rm VM} = \cM_K(\CY)$.
However, the prepotential \eqref{Kprepot} is only the large volume approximation. Its the correct form can be derived by exploiting mirror symmetry.
In particular, since the moduli spaces of type IIA and type IIB theories should coincide provided they are compactified on a mirror pair of Calabi-Yaus, we have 
\bea
\cM^{A/B}_{\text{VM}}(\CY) = \cM^{B/A}_{\text{VM}}(\CYm), \,\,\,\,\,
\cM^{A/B}_{\text{HM}}(\CY) = \cM^{B/A}_{\text{HM}}(\CYm).
\eea
Therefore, the prepotential on the type IIA side can be computed from the exact prepotential on the type IIB side. The result of such computation turns out to be
\bea
\begin{split}
F(X) &= -\kappa_{ijk}\,\frac{X^iX^jX^k}{6X^0} + A_{IJ} X^I X^J + \chi_{\CY} \frac{\zeta(3) (X^0)^2}{2(2\pi\I)^3}
\\
&  - \frac{(X^0)^2}{(2\pi\I)^3}  \sum_{k_i\gamma^i \in H_2^+(\CY)} \, n^{(0)}_{k_i} \Li_3 (e^{2\pi\I k_i X^i/X^0}).
\label{vectIIAprepot}
\end{split}
\eea
The last two terms arise due to $\alpha'$-corrections and can not be derived by Kaluza-Klein reductions. The term involving
$\zeta(3)$ comes from perturbative corrections due to $\alpha'$ and the last term comes from corrections due to worldsheet
instantons. The $\Li_3$ arises due to multicovering effects and $n^{(0)}_{k_i}$ are genus zero Gopakumar-Vafa (GV) invariants.
The procedure is described in \cite{Hosono:1993qy,Candelas:1990rm}. The second term which is a quadratic one with real symmetric matrix $A_{IJ}$, is usually omitted
in the literatures because it does not affect the K\"ahler potential. However in later stages it will be crucial for our discussion and we include it from the very beginning.

The vector multiplet side is completely under control once the prepotential is determined, because the moduli space is a special K\'ahler
manifolds, as had been mentioned before. On the other hand, the situation for the hypermultiplets is not as simple. $\cM_{\rm HM}$ can
have corrections in $g_s$ \cite{Becker:1995kb} and furthermore, the QK geometry is much more complicated. In chapter \ref{chapter3}, we will address this issue
using twistor techniques.   

\section{Hypermultiplet moduli space}

This section is devoted to description of the classical HM moduli space and its symmetries.
We will also describe how the quantum corrections break the continuous symmetries to discrete ones.

\subsection{Classical hypermultiplet metric}

$\cM_{\rm HM}$ always contains field parametrizing either  $\cM_C$ or $\cM_K$ depending on whether we are
on type IIA or type IIB side respectively. 
As a consequence of classical mirror symmetry the metrics of both sides should coincide. 
It is actually known that to every projective (local) special K\"ahler
manifold one can associate a quaternion K\"ahler manifold \cite{Cecotti:1989qn,Ferrara:1989ik}, so that the latter are still described only by
prepotential. 
This construction realizes the classical metric on $\cM_{\rm HM}$.
The procedure is known as c-map and correspondingly the metric \eqref{treemet} is said to be in the image of it. 

We now turn to understanding the physical origin of this c-map. Let us consider type IIA and type IIB theories compactified on the
same Calabi-Yau threefold times a circle. T-duality along the circle implies that
\bea
{\rm IIA}/(\CY\times S^1_R) \cong {\rm IIB}/(\CY \times S^1_{1/R}).
\label{Tduali}
\eea 
On the other hand, their low energy descriptions can be obtained if one compactifies the four dimensional theory
from the previous subsection on a circle. This leads to two massless scalars corresponding to each of the gauge fields.
One of them comes from the component of the gauge field along the circle and the other appears by dualization of the
remaining three dimensional component of that gauge field. In particular, it means that a vector multiplet in four dimensions 
leads to a hypermultiplet in $3d$ additionally to the hypermultiplet associated with the metric. the corresponding three dimensional
effective action is then a non-linear $\sigma$-model whose target space is a product of two QK spaces.
 For either of type IIA or IIB theories the full three dimensional
moduli space is given by,
\bea
\cM_{\rm 3d} = \cQ_C(\CY) \times \cQ_K(\CY),
\eea 
which are extensions of complex structure and complexified K\'ahler class moduli spaces of the Calabi-Yau, respectively. 
One of the factor come from the HM moduli space $\cM_{\rm HM}$ and the other comes from vector and gravitational
multiplet in the way which we have just described above. Then \eqref{Tduali} simply exchanges the two factors. 

At the classical level, where one ignores instanton corrections coming from the winding modes of the circle, the 
Kaluza-Klein reduction is sufficient. Then it is possible to find the metric on $\cM^{\rm A/B}_{\rm HM}$ from the metric on $\cM^{\rm B/A}_{\rm VM}$,
explaining why the hypermultiplet moduli spaces are still given by the same holomorphic prepotentials.  

In the following we present the c-map metric using the fields adapted to the type IIA side only. To find the metric on the type IIB side, what
one needs is a simple application of classical mirror symmetry which we will explain elaborately in chapter 5. The result on the type IIA side
reads as follows,
\bea
\begin{split}
\de s^2 & = \frac{\de r^2}{r^2} - \frac{1}{2r} (\Im \cN)^{\Lambda\Sigma} \(\de\tzeta_\Lambda - \cN_{\Lambda\Lambda'}\de \zeta^{\Lambda'}\)
 \(\de\tzeta_\Sigma - \cN_{\Sigma\Sigma'}\de \zeta^{\Sigma'}\)
\\ &
+ \frac{1}{16r^2} \(\de\sigma +\tzeta_\Lambda\de\zeta^\Lambda - \zeta^\Lambda\de\tzeta_\Lambda\)^2 + 4\cK_{a \bar{b}} \de z^a \de \bz^{\bar{b}},
\label{treemet}
\end{split}
\eea
where $r= e^\phi \propto 1/g^2_{(4)}$ and the K\"ahler potential is given in terms of \eqref{cxKahler}.

Let us present the topological structure of the space described by the classical metric presented above. On top of 
the complex structure moduli space$\cM_C(\CY)$ parametrized by $z^a$, there is a torus bundle on the intermediate Jacobian,
$\cJ_C(\CY)$ with the fiber parametrized by the RR scalars, as $T_{\zeta,\tzeta} = H^{3}(\CY,\IR)/H^3(\CY,\IZ)$. 
One can choose Griffiths or Weil complex structure on the intermediate Jacobian, where the periods are related by \eqref{periodrel}.
On top of it, there is still a circle bundle parametrized by the NS-axion $\sigma$ and the dilaton appears as an overall factor. The structure then is that of 
a two-staged fibration,
\be
\label{doublefib2}
\IR^+_r \ \times\ \(\begin{array}{rc}
S^{1}_\sigma\ \longrightarrow &\cC(r)
 \\
&\downarrow
\\
\cT_{\zeta,\tzeta}\ \longrightarrow &\cJ_c(\CY)
\\
&\downarrow
\\
&  \cM_C(\CY)
\end{array}\).
\ee

\subsection{Symmetries of the classical metric}

Before we embark on to the study of the quantum corrections to the hypermultiplet moduli space, it is important to 
ponder over the symmetries at the classical level. In the absence of explicit microscopic computations, these symmetries are the
only tools that allow us to access the quantum corrected metric. At this stage, let us remind the reader for convenience the field content in the hypermultiplet 
sectors of the cases for type IIA and type IIB theories.
\bea
\begin{split}
& {\rm Type \, IIA} : z^a,\phi,\sigma \,\,{\rm and}\,\, \zeta^\Lambda,\tzeta_\Lambda.
\\ &
{\rm Type \, IIB}    : v^i \equiv b^i + \I t^i, \phi,\psi \,\,{\rm and} \,\,c^0,c^i,c_i,c_0.
\end{split}
\eea

\begin{itemize}
\item Symplectic invariance

Type IIA construction has manifest symplectic invariance because all the four dimensional fields are defined with respect to
a symplectic basis of 3-cycles $(\cA^\Lambda,\cB_\Lambda)$. Although \eqref{treemet} is written down when a particular
choice of such cycles is made, there is no canonical way of doing it. It is always possible to perform a symplectic
rotation by a matrix $\cO \in Sp(2h^{2,1} +2,\IZ)$ and as the physical results must not be dependent on the symplectic marking,
in particular the metric, they should be manifestly invariant under such transformations.
The symplectic rotation acts as the following on all symplectic vectors,
\bea
\cO : \begin{pmatrix} \cA^\Lambda \\ \cB_\Lambda \end{pmatrix}
\,\, \mapsto \,\, \begin{pmatrix} {\bf{a}} & {\bf{b}} \\ {\bf{c}} & {\bf{d}} \end{pmatrix}
\,   \begin{pmatrix} \cA^\Lambda \\ \cB_\Lambda \end{pmatrix} ,
\eea
where ${\bf{a,b,c,d}}$ are $(h^{2,1}+1) \times (h^{2,1}+1)$ matrices obeying
\bea
{\bf a}^T {\bf c} - {\bf c}^T {\bf a} = 0 = {\bf b}^T {\bf d} - {\bf d}^T {\bf b},
\qquad
{\bf a}^T {\bf d} - {\bf c}^T {\bf b} = 1.
\eea

The scalars $r$ and $\sigma$ are invariant under symplectic transformation. 
To see that the metric is symplectically invariant, let us start by considering the second term.
Now noticing \cite{deWit:1995jd},
\be
\begin{split}
\cN\ \mapsto&\ ({\bf{c}}+{\bf{d}}\cN)({\bf{a}}+{\bf{b}}\cN)^{-1},
\\
\Im \cN\ \mapsto&\ ({\bf{a}}+{\bf{b}}\cN)^{-T}\Im \cN({\bf{a}}+{\bf{b}}\bar\cN)^{-1},
\end{split}
\label{transfN}
\ee
we find that this term is indeed invariant.

The transformation of $z^a$ is induced by symplectic transformation of the vector $(X^\Lambda,F_\Lambda)$,
built from the homogeneous coordinate and the prepotential. This fact allows one verify explicitly that the last term
in  \eqref{treemet} is invariant under symplectic transformations. Altogether, these ensure that the metric is symplectically 
invariant.

\item S-duality

Moreover the type IIB string theory is invariant under S-duality \cite{Hull:1995np}.
In absence of $\alpha'$ or $g_s$-corrections at large volume and small string coupling limit, this 
symmetry is realized by continuous action of $SL(2,\IR)$ group. 
Its action on the coordinates adapted to the type IIB side (for manifest invariance under $SL(2,\IR)$) is,
\be\label{SL2Z}
SL(2,\IR)\ni \gl{}\ :\quad
\begin{array}{c}
\displaystyle{
\tau \mapsto \frac{a \tau +b}{c \tau + d} \, ,
\qquad
t^a \mapsto t^a |c\tau+d| \, ,
\qquad
c_a\mapsto c_a \,  ,}
\\
\displaystyle{
\begin{pmatrix} c^a \\ b^a \end{pmatrix} \mapsto
\begin{pmatrix} a & b \\ c & d  \end{pmatrix}
\begin{pmatrix} c^a \\ b^a \end{pmatrix}\, ,
\qquad
\begin{pmatrix} c_0 \\ \psi \end{pmatrix} \mapsto
\begin{pmatrix} d & -c \\ -b & a  \end{pmatrix}
\begin{pmatrix} c_0 \\ \psi \end{pmatrix},}
\end{array}
\ee
with $ad-bc=1$,
where the relations between fields in type IIA and type IIB are given by\footnote{When we use $a$ to label the indices of type IIB fields, we tacitly assume that
we are working on the mirror Calabi-Yau, so that $h^{1,1}(\CYm) = h^{2,1}(\CY)$, and hence
is not a contradictory notation.},
\be
\label{symptobd}
\begin{split}
z^a & =b^a+\I t^a\, ,
\qquad\ \
\zeta^0=\tau_1\, ,
\qquad\
\zeta^a = - (c^a - \tau_1 b^a)\, ,
\\
\tzeta_a &=  c_a+ \frac{1}{2}\, \kappa_{abc} \,b^b (c^c - \tau_1 b^c)\, ,
\qquad
\tzeta_0 = c_0-\frac{1}{6}\, \kappa_{abc} \,b^a b^b (c^c-\tau_1 b^c)\, ,
\\
\sigma &= -2 (\psi+\frac12  \tau_1 c_0) + c_a (c^a - \tau_1 b^a)
-\frac{1}{6}\,\kappa_{abc} \, b^a c^b (c^c - \tau_1 b^c)\, ,
\end{split}
\ee

However, after incorporation of quantum corrections one expects it
to be broken down to a discrete subgroup of SL(2,$\IR$).
We will get back to its discussion in chapter \ref{chapter5}, and it will be our guiding principle for
deriving results for the ``fivebrane instantons".

\item Heisenberg symmetry

The RR scalars and the NS scalar in the metric \eqref{treemet} are actually obtained
as either periods or duals to gauge fields present in ten dimensional superstring theories. This is
the origin of the Heisenberg symmetry, as it is a consequence of the gauge symmetries.
\cite{dw:1992np} discusses this situation and they are examples of the so called Peccei-Quinn symmetries. Its action
 is most conveniently realized in terms of type IIA fields :
\bea
T_{H,\kappa} : (\zeta^\Lambda,\tzeta_\Lambda,\sigma) \, \mapsto
(\zeta^\Lambda + \eta^\Lambda, \tzeta_\Lambda + \tilde{\eta}_\Lambda, \sigma + 2\kappa
- \tilde{\eta}_\Lambda \zeta^\Lambda + \eta^\Lambda \tzeta_\Lambda).
\label{Heis0}
\eea
Furthermore, the composition of two such transformations is given by
\bea
T_{H_1,\kappa_1}\cdot T_{H_2,\kappa_2} =T_{H_1+H_2, \kappa_1+\kappa_2 +\frac12\, \<H_1,H_2\>}.
\eea
Here $H=(\eta^\Lambda,\tilde{\eta}_\Lambda)$ is a symplectic vector and
the symplectic invariant inner product is given by
$\<H,H'\> = \tilde{\eta}_\Lambda {\eta'}^\Lambda - \eta^\Lambda {\tilde\eta}'_\Lambda .$
At this stage $H \in \IR^{b_3}$ and $\kappa \in \IR$. In the following section we will discuss how
non-perturbative quantum corrections due to instantons break them down to integers.

\item Symmetry around the large volume point

There is another kind of Peccei-Quinn symmetry which can be easily understood from the type IIB side.
When the scalars $b^a$ dual to the two form field $B_2$ is in the hypermultipler sector, this symmetry
is inherited from the gauge symmetry of the latter. 
In the prepotential \eqref{vectIIAprepot} one can actually drop the last two terms in the large volume limit, where $\alpha'$-corrections are
absent. In that case it is a continuous symmetry, whereas in presence of $\alpha'$-corrections it becomes discrete.
Furthermore such shift should be accompanied by transformation of the RR scalars as the following,
\be
\label{bjacr}
M_{\epsilon^a}\ :\quad  \begin{array}{c}
\displaystyle{b^a\mapsto b^a+\epsilon^a\, ,
\!\!\!\!\!\!\!\!\qquad
\zeta^a\mapsto \zeta^a + \epsilon^a \zeta^0\, ,
\!\!\!\!\!\!\!\!\qquad
\tzeta_a\mapsto \tzeta_a -\kappa_{abc}\zeta^b \epsilon^c
-\frac12\,\kappa_{abc} \epsilon^b \epsilon^c \zeta^0\, ,}
\\
\displaystyle{\tzeta_0\mapsto \tzeta_0 -\tzeta_a \epsilon^a+\frac12\, \kappa_{abc}\zeta^a \epsilon^b \epsilon^c
+\frac16\,\kappa_{abc} \epsilon^a \epsilon^b \epsilon^c \zeta^0\, .}
\end{array}
\ee

\item Mirror symmetry

As we have already seen, the moduli spaces of type IIA and type IIB sides are identical, provided
they are obtained by compactfication on mirror Calabi-Yau threefolds. These different facets of the same theory
are useful to deal with various types of symmetries which are most naturally realized in one or the other. For example,
whereas S-duality is a manifest symmetry in the type IIB side, type IIA formulation is ideally suited for symplectic invariance.
The advantage of mirror symmetry is that it allows to pass from one to the other.

\end{itemize}

\section{Qualitative nature of quantum correction}

In string theory there are two sources of quantum corrections to the results obtained by Kaluza-Klein reduction
: they can come from two coupling constants, $\alpha'$ and $g_s$, and can be both perturbative and non-perturbative. 
The $\alpha'$-corrections can appear only where the K\"ahler moduli present,
for example this is the reason why type IIA vector multiplets receive such corrections. Due to the c-map the situation is reversed for the hypermultiplet sector.
Whereas $\cM_{\rm HM}$ is $\alpha'$-exact, it receives corrections from them on the type IIB side. However, for both of these cases
in the lowest order in $g_s$
the metric is determined completely by the prepotential by \eqref{treemet}.

The $g_s$ corrections appear in both $\cM_{\rm HM}^{\rm A}$ and $\cM_{\rm HM}^{\rm B}$
as the dilaton is a part of hypermultiplet in both type IIA and type IIB theories.
On the other hand for
the vector multiplets the metric is tree level exact in $g_s$. 
Below, we discuss qualitative effects of the $g_s$ corrections to the geometry of $\cM_{\rm HM}$ both perturbatively
and non-perturbatively for the case of hypermultiplets. 

\begin{itemize}

\item  {\bf Perturbative corrections}

Perturbative corrections should be obtained by expanding in powers of dilaton which is the counting parameter
for the string loop expansion. It is important that in the perturbative approximation the metric must retain the continuous Heisenberg 
symmetry \eqref{Heis0} \cite{Antoniadis:1997eg,Antoniadis:2003sw}..

\item {\bf Non-perturbative corrections}

Non-perturbative corrections emanate from Euclidean branes wrapping different non-contractible cycles of the Calabi-yau threefold \cite{Becker:1995kb}.
Such corrections are called instantons because these objects seem to be point-like objects from the four dimensional perspective. 
In the case of of string theory two of such objects generate instanton contributions, 

\begin{itemize}

\item
The first of them are so called D-instanton effects coming from D-branes. Physically they can
be interpreted as arising from D$p$-branes with $(p+1)$-dimensional worldvolume, wrapping appropriate cycles in the Calabi-Yau. For type IIA, $p$ is even, whereas for
type IIB $p$ is odd.

\vspace{ 1.5 mm}

 Since a Calabi-Yau threefold does not admit any non-trivial one or five cycle, there are only D2-instantons in the type IIA side wrapping
3-dimensional cycles in $\CY$. They are labeled by the charge vector $\gamma = (p^\Lambda,q_\Lambda)$, representing the cycle wrapped,
$q_\Lambda \cA^\Lambda + p^\Lambda \cB_\Lambda \in H_3(\CY)$. 

In the case of type IIB side all instantons wrapping different even-dimensional cycles are present. 
The ones that
come from wrapping $0$ and $2-$dimensional cycles are called D(-1) and D1 instantons. Through mirror symmetry they are related to the A-D2 instantons ($q_\Lambda \ne 0,
p^\Lambda  = 0$ coming from D-branes wrapping $\cA$-cycles in type IIA side).
The other two D3 and D5 instantons are mapped to the B-D2 instantons (coming from D-branes wrapping $\cB$-cycles in type IIA) with non-vanishing $p^a$ and $p^0$ respectively.

Schematically the D-instanton corrections take the following forms  \cite{Becker:1995kb},
\be
\label{d2quali}
\delta \de s^2\vert_{\text{D-inst}} \sim \,\Omega(\gamma;z)\,
 e^{ -2\pi|Z_\gamma|/g_s
- 2\pi\I (q_\Lambda \zeta^\Lambda-p^\Lambda\tzeta_\Lambda)} ,
\ee
where it depends on some moduli dependent functions giving strengths to the instantons, known as central charges $Z_\gamma$
and are given by
\bea
Z_\gamma = q_\Lambda z^\Lambda - p^\Lambda F_\Lambda.
\eea
The BPS indices $\Omega(\gamma;z)$ are some moduli
and charge dependent coefficients. They give rise to the instanton measures and it is well known that
they are related to the counting states for the case of black holes. 
They are piecewise constant function on the moduli $z^a$, where the jumps occur across codimension one 
surfaces, the so called walls of marginal stability about which we will discuss in chapter 4. 

It is clear that the RR-fields ($\zeta^\Lambda,\tzeta_\Lambda$) , in the D-instanton action break the continuous Heisenberg symmetries \eqref{Heis0}
to discrete ones. After incorporation of all sorts of D-instantons all shift symmetries along the RR-fields become discrete. For the type IIB side,
the effect is similar, the theta-angles are now the fields $c^0, c^i, c_0,c_i$.

\item
There is another non-perturbative obejact called NS5-branes, that are magnetic dual to fundamental strings. Since they have six dimensional
worldvolume they can wrap the entire Calabi-Yau manifold. They are responsible for the breaking down of the final continuous symmetry which is
continuous translation along NS axion $\sigma$. 
Schematically, their action is given by, 
\be
\label{ns5quali}
\delta \de s^2\vert_{\text{NS5}} \sim \,
 e^{ -2\pi|k| \cV/g_s^2- \pi\I k\sigma} .
\ee

Taking all non-perturbative effects into account, leads to the hypermultiplet moduli space devoid of all continuous isometries, while 
only a discrete subgroup of them is retained. 

\end{itemize}

\end{itemize}

\section{Present status }

Recent years have marked a lot of progress in the understanding the hypermultiplet metric for the type II string theories.
We presently have a neater description of QK geometries using twistor techniques, understanding which will be the main focus of the following chapter.
The inclusion of D-instantons on type IIA side consistently with the symplectic invariance had been already achieved
in \cite{Alexandrov:2008gh,Alexandrov:2009zh}. The D(-1) and D1 instantons on the type IIB side had also been accounted for in
\cite{Alexandrov:2009qq}. Furthermore initial steps of including NS5 instantons have also been taken in \cite{Alexandrov:2010ca}, where
also several interesting and connected problems are addressed, for example some relation to topological strings. The question of D3-instantons on type IIB  has
also been attempted in \cite{Alexandrov:2012au} although its understanding still remains incomplete as the analysis has been done in the large volume limit and
only upto first order in instantons. Here we go beyond these results. 
In particular, now we have understood the NS5 instantons on the type IIB side mostly, barring a few subtleties,
\cite{Alexandrov:2014mfa,Alexandrov:2014rca}. Furthermore an explicit expression for the metric has been calculated in \cite{Alexandrov:2014sya} in
presence of A-D2 instantons on the type IIA side. We reiterate that all of these were achieved by invoking different symmetries of the type IIA and IIB formulations.
Before coming to these points in later chapters, we need to first review the twistor techniques for the QK manifolds, which we are going to do next.

\newpage

\chapter{Twistorial description of Quaternion K\"ahler manifolds}
\label{chapter3}


In the preceding chapter we found that compactifications of type II string theories on Calabi-Yau threefolds lead us
to study hypermultiplet moduli spaces of four dimensional $\cN=2$ supergravities. Moreover, we also saw that these
hypermultiplet moduli spaces are quaternion-K\"ahler manifolds. The physical problem that we want to address is describing
these hypermultiplet moduli spaces when all possible quantum corrections are present.
But it is not easy to deal with general QK manifolds.
The difficulty of handling them stems from the fact
that they are rather complicated geometric objects. This fact renders the problem of including instanton corrections directly at
the level of metric to be extremely difficult. Fortunately there is a way to circumvent this problem .

Let us recall why special K\"ahler manifolds are comparatively easier geometric objects.
The reason is that there exists a holomorphic prepotential which
completely  encodes their geometry.
In the case of QK manifolds we would also like to have such a nice, elegant and substantially more convenient description.
The twistor construction that we are going to present in this chapter actually does the job.
It turns out that such construction leads to similar objects through which the geometry of QK space can be completely encoded.
In the following we are going to recapitulate the essence of these techniques and we will find out how certain holomorphic transition functions on the twistor space
will serve our purpose.

\section{Quaternion K\"ahler manifolds and their twistor spaces}

Our discussion closely follows \cite{Alexandrov:2008nk} and the review \cite{Alexandrov:2011va}.
A QK space $\cM$ is a $4n$ real dimensional Riemannian manifold with the holonomy group contained in $USp(n) \times SU(2)$ \cite{MR1327157}.
The $n=1$ case is the simplest and special in the sense that the QK manifold is then defined as a self-dual Einstein manifold.

The QK space carries a triplet of non-integrable almost complex structures $J^i$ satisfying the quaternionic algebra. They define the
quaternionic two-forms by $\omega^i(X,Y) = g(J^i X,Y)$. These two forms are covariantly closed with respect to the $SU(2)$  part
$\vec{p}$ of the Levi-Civita connection and are proportional to the curvature of $\vec{p}$,
\bea
\de \vec{\omega} + \vec{p}\times\vec{\omega} = 0 ,
\qquad
\de \vec{p} + \frac12 \vec{p}\times\vec{p} = \frac{\nu}{2} \vec{\omega},
\label{Kformdef}
\eea
where $\nu$ is related to the Ricci curvature $\nu = \frac{R}{4n(n+2)}$.
The major problem with such spaces is that they are not even complex manifolds. However,
it is still possible to describe the geometry of QK spaces analytically by passing to its twistor space.

\vspace{2 mm}

The twistor space $\cZ$ of the QK manifold $\cM$ is the total space of a $\IC P^1$ bundle on $\cM$.
This $\IC P^1$ is the sphere defined by the triplet of almost complex structures.
In contrast to the QK space, this twistor space is a complex manifold.
Furthermore it carries an integrable canonical complex contact structure given by the kernel of the $\cO(2)$-twisted
(1,0)-form
\bea
\De t = \de t + p_+ - \I p_3 t + p_- t^2,
\eea
where $p_{\pm}= -\frac12\, (p_1 \mp \I p_2)$ and $t$ is the stereographic coordinate of $\IC P^1$.

There is a more convenient way to describe this complex contact structure via a holomorphic one-form
$\cX$, known as contact one-form. Locally there exists a function, the so called contact potential $\Phi^{[i]}$,
that allows us to construct it from the $(1,0)$-form $\De t$,
\bea
\cX^{[i]} = \frac{4}{\I t} \,e^{\Phi^{[i]}} \,\De t,
\label{defcontpot}
\eea
where the index indicates the fact that we are working only locally in a patch $\cU_i \subset \cZ$.
This contact potential is a quantity that will play a very important role in future. For instance
it determines the K\"ahler potential on $\cZ$ by
\bea
\cK_{\cZ}^{[i]} = \, \log \frac{1+t\bar{t}}{|t|}  +  \Re \Phi^{[i]}.
\eea

The advantage of the contact one-form $\cX^{[i]}$ becomes manifest when one introduces holomorphic Darboux coordinates.
Namely, it is always possible to choose complex coordinates $(\xii{i}^\Lambda,\txii{i}_\Lambda,\ai{i})$ on a patch $\cU_i$ on the
twistor space such that $\cX^{[i]}$ can be locally trivialized as \cite{Neitzke:2007gh, Alexandrov:2008nk}
\bea
\cX^{[i]} = \de \ai{i} + \xii{i}^\Lambda \, \de \txii{i}_\Lambda.
\label{one-form}
\eea
The contact structure is completely determined by the transformations relating the Darboux coordinates in the intersection
of two different patches $\cU_i \cap \cU_j$. These ``contactomorphisms" should preserve the contact one-form up to a
holomorphic factor,
\bea
\cX^{[j]} = \hat{f}^2_{[ij]} \cX^{[i]}.
\label{Gluingcont1form}
\eea

The contactomorphisms defined above can be parametrized by the transition functions $\Hij{ij} \in H^1(\cZ,\cO(2))$, which
are holomorphic functions of Darboux coordinates $\xii{i}^\Lambda$ on one patch and $\txii{j}_\Lambda,\ai{j}$ on the other.
The gluing conditions that relate Darboux coordinates on two different patches turn out to be
\bea
\begin{split}
\xii{j}^\Lambda & = \xii{i}^\Lambda - \p_{\txii{j}_\Lambda} \Hij{ij} + \xii{j}^\Lambda \p_{\ai{j}} \Hij{ij},
\\
\txii{j}_\Lambda & = \txii{i}_\Lambda + \p_{\xii{i}^\Lambda} \Hij{ij},
\\
\ai{j} & = \ai{i} + \Hij{ij} - \xii{i}^\Lambda \p_{\xii{i}^\Lambda} \Hij{ij},
\label{Gluingcond}
\end{split}
\eea
using which the holomorphic rescaling factors in \eqref{Gluingcont1form} becomes
\bea
\hat{f}^2_{[ij]}  = 1- \p_{\ai{j}} \Hij{ij},
\eea
and is completely determined by $\Hij{ij}$.

These gluing conditions appended with appropriate regularity conditions can be rewritten as a set of integral equations for the Darboux coordinates
\cite{Alexandrov:2008nk}. The Darboux coordinates here are functions
of the QK coordinates $(x^\mu)$ as well as the $\IC P^1$-fiber coordinate $t$. We require that the coordinate $\xii{i}^\Lambda$ has only simple poles at $t=0,\infty$,
and allow for the coordinate $\ai{i}$ to have logarithmic singularities at $t=0,\infty$ .\footnote{Actually one can allow for logarithmic singularities for both $\txii{i}_\Lambda$
and $\ai{i}$. They give rise to the so called ``anomalous dimensions" $c_\Lambda$ and $c_\alpha$  \cite{Alexandrov:2008nk}. In what follows, $c_\Lambda$'s are not
relevant for our considerations. But $c_\alpha$ is, because it encodes the one loop correction in $g_s$.}
\bea
\begin{split}
\xii{i}^\Lambda (x^\mu,t) & = A^\Lambda + t^{-1} Y^\Lambda - t \bY^\Lambda - \frac12\, {\sum_{j}} \oint_{C_j} \frac{\de t'}{2\pi\I t'} \, \frac{t'+t}{t'-t}\, \(\p_{\txii{j}_\Lambda} \Hij{ij} -
 \xii{j}^\Lambda \p_{\ai{j}} \Hij{ij}\),
\\
\txii{i}_\Lambda(x^\mu,t) & = B_\Lambda + \frac12\, {\sum_{j}} \oint_{C_j} \frac{\de t'}{2\pi\I t'} \, \frac{t'+t}{t'-t}\, \p_{\xii{i}^\Lambda} \Hij{ij},
\\
\ai{i}(x^\mu,t) & = B_\alpha - 2\I c_\alpha \log t + \frac12\, {\sum_{j}} \oint_{C_j} \frac{\de t'}{2\pi\I t'} \, \frac{t'+t}{t'-t}\, \(\Hij{ij} - \xii{i}^\Lambda \p_{\xii{i}^\Lambda} \Hij{ij}\).
\label{Darbouxinteq}
\end{split}
\eea

Here $t \in \hat{\cU}^{[i]}$, where $ \hat{\cU}^{[i]}$ is the projection of $ \cU^{[i]}$ on the $\IC P^1$
and $C_i$'s are contours that surround $ \hat{\cU}^{[i]}$. The complex variables $Y^\Lambda$ and the real variables $A^\Lambda,B_\Lambda,B_\alpha$ play
the roles on coordinates of the QK manifold.
Counting the number of parameters we find $4n+1$, where $n$ is the quaternionic dimension of $\cM$.
They contain one parameter extra because the overall phase rotation of $Y^\Lambda$ can be absorbed by redefining the $\IC P^1$ coordinate $t$.
In mathematical parlance the procedure of solving these integral equations is known as ``parametrizing the twistor lines".
There is an important thing to notice in the above integral equations : the index $i$ inside the integral can be replaced with any other index for any other
patch in the twistor space as the transition functions satisfy certain cocycle conditions. This fact is important in computations presented later.

\vspace{2mm}

Finally one can find the gluing conditions for the contact potential,
\bea
\Phi^{[i]} - \Phi^{[j]} = \log (1-\p_{\ai{j}} \Hij{ij}).
\eea
This can be solved to find  \cite{Alexandrov:2008nk}
\bea
\Phi^{[i]} = \phi -  \frac12\, {\sum_{j}} \oint_{C_j} \frac{\de t'}{2\pi\I t'} \,\log (1-\p_{\ai{j}} \Hij{ij}),
\label{contpot}
\eea
where the constant part is
\bea
e^\phi = e^{\Re\[ \Phi^{[+]} (t=0) \]}
= \frac{\frac{1}{8\pi}  {\sum_{j}} \oint_{C_j} \frac{\de t}{t} \, \( t^{-1} Y^\Lambda - t \bY^\Lambda\)\, \p_{\xii{i}^\Lambda} \Hij{ij} + c_\alpha}
{2\cos\[\frac{1}{4\pi}  {\sum_{j}} \oint_{C_j} \frac{\de t}{t}\,  \log (1-\p_{\ai{j}} \Hij{ij})\]},
\eea
and the index $[+]$ denotes the patch surrounding the north pole ($t=0$) on $\IC P^1$.

One more important fact to note that all the quantities such as contact one-form, Darboux coordinates, transition functions and contact potential satisfy reality conditions
under the combined action of complex conjugation and the antipodal map,
$\varsigma[t] = -1/\bar{t}$, provided that the covering of $\cZ$ by patches $\{\cU_i\}$ remains invariant, i.e. $\cU_i$ is mapped to $\cU_{\bar{i}}$ and back.
In particular, the transition functions satisfy
\bea
\overline{\varsigma[{\Hij{ij}}]} = \Hij{\bar{i}\bar{j}}.
\eea

\section{Contact bracket formalism}

There is a further simplification if one introduces certain objects called ``contact hamiltonians". They were introduced in
\cite{Alexandrov:2014mfa,Alexandrov:2014rca}. The difficulty of parametrizing the contact transformations with usual
transition functions arises from the fact that $\Hij{ij}$ are functions of Darboux coordinates in two different patches.
Due to this the apparently simple gluing conditions may be generated by complicated transition functions. In particular
when one discusses symmetries this creates a big obstacle. Typically such symmetries  define some action locally at each
patch $\cU_i$ on the twistor space by their action on the Darboux coordinates. These actions can take a highly
nonlinear form if one tries to realize such symmetries in terms of $\Hij{ij}$. We will see examples of such situations later.

Before, we introduce the contact hamiltonian,  we would like to recapitulate certain facts about contact geometry,
the relevant geometry for the twistor spaces. Let us consider a contact vector field $X_h$ associated with a function $h$.
It is determined by
\bea
\iota_{X_h} \de \cX = -\de h + R(h) \cX, \qquad
\iota_{X_h} \cX = h.
\label{contact-def}
\eea
Here $\iota_X$ denotes the contraction with a vector field and $R$ is the so called Reeb vector determined as the unique
element of the kernel of $\de \cX$ satisfying $\cX(R) = 1$.

Furthermore, we define the contact bracket which is an extension of Poisson bracket in the realm of contact geometry. Its definition for two
objects $\mu_1 \in \cO(2m)$ and $\mu_2 \in \cO(2n)$ is as a map to $\cO(2(m+n-1))$ given by \cite{Alexandrov:2008gh},
\be
\label{poisson}
\begin{split}
\{ \mu_1, \mu_2 \}_{m,n}= &\,
\pa_{\xi^\Lambda}  \mu_1 \p_{\txi_\Lambda}  \mu_2 +
\(m \mu_1  -\xi^\Lambda \p_{\xi^\Lambda} \mu_1\)\p_\alpha\mu_2
\\
&\, - \p_{\xi^\Lambda}  \mu_2 \p_{\txi_\Lambda}  \mu_1
-\(n \mu_2-\xi^\Lambda \p_{\xi^\Lambda} \mu_2\) \p_\alpha \mu_1 .
\end{split}
\ee
From the definition the skew-symmetry is obvious, in particular
\bea
\{\mu,\mu\}_{m,n} = (m-n) \mu\p_\alpha \mu.
\label{sameskew}
\eea
\eqref{poisson} also satisfies the Leibnitz rule and the Jacobi identity.
For some section $\mu_3 \in \cO(2k)$ the Leibnitz rule becomes
\bea
\{\mu_1\mu_2,\mu_3\}_{m+n,k} = \mu_1\{\mu_2,\mu_3\}_{n,k} + \mu_2 \{\mu_1,\mu_3\}_{m,k}.
\label{skew-symm}
\eea
The Jacobi identity turns out to be the following relation,
\bea
\begin{split}
&\{\mu_1,\{\mu_2,\mu_3\}_{n,k}\}_{m,n+k-1} - \{\mu_2,\{\mu_1,\mu_3\}_{m,k}\}_{n,m+k-1}
\\ & \,\,\,\,\,\,\,\,
- \{\{\mu_1,\mu_2\}_{m,n},\mu_3\}_{m+n-1,k} =0.
\label{Jacobi}
\end{split}
\eea

Since we will mostly need the case of $\{,\}_{1,0}$ below, to reduce the cumbersomeness of the notation we will drop the indices
if no confusion arises. It provides the action of
a vector field $X_{\mu_1}$ with the (generalized) moment map $\mu_1$ on a local complex function $\mu_2$ \cite{MR872143}.
In particular given such $\mu_1$ and $\mu_2$, the contact bracket is defined by the action of the vector field associated to 
$\mu_1$ as 
\bea
\{\mu_1,\mu_2\} = X_{\mu_1} \mu_2.
\eea
Setting $\mu_1=\hH$ and $\mu_2$ to be one of the Darboux coordinates, one explicitly finds
\be
\begin{split}
\{\hH,\xi^\Lambda\}=&\, -\p_{\txi_\Lambda} \hH+\xi^\Lambda\p_\alpha \hH,
\qquad
\{\hH,\txi_\Lambda\}=\p_{\xi^\Lambda} \hH,
\\
&\qquad
\{\hH,\alpha\}=\hH-\xi^\Lambda\p_{\xi^\Lambda} \hH.
\end{split}
\label{contbr}
\ee

If the same function $h$ is used to represent sections of two different bundles $\cO(2)$ and $\cO(0)$, one finds
that the bracket is not antisymmetric, the fact which follows from \eqref{sameskew}. Instead it satisfies,
\bea
\{h,h\} =h\p_\alpha h =  h R(h),
\eea
which can be obtained by applying $\iota_{X_h}$ on the first equation of \eqref{contact-def} and
of course, without loosing generality we can choose the Reeb vector field $R(h) = \p_\alpha h$.

A crucial property of the contact bracket following from its definition is that it generates infinitesimal
contact transformation. This means
\bea
\cL_{X_h} \cX = \de \iota_{X_h} \cX + \iota_{X_h} \de \cX = R(h) \cX.
\eea
Thus exponentiating the action of the contact bracket one generates a finite contact transformation.
This implies that the gluing conditions can be rewritten as,
\bea
\Xi^{[j]} = e^{\{\hHij{ij},\cdot\}} \Xi^{[i]},
\eea
for some function $\hHij{ij}$.

The action on each of the Darboux coordinates is given by
\be
\begin{split}
\xii{j}^\Lambda = &\,e^{\{\hHij{ij},\, \cdot\,\}}\,\xii{i}^\Lambda=\, \xii{i}^\Lambda-\p_{\txii{j}_\Lambda }\Hij{ij}
+\xii{j}^\Lambda \, \p_{\ai{j} }\Hij{ij}  ,
\\
\txii{j}_\Lambda = &\,e^{\{\hHij{ij},\, \cdot\,\}}\,\txii{i}_\Lambda = \,
\txii{i}_\Lambda+ \p_{\xii{i}^\Lambda } \Hij{ij} ,
\\
\ai{j} = &\,e^{\{\hHij{ij},\, \cdot\,\}}\,\ai{i}=\, \ai{i}+
\Hij{ij}- \xii{i}^\Lambda \p_{\xii{i}^\Lambda}\Hij{ij}.
\label{gluing-contbr}
\end{split}
\ee
From the above we can also find an expression for the usual transition function $\Hij{ij}$ in terms of $\hHij{ij}$ as
\be
\Hij{ij}=\(e^{\{\hHij{ij},\, \cdot\,\}}-1\)\ai{i}+\xii{i}^\Lambda\( e^{\{\hHij{ij},\, \cdot\,\}}-1\)\txii{i}_\Lambda.
\label{relHH}
\ee

The important point here is that rewriting the gluing conditions by \eqref{gluing-contbr} avoids the complication
of having to compute derivatives only after transferring to the patches $\cU_j$ for the Darboux coordinates $\txii{j}_\Lambda$ and $\ai{j}$,
which are arguments of $\Hij{ij}$. 
Indeed, if one uses the contact hamiltonians, the derivatives that enter the gluing conditions can be taken treating
the Darboux coordinates on the same patch as independent variables.
This simplification will play an important role when we will discuss QK manifolds with symmetries and their twistor spaces.
Also it is useful to note that in the case where $\Hij{ij}$ is dependent on $\xii{i}^\Lambda$ only
the contact hamiltonian $\hHij{ij}$ coincides with the transition function, $\Hij{ij} (\xii{i}^\Lambda) = \hHij{ij}(\xii{i}^\Lambda)$.

\section{Gauge transformation}

The transition functions are actually not uniquely determined. It is still possible to perform local contact transformations on each of the
patches of the twistor space $\cU_i \in \cZ$ \cite{Alexandrov:2008nk}.
Since these gauge transformations are contact transformations after all, the formalism
of the contact brackets can be utilized in this case as well\cite{Alexandrov:2014rca}. We will call the contact hamiltonians that generate these local contact
transformations as ``gauge transformation contact hamiltonians". A crucial difference between the transition functions that generate
contact transformations between two different patches and the generators of gauge transformations is that the latter has to be regular in
the patches on which they act ensuring that the regularity of Darboux coordinates on their patches of definition is not spoiled.
In the following, we denote the gauge transformation by $\gi{i}$ and as explained they are regular in $\cU_i$.

The effect of a general gauge transformation is captured by the following formula,
\be
\exp\(X_{\hHij{ij}_g}\)
= e^{-X_{\gi{i}} } \, \exp\Bigl(X_{\hHij{ij}} \Bigr)\,e^{X_{\gi{j}} },
\label{gaugetr}
\ee
and can be computed in principle by using the Baker-Campbell-Hausdorff formula.
Later we will find that the relevant gauge transformation contact hamiltonians take the same functional
form in each of the patches and will depend on only the Darboux coordinates $\xii{i}^\Lambda$, i.e. $\gi{i} = g(\xii{i})$.
The Jacobi identity \eqref{Jacobi} in this case implies
\be
\[X_{g},X_{h}\]=X_{\{g,h\}_{1,1}^{}}.
\label{commXXh}
\ee
Using this we can find explicitly how the contact hamiltonian $\hHij{ij}$ changes to $\hHij{ij}_g$ due the gauge tranformation induced by $\gi{i}$ as
\be
\hHij{ij}_{g}=e^{-\{ g,\,\, \cdot\,\, \}_{1,1}^{}}\cdot \hHij{ij},
\ee
Then the above entails into the following simple relation,
\be
\hHij{ij}_g=\hHij{ij}\(\xi^\Lambda\, ,\, \txi_\Lambda - \p_{\xi^\Lambda} g \, ,\, \alpha-g+\xi^\Lambda\p_{\xi^\Lambda} g \),
\label{gaugehHxi}
\ee
affecting the contact hamiltonian $\hHij{ij}$ by only linear shifts in the arguments.

\section{QK manifold with symmetries }

In this section we will discuss the general procedure for dealing with QK manifolds with symmetries.
As was mentioned in the previous chapter, it is not possible to perform detailed microscopic computations
explicitly in many occasions. In such situations the only guiding principles are the symmetries of the metric on the
moduli space. Remarkably, the symmetries turn out to be sufficient to determine the non-perturbative structure of $\cM_{\rm{ HM}}$
entirely. More precisely, in terms of some Calabi-Yau data such as Donaldson-Thomas invariants, intersection numbers etc.
the symmetries impose strong enough requirements from which in principle the metric is computable.

For having better understanding of such symmetries we will again use the twistor formulation.
It is well known that all symmetries of QK spaces can be lifted to their twistor spaces
\cite{MR1327157}. Apart from the symmetries there is now an extra
advantage of holomorphicity whose virtue will be extensively used later.

The lift of the symmetries to the twistor space $\cZ$ is defined by the action on the Darboux coordinates.
Naturally, since the transition functions depend on the Darboux coordinates, they can not be arbitrary 
in presence of such symmetries. Instead the defining action on the Darboux coordinates will give rise to 
certain constraints on the transition functions. Furthermore for a proper lift, the metric on the twistor space 
should also remain invariant under the symmetry. This means that the stereographic coordinate $t$ must also
transform in a specific way. Indeed, this is the case because if we look at the integral equations for the Darboux
coordinates, we find that there are explicit dependences on $t$. If the lift is consistent, one should be able to rederive
the transformation properties of the Darboux coordinates, provided the transformation of the fiber coordinate. 
Finally, since the transition functions are associated to the patches covering the twistor space, constraints on the
transition functions also fix how the patches should reshuffle. More precisely, under the action of symmetries, 
the transition functions and the patches to which they are associated with, transform in such a way that the covering of $\cZ$
remains invariant. At this stage, it is also important to note that the symmetries render the contact structure on $\cZ$ invariant. In particular,
the contact one-form gets only rescaled by a holomorphic factor under the action of symmetries. Let us investigate the effect of this a bit elaborately. 

Detailed discussions of these lifting problems already exist in the literatures \cite{Alexandrov:2008gh,Alexandrov:2010ca,Alexandrov:2012au,Alexandrov:2012bu,
Alexandrov:2013mha,MR872143}. As was remarked before that crucial simplifications can be achieved if one formulates construction exploiting the contact
bracket formulation, we will focus on this only here following \cite{Alexandrov:2014mfa,Alexandrov:2014rca}.
We want to find how the contact bracket, \eqref{contbr} behaves under contact transformation $\vrh : \cX \mapsto \lambda \cX$, where
$\lambda$ is a holomorphic factor.
For this we will prove the following,
\be
\vrh \cdot X_{h}=X_{\lambda^{-1} \vrh\cdot h}.
\label{transf}
\ee
Let us denote the transformed quantities by primes.
We have \eqref{contact-def} at our disposal. Instead of proving \eqref{transf}, we will prove the following,
\be
\iota_{X_{\lambda^{-1}h'}}\de\(\lambda \cX\)=-\de h'+ R'(h') \lambda\cX,
\qquad
\iota_{X_{\lambda^{-1}h'}}\(\lambda\cX\)=h'.
\label{defXh-prime}
\ee
To show \eqref{transf} it is enough to prove,
\bea
\begin{split}
\lambda R(\lambda^{-1} h')\cX
& = \lambda \iota_{X_{\lambda^{-1}h'}} \de \cX + \lambda \de (\lambda^{-1} h')
\\
& = \lambda \iota_{X_{\lambda^{-1}h'}} \de \cX + \de h' - (\iota_{X_{\lambda^{-1} h'}} \cX) \de \lambda,
\end{split}
\eea
from \eqref{contact-def}.
The second equation in \eqref{defXh-prime} is satisfied trivially. The first one can be rewritten using the above as the following,
\be
\begin{split}
&\,
\(\iota_{X_{\lambda^{-1}h'}}\de\lambda\) \cX-\(\iota_{X_{\lambda^{-1}h'}}\cX\) \de\lambda+\lambda\,\iota_{X_f{\lambda^{-1}h'}}\de\cX
+\de h'- \lambda  R'(h') \cX
\\
&\, =
\lambda\(\iota_{X_{\lambda^{-1}h'}}\de\log\lambda
+ R(\lambda^{-1}h')- R'(h')\) \cX=0.
\end{split}
\label{prop-to-prove}
\ee
Applying the map $\vrh$ to  \eqref{contact-def} we obtain
\be
\cX(R')=\lambda^{-1},
\qquad
\iota_{R'}\de\(\lambda\cX\)=\(\iota_{R'}\de\lambda\)\cX-\de\log\lambda+\lambda\,\iota_{R'}\de\cX=0.
\ee
Contracting the second equation with $X_{\lambda^{-1}h'}$ we get
\bea
\begin{split}
&\,
\lambda^{-1}h' R'(\lambda)-\iota_{X_{\lambda^{-1}h'}}\de\log\lambda-\lambda\,\iota_{R'}\iota_{X_{\lambda^{-1}h'}}\de\cX
\\
&\, =R'(h')-\iota_{X_{\lambda^{-1}h'}}\de\log\lambda-R(\lambda^{-1} h')=0,
\end{split}
\eea
completing the proof of \eqref{prop-to-prove}.

Thus we know how $\rho$ acts on the contact brackets from \eqref{transf}. It is given by
\be
\vrh\cdot \{h,f\}=\{ \lambda^{-1}\vrh\cdot h, \vrh\cdot f\}.
\label{trans-contact}
\ee
This property generalizes the familiar invariance of the Poisson bracket under canonical transformations
to the realm of contact geometry. The usefulness of of the property in \eqref{trans-contact} will be revealed
in due course of the thesis, in particular, when we have to deal with S-duality symmetry on the type IIB side.

\section{Evaluation of the metric on the QK manifolds}

\label{compmet}

This section is devoted towards explaining the procedure of computing the metric on QK space
starting from the data on its twistor space. The discussion is along the lines of \cite{Alexandrov:2008nk, Alexandrov:2014sya}.
In the next chapter we present the computation of an explicit expression for the metric to all orders in instanton expansion
for a special case.

Since later we will derive the metric in the case where the transition functions $\Hij{ij}$ are
independent of the coordinate $\alpha$, in the following formulas we drop this dependence from the very beginning.
For this case moreover the contact potential becomes globally well defined.
Our interest lies in the case when the curvature $\nu$ is negative.
Hence we set $\nu = -1$, as it can be adjusted by changing the value of the cosmological constant affecting the metric by an overall
factor only.

First one needs to expand the Darboux coordinates around the north pole in the $\IC P^1$,
\be
\begin{split}
\xii{+}^\Lambda=&\,\xii{+}^{\Lambda,-1}\varpi^{-1}+\xii{+}^{\Lambda,0}+O(\varpi),
\\
\txii{+}_\Lambda=&\,\txii{+}_{\Lambda,0}+O(\varpi),
\\
\ai{+}=&\,-2\I c_\alpha\log \varpi+\ai{+}_{0}+O(\varpi),
\end{split}
\label{expDc}
\ee
From \eqref{defcontpot} we can compute the Levi-Civita part of the $SU(2)$ connection,
\be
\begin{split}
p^+ &=\frac{\I}{4}\, e^{-\phi}
\, \xi^{\Lambda,-1}_{[+]}  \de\txi^{[+]}_{\Lambda,0},
\\
p^3 &= -\frac{1}{4}\, e^{-\phi} \left( \de\alpha^{[+]}_0 +
\xi^{\Lambda,0}_{[+]}  \de\txi^{[+]}_{\Lambda,0} +
\xi^{\Lambda,-1}_{[+]}  \de\txi^{[+]}_{\Lambda,1}  \right) ,
\end{split}
\label{connection}
\ee
from which we determine the 2-form $\omega^3$ by the following,
\be
\omega^3 = -2{\rm d} p^3+ 4\I  p^+ \wedge p^-.
\label{Kform}
\ee
The next aim is to reexpress this $\omega^3$ in terms of a basis of $(1,0)$-forms. Such a basis was provided
in \cite{Alexandrov:2008nk, Alexandrov:2014sya} and reads as
\be
\label{defPi}
\pi^a =\de \(\xii{+}^{a,-1}/\xii{+}^{0,-1}\) ,
\qquad
\tilde\pi_\Lambda= \de\txii{+}_{\Lambda,0}  ,
\qquad
\tilde\pi_\alpha = \frac{1}{2\I}\,\de\ai{+}_0 +2c \,\de\log\xii{+}^{0,-1}.
\ee
This basis specifies the almost complex structure $J^3$.
Finally the metric can be computed from  $g(X,Y) = \omega^3(X,J^3 Y)$.
To do this in practice, one should rewrite $\omega^3$, computed by \eqref{Kform} in terms of differentials of (generically real)
coordinates on $\cM$.
Then in terms of the basis $\pi^X=(\pi^a,\tilde\pi_\Lambda,\tilde\pi_\alpha)$ given in \eqref{defPi}. The final result should look like
\be
\omega^3=\I g_{X\bY} \pi^X\wedge \bar\pi^{Y},
\label{metom}
\ee
which is explicitly of (1,1) Dolbeault type.
From \eqref{metom} the metric readily follows as $\de s^2 =2 g_{X\bY} \pi^X \otimes \bar\pi^{Y}$.
In the next chapter we will discuss implementation of this procedure explicitly in some special cases.

\newpage

\chapter{Quantum corrections to the hypermultiplet moduli space and D-instantons}
\label{chapter 4}


In this chapter we initiate the analysis of including quantum corrections to the hypermultiplet moduli space $\cM_{\rm{ HM}}$. As was
mentioned, these quantum corrections can be of two types, perturbative and non-perturbative.
First, we will find out how $\cM_{\rm{HM}}$ gets deformed by inclusion of perturbative corrections in string coupling constant $g_s$. Then
we will proceed on to finding  its description when D-instantons are present restricting to the type IIA formulation. We will do this by exploiting
the twistor formalism that has been discussed in chapter \ref{chapter3}. After that we will compute an explicit expression
for the metric when a particular subset of the D-instantons is present \cite{Alexandrov:2014sya}. We will also consider in detail
the four dimensional moduli space resulting from the compactification  performed on a rigid Calabi-Yau manifold and its relation with Toda equation.
We will finish this chapter by discussing what happens to the curvature singularity appearing in the perturbative metric
\eqref{1lmetric}, in the presence of D-instantons.

\section{One loop corrected metric}

At the one-loop level the metric on $\cM_{\rm HM}$ is well known  \cite{Antoniadis:1997eg,Gunther:1998sc,Antoniadis:2003sw,Rocek:2006xb,Alexandrov:2007ec}.
It has also been reproduced using twistorial analysis in \cite{Alexandrov:2008nk}. 
The result is,
\be
\begin{split}
\de s^2 =&\, \frac{r+2c}{r^2(r+c)}\, \de r^2
-\frac{1}{r} \(N^{\Lambda\Sigma} - \frac{2(r+c)}{rK}\, z^\Lambda \bz^\Sigma\) \(\de \tzeta_\Lambda - F_{\Lambda\Lambda'} \de \zeta^{\Lambda'}\)
\(\de\tzeta_\Sigma -\bF_{\Sigma\Sigma'} \de \zeta^{\Sigma'}\)
\\
&\,
+\frac{r+c}{16 r^2(r+2c)} \(\de\sigma + \tzeta_\Lambda \de \zeta^\Lambda - \zeta^\Lambda \de\tzeta_\Lambda + 8c \cA_K\)^2
+ \frac{4(r+c)}{r}\, \cK_{a\bar{b}} \de z^a \de \bz^b,
\end{split}
\label{1lmetric}
\ee
where as was introduced before, $N_{\Lambda\Sigma} = -2\Im F_{\Lambda\Sigma}$ is an invertible matrix with indefinite signature ($N^{\Lambda\Sigma}$ being the inverse) on the special K\"ahler base parametrized by $z^a$, $\cA_K$ is the K\"ahler connection  $\cA_K = \frac{\I}{2} \(\cK_a \de z^a - \cK_{\bar{a}} \de \bz^{\bar{a}}\)$.
The one-loop correction is parametrized by $c$ which is $c = -\frac{\chi_{\CY}}{192\pi}$,
where $\chi_{\CY} =  2\(h^{1,1}(\CY) - h^{2,1}(\CY) \)$ is the Euler class for the Calabi-Yau threefold $\CY$. Topologically the one-loop deformed metric still has the structure of \eqref{doublefib2}.

It is well known that the perturbative metric does not allow corrections higher than one loop in $g_s$ . The evidence
for it can be understood by analysis of the structure of \eqref{1lmetric}.
We observe that the one-loop  parameter $c$ corrects various terms which depend on the dilaton $r$ as well as the connection of the $\sigma$
circle bundle by introducing a term proportional to the K\"ahler connection $\cA_K$,
\bea
\De \sigma = \de \sigma + \tzeta_\Lambda \de \zeta^\Lambda - \zeta^\Lambda \de \tzeta_\Lambda + 8c \cA_K.
\eea
The curvature of the circle bundle of the axion $\sigma$ in  \eqref{doublefib2}
takes the following form
\bea
\de (\frac12\, \De\sigma) = \omega_{\cT} - \frac{\chi_{\CY}}{48\pi} \de \cA_K,
\label{curvpert}
\eea
where $\omega_{\cT} = \de\tzeta_\Lambda \wedge \de\zeta^\Lambda$ is the curvature form of
the torus $\cT_{\zeta,\tzeta} = H^3(\CY,\IR)/H^3(\CY,\IZ)$ parametrizing the fiber of the intermediate Jacobian $\cJ_c(\CY)$.
Perturbative corrections beyond one-loop will presumably
lead to presence of positive power terms in $r$ in the connection of the $\sigma$ circle bundle conflicting with the quantization condition on
the first Chern class \cite{Alexandrov:2010ca}. In particular, this argument indicates why higher loop corrections than one-loop are forbidden.

\subsection{Twistor description of one-loop corrected HM moduli space}

Let us now give the twistorial construction of the one-loop corrected hypermultiplet moduli space described by the metric \eqref{1lmetric}.
As we have already learned in the previous chapter, 
it is sufficient to provide a covering and the set of transition functions associated to it. In practice,
we will only work locally in the moduli space. This means that we should only consider the covering of the $\IC P^1$ fiber and
the Darboux coordinates can develop singularities only in $t$.

Thus we have to consider the Riemann sphere of the $\IC P^1$. Let us cover it three patches $\cU_+,\cU_-$ and $\cU_0$, where
the first two cover the north pole and the south pole respectively and $\cU_0$ covers the equatorial region. Clearly one requires two independent
transition functions for the corresponding intersections of $\cU_0$ with $\cU_{\pm}$. It turns out that if we choose the following transition functions,
\bea
\Hij{+0} = F(\xi), \qquad \Hij{0-} = \bF(\xi),
\eea
one recovers the expression for the classical metric \eqref{treemet}. To incorporate one-loop correction, one should add the anomalous dimensions 
$c_\alpha = -2c$. Then it is straightforward to recover \eqref{1lmetric} following the procedure presented in \ref{compmet}.

This is an example of toric QK manifold (that has $n+1$ commuting isometries) for which the Darboux coordinate $\xi^\Lambda$ is globally well-defined. For this reason
we abstain from writing the patch index for them in the above equation.

We also write down the Darboux coordinates on the patch $\cU_0$,
\bea
\begin{split}
\xii{0}^\Lambda &= \zeta^\Lambda + \cR(t^{-1} z^\Lambda - t \bz^\Lambda),
\\
\txii{0}_\Lambda &= \tzeta_\Lambda + \cR(t^{-1} F_\Lambda (z) - t \bF_\Lambda (\bz)),
\\
\tai{0} &= \sigma + \cR (t^{-1} W (z) - t \bW (\bz)) - 8\I c \log t,
\label{1lDarboux}
\end{split}
\eea
where $\tai{0} = -2\ai{0} - \xii{0}^\Lambda \txii{0}_\Lambda$ is a symplectic invariant object constructed from $\ai{0}$,
and
\bea
W(z) = F_\Lambda(z) \zeta^\Lambda - z^\Lambda \tzeta_\Lambda.
\eea
Furthermore $\cR$ is related to the dilaton. More precisely, the dilaton is identified with the contact potential which for this
case is globally well-defined as we are working with a toric QK manifold,
\bea
e^{\Phi} = e^\phi = \frac{\cR^2}{4} K(z,\bz) - c.
\eea

\section{D-instantons in twistor space}

The metric \eqref{1lmetric} given in the previous section provides the complete description of perturbatively corrected
hypermultiplet moduli space. The next step is to include instantons. We know already from chapter \ref{chapter2} that
they can be of the two types : D-instantons and NS5-instantons. The latter are exponentially suppressed compared to the former 
(we assume that we are working in the region of the moduli space where the string coupling is small)
and will be ignored in this chapter. Besides, in this section we will be working in the type IIA
side. 

The construction of the twistor space in presence of D-instantons had been found in \cite{Alexandrov:2008gh,Alexandrov:2009zh}.
 Since we are working on the type IIA side, the D-instantons can possibly wrap
only non-contractible three cycles. It should be a special Lagrangian submanifold in the homology class
$q_\Lambda \gamma^\Lambda - p^\Lambda \gamma_\Lambda \in H_3(\CY,\IZ)$, where the vectors $\gamma = (p^\Lambda,q_\Lambda)$
are the charges of the D-branes associated with them. Our goal here is to describe corrections that appear to $\cM_{\rm{HM}}$ owing to the
presence of such BPS objects. Our construction crucially hinges on the techniques developed in the previous chapter. Introduction of the D-instantons
can be viewed as a refinement of the covering of the twistor space compared to the one-loop corrected case. Once the final description is obtained, it
is in principle possible to compute the metric on the D-instanton corrected HM moduli space following the procedure in \ref{compmet}.

\vspace{2mm}

So the first task is to describe this new covering. To begin let us fix particular value for the complex structure
moduli $z^a$. Then the phase of the central charge function can be used to determine the direction of the so called BPS rays $\ell_\gamma$ which
are lines on the $\IC P^1$ that run from the south to the north pole defined by,
\bea
\ell_\gamma = \{t : Z_\gamma/{\I t} \in \I \IR^{-}\}.
\eea
For each $\ell_\gamma$ there is an anti-brane BPS ray corresponding to the charge $-\gamma$.
$\ell_{-\gamma}$ can be obtained by taking antipodal image of the former.  Together they form a circle which divides the equatorial
patch $\cU_0$ into two parts. Introduction of all possible D-instantons then amounts to subdividing $\cU_0$ by introducing a countably infinite
number of such BPS rays indexed by their associated charges.

After this one needs to give the transition functions for these rays. They are
given  by
\bea
\Hij{ij} = H_\gamma (\Xi^{[i]}_\gamma) - \frac12\, q_\Lambda p^\Lambda (H'_\gamma(\Xi^{[i]}_\gamma))^2,
\label{Dinst-trans}
\eea
where 
\bea
\label{Dinsttrans}
H_\gamma\(\Xi_\gamma\)
= \frac{\Omega(\gamma)}{4\pi^2}\,\Li_2\(\sigma_D(\gamma )\, e^{-2\pi\I \Xi_\gamma}\) ,
\eea
and $\Xi^{[i]}_\gamma = q_\Lambda\xii{i}^\Lambda - p^\Lambda\txii{i}_\Lambda$, $\sigma_D(\gamma)$ and
$\Omega(\gamma)$ are the so called quadratic refinements and the Donaldson-Thomas invariant respectively,
 which we define below. The quadratic refinement is a phase assignment
to the charge lattice $\Gamma$ required for the consistency with the wall-crossing which we shall discuss shortly.
It should satisfy the following relation,
\bea
\label{qrrel}
\sigma_D(\gamma + \gamma') = (-1)^{\langle\gamma,\gamma'\rangle} \sigma_D(\gamma)\sigma_D(\gamma'),
\eea
where $\langle,\rangle$ denotes the standard symplectic pairing. The most general solution to
the above condition reads as
\bea
\label{qrD}
\sigma_D(\gamma) = \exp \[2\pi\I \(-\frac12\, q_\Lambda p^\Lambda + q_\Lambda \theta^\Lambda_D - p^\Lambda \phi^D_\Lambda\)\],
\eea
where $(\theta^\Lambda_D,\phi_\Lambda^D)$ are the so called characteristics. If they are half integers, the quadratic refinement
becomes just a sign factor.
The numerical factors $\Omega(\gamma)$'s are some topological invariants associated to the Calabi-Yau manifolds. When the magnetic
charges are absent, i.e. $p^\Lambda = 0$, they reduce to the genus zero Gopakumar-Vafa invariants that are present in the in the instanton part of the
holomorphic prepotential in \eqref{vectIIAprepot},
\bea
\Omega(0,q_\Lambda) = n^{(0)}_{q_a}, \qquad
\Omega(0,0,q_0) = \chi_{\CY}.
\eea

One should also worry about the transition functions associated to the patches that surround the north and south poles $\cU_{\pm}$,
\bea
\Hij{0+}= F(\xii{+}) + G_\gamma, \qquad
\Hij{0-} = \bF(\xii{-}) - G_\gamma.
\eea
The explicit expression of $G_\gamma$ can be found in \cite{Alexandrov:2009zh}.
Here we just want to mention that $G_\gamma$ has two branch cuts from $t=0$ and $t=\infty$ to the two roots of $\Xi^{[i]}_\gamma(t) = 0$,
such that the jumps in the Darboux coordinates get cancelled by the ones in $\Hij{0+}$ and $\Hij{-0}$ leading to regular Darboux coordinates in $[\pm]$ patches.

At this stage we would like to show an instance of the virtue of the contact bracket formulation that was described in the previous chapter.
The D-instanton transition function in \eqref{Dinst-trans} can be reproduced using \eqref{relHH} if one chooses the contact hamiltonians to be
\be
\hHij{\gamma}(\xi,\txi)=H_{\gamma}(\Xi_\gamma).
\label{Hgam}
\ee
Notice that already for this relatively simpler case we see that we can get rid of the non-linear term in \eqref{Dinst-trans} using our new formalism
of contact brackets.

We also write down the expression for the Darboux coordinates  here as they will be required for computation
of the metric which we present later,
\be
\label{exline}
\begin{split}
\xii{\gamma}^\Lambda =&\, \zeta^\Lambda + \cR \left(
\varpi^{-1} z^\Lambda - \varpi \, \bz^\Lambda\right) +
\frac{1}{8\pi^2}\sum_{\gamma'} \Om{\gamma'} p'^\Lambda \Iggp(\varpi) ,
\\
\txii{\gamma}_\Lambda =&\,
\tzeta_\Lambda
+\cR \left( \varpi^{-1} F_\Lambda - \varpi \, \bF_\Lambda \right)
+\frac{1}{8\pi^2}\sum_{\gamma'} \Om{\gamma'} q'_{\Lambda}\Iggp(\varpi)  ,
\\
\ai{\gamma}=&\,4\I c \log \varpi
-\hf\,\sigma-\frac{ \cR}{2}\,\(t^{-1} W-t \bar W\)
+\frac{\cR}{16\pi^2}\sum_{\gamma'} \Om{\gamma'}\(\varpi^{-1}Z_{\gamma'}+\varpi\bar{Z}_{\gamma'} \)\Iggp(0)
\\
&\,
-\frac{\I}{16\pi^3} \sum_{\gamma'} \Om{\gamma'} \int_{\ellg{\gamma'}}\frac{\d \varpi'}{\varpi'}\,
\frac{\varpi+\varpi'}{\varpi-\varpi'} \,L_{\qrp}\(e^{-2\pi \I \Xigi(\varpi')}\)
-\hf\, \xii{\gamma}^\Lambda(\varpi)\, \txii{\gamma}_\Lambda(\varpi),
\end{split}
\ee
where
\be
\label{defWnonThkl}
W(z) \equiv  F_\Lambda(z) \zeta^\Lambda - z^\Lambda \tzeta_\Lambda
\ee
and we introduced two functions\footnote{The first function is a variant of the Roger dilogarithm which satisfies
the famous {\it pentagon identity} and plays an important role in integrability \cite{Zagier-dilog}.}
\be
\begin{split}
L_\epsilon(z) =&\,  \Li_2 (\epsilon z) + \frac12\, \log z \log (1-\epsilon z),
\\
\Igg(\varpi)=&\, \int_{\ellg{\gamma}}\frac{\d \varpi'}{\varpi'}\,
\frac{\varpi+\varpi'}{\varpi-\varpi'}\,
\log\(1-\qr e^{-2\pi \I \Xigi(\varpi')}\).
\end{split}
\label{newfun}
\ee

\subsection{Relations to wall-crossing phenomena}

Till this point we confined our attention to a fixed value of the complex structure moduli. Let us now vary it slowly across the
moduli space. As a result, the central charges $Z_\gamma(z)$ corresponding to the charges change and consequently at some point
for two different charges they align (i.e. their phases become the same)  after which they
exchange their relative positions. But the contactomorphisms across such BPS rays do not commute provided the charges are mutually non-local,
$\langle\gamma,\gamma'\rangle \ne 0$. Thus it means that two different orders of BPS rays lead to two different solutions for the Darboux coordinates
leading to two different metrics for them. This is an apparent contradiction as the metric on the moduli space is supposed to be smooth everywhere
from physical expectations.

To reconcile with the continuity of the metric, one must take into consideration the wall-crossing phenomena. Indeed, in the context of gauge theories and supergravity
it is well-known that there exist codimension 1 lines of marginal stability (LMS) in the moduli space \cite{Ferrari:1996sv,Denef:2000nb}. Across the LMS the spectrum of BPS states jumps discontinuously.
This means that crossing the wall either the BPS states decay or recombine to form a bound state. For example, it is known that the distance between 
the centers of a two-centerd
black hole is given by \cite{Denef:2000nb}
\bea
r_{12} = \frac12\, \frac{\langle\gamma_1,\gamma_2\rangle\, |Z_{\gamma_1}+Z_{\gamma_2}|}{\Im (Z_{\gamma_1}\bZ_{\gamma_2})}.
\eea
The presence of the denominator implies that the moduli space is split into two chambers by the codimension 1 LMS for which central
charges of $\gamma_1$ and $\gamma_2$ align. These two regions correspond to the cases $r_{12} > 0$ and $r_{12} < 0$. Since a negative distance is
meaningless physically, it indicates that the two-centered solution is unstable and it decays.

Thus we see that LMS appears when the central charges align. In the language of twistor construction this means that the BPS rays $\ell_\gamma$'s align.
Furthermore, the information about the single particle spectrum is contained in the BPS indices $\Omega(\gamma)$ and they appear explicitly in the transition functions.
They also depend on the moduli $z^I$, but since they are piecewise constant we drop this dependence if no confusion arises. 
It turns out that to take jumps of the BPS indices into account is important. Because this the only possibility through which we can cancel the discontinuity arising from reordering of the BPS rays.
That this is indeed the case has been established in \cite{Gaiotto:2008cd}.

Note that the operators $e^{X_{\hHij{\gamma}}} = e^{\{\hHij{\gamma},\cdot\}}$ generating the contact transformations induced by \eqref{Hgam}
are nothing else but a lift to the contact geometry
of the Kontsevich-Soibelman (KS) operators
$U_\gamma^{\Omega(\gamma)}$
satisfying the wall crossing formula \cite{ks}. It dictates how the DT invariants
change after crossing a wall of marginal stability in the  moduli space of the complex structure deformations parametrized by $z^a$
and ensures the smoothness of the moduli space metric across the walls \cite{Gaiotto:2008cd}.
Provided $\Gamma(z)$ is a set of charges for which $Z_\gamma(z)$ become aligned at point $z^a$ and
$\Omega^\pm(\gamma)$ are the  DT invariants on the two sides of the wall,
the KS formula states that
\be
\label{ewall}
\prod^\ccwarrow_{\gamma\in \Gamma(z)}
U_{\gamma}^{\Omega^-(\gamma)}
=
\prod^\cwarrow_{\gamma\in \Gamma(z)}
U_{\gamma}^{\Omega^+(\gamma)},
\ee
where the two products are taken in the opposite order. (In both cases the order corresponds to decreasing the phase of $Z_\gamma$
at a given point in the moduli space.) The operators $U_\gamma$ are defined as 
\bea
U_\gamma = \exp \(\sum_{n=1}^\infty \frac{e_{n\gamma}}{n^2}\),
\eea 
and 
$e_\gamma$ satisfies the algebra of the complex torus
\bea
\[e_\gamma,e_{\gamma'}\] = (-1)^{\langle\gamma,\gamma'\rangle} \langle\gamma,\gamma'\rangle e_{\gamma+\gamma'}.
\eea
Identifying $e_\gamma = \sigma_D(\gamma)e^{-2\pi I \Xi_\gamma}$, one recognizes in the operators $U_\gamma^{\Omega(\gamma)}$ as the 
exponentiated transition functions $H_\gamma$ and the adjoint action on $e_{\gamma'}$ generates the symplectomorphisms. 
Actually for this chapter the above discussion suffices as the D-instantons induce symplectomorphism 
in the $(\xi,\txi)$ subspace.  
However, for the most general case, one needs to still extend the KS formula to the realm of contact geometry which was achieved in
 \cite{Alexandrov:2011ac}  and was done through use of certain dilogarithm identities. 

This brings out the fact that the smoothness of the moduli space $\cM_{\rm HM}$ crucially depends on the existence of
the lines of marginal stability. Across them the spectrum of BPS particles change and the way it does is dictated by the KS wall-crossing formula \eqref{ewall}.
The realization of this in the twistorial language then has the simple interpretation that crossing the wall leads to reshuffling of the contours on the $\IC P^1$
in such a way that discontinuity of the Darboux coordinates and the jumps of the BPS indices precisely cancel each other. This is actually a restatement of the KS formula
in our context.

\section{Explicit expression for the metric in the mutually local case}

In this section we will construct an explicit expression for the metric for the case of mutually local charges $\langle\gamma,\gamma'\rangle$.
In the most general case implementing the procedure of  section \ref{compmet} amounts to solving the integral equations for the Darboux coordinates
explicitly, which can be attained order by order only. For the case of D-instanton corrected metric 
non mutually local charges finding an expression that is valid for all orders in the instanton expansion is impossible
as the equations take the form of a set of TBA equations \cite{Alexandrov:2010pp}, which are known to be impossible to solve exactly. 
Due to this reason, we restrict to the case where the D-instanton charges satisfy $\langle \gamma,\gamma'\rangle = 0$. 

The first step is to extract the relevant Laurent coefficients for the Darboux coordinates,
\bea
\xii{+}^{\Lambda,-1} &=& \cR z^\Lambda,
\\
\xii{+}^{\Lambda,0}&=&\zeta^\Lambda-\frac{1}{8\pi^2}\sum\limits_{\gamma} \hng{} p^\Lambda \Igg{},
\\
\txii{+}_{\Lambda,0}&=&\tzeta_\Lambda-F_{\Lambda\Sigma}\zeta^\Sigma
-\frac{1}{8\pi^2}\sum\limits_{\gamma} \hng{} V_{\gamma\Lambda}\Igg{},
\label{Lexptxi}
\\
\txii{+}_{\Lambda,1}&=&-\I\cR\bz^\Sigma N_{\Lambda\Sigma}
-\frac{1}{2\cR}\, F_{\Lambda\Sigma\Theta}\zeta^\Sigma\zeta^\Theta
-\frac{1}{4\pi^2}\sum\limits_{\gamma} \hng{} \bigg[V_{\gamma\Lambda}\Igamp{\gamma}
\nn\\
&&
-\frac{1}{2\cR}\, F_{\Lambda\Sigma\Theta}p^\Sigma\zeta^\Theta\Igg{}
+\frac{1}{32\pi^2\cR}\, F_{\Lambda\Sigma\Theta}p^\Sigma\sum_{\gamma'} \Om{\gamma'} p'^\Theta \Igg{}\Igam{\gamma'}\bigg],
\\
\alpi{+}_{0}&=&
-\hf\(\sigma+\zeta^\Lambda\tzeta_\Lambda-F_{\Lambda\Sigma}\zeta^\Lambda\zeta^\Sigma\)+2\I\(r+c\)
\\ &&
-\frac{1}{8\pi^2}\sum\limits_{\gamma} \hng{} \[
\frac{1}{2\pi\I }\int_{\ellg{\gamma}}\frac{\d \varpi'}{\varpi'}\,
\Li_2\(\qr e^{-2\pi \I \Xigi{}(\varpi')}\)
\right.
\nn\\
&& \left.
- V_{\gamma\Lambda}\zeta^\Lambda \Igg{}
-\cR\Zg{}\Igp
+\frac{1}{16\pi^2}\, p^\Lambda\Igg{}\sum_{\gamma'}\Om{\gamma'} V_{\gamma'\Lambda}\Igam{\gamma'}\],
\label{LaurentDarboux}
\eea
where we used the following abbreviations for the integrals
\bea
&&
\Igg{} =
\int_{\ellg{\gamma}}\frac{\de \varpi}{\varpi}\,
\log\(1-\qr e^{-2\pi \I \Xi_\gamma(\varpi)}\),
\\ &&
\rIg=
 \int_{\ellg{\gamma}}\frac{\de \varpi}{\varpi}\,
\frac{1}{\qr e^{2\pi \I \Xi_\gamma(\varpi)}-1},
\\ &&
\Igpm  =
\pm\int_{\ellg{\gamma}}\frac{\de \varpi}{\varpi^{1\pm 1}}\,\log\(1-\qr e^{-2\pi \I \Xi_\gamma(\varpi)}\),
\\ &&
\rIgpm =
 \pm \int_{\ellg{\gamma}}\frac{\de \varpi}{\varpi^{1\pm 1}}\,
\frac{1}{\qr e^{2\pi \I \Xi_\gamma(\varpi)}-1}.
\label{newfun-expand}
\eea
We also introduced two useful notations $\Xi_\gamma(t) = q_\Lambda \zeta^\Lambda - p^\Lambda \tzeta_\Lambda + \cR(t^{-1} Z_\gamma - t \bZ_\gamma)
\equiv \Theta_\gamma + \cR(t^{-1} Z_\gamma - t \bZ_\gamma)$ and $V_{\gamma\Lambda} = q_\Lambda - F_{\Lambda\Sigma} p^\Sigma$.

\vspace{2mm}

Lastly the contact potential becomes
\be
r = \frac{\cR^2}{4}\, K(z,\bz)-c
-\frac{ \I \cR}{32\pi^2}\sum\limits_{\gamma} \hng
\int_{\ellg{\gamma}}\frac{\d \varpi}{\varpi}\,
\( \varpi^{-1}\Zg -\varpi\bZg\)
\log\(1-\qr e^{-2\pi \I \Xigi(\varpi)}\).
\label{phiinstmany}
\ee

\vspace{2mm}

Now let us construct the basis of one-forms using \eqref{defPi}. This basis can certainly be further simplified and are given by,
\be
\begin{split}
\de z^a, &
\\
\cY_\Lambda = &\,
\de\tzeta_\Lambda-F_{\Lambda\Sigma}\de\zeta^\Sigma
-\frac{1}{8\pi^2}\sum\limits_{\gamma} \hng{} \(q_{\Lambda}-p^\Sigma F_{\Lambda\Sigma}\)\de \Igg{},
\\
\Sigma=&\, \de r +2c\,\de \log\cR-\frac{\I}{16\pi^2}\sum\limits_{\gamma} \hng{} \(\cR\Zg{} \de\Igp-\Igm \de\(\cR\bZg{}\)\)
\\
&\,
+\frac{\I}{4}\(\de \sigma+\tzeta_\Lambda\de \zeta^\Lambda-\zeta^\Lambda\de \tzeta_\Lambda\).
\label{holforms}
\end{split}
\ee

As was mentioned the goal is to rewrite the quaternionic 2-form $\omega^3$ as \eqref{metom} using the $(1,0)$-forms in \eqref{holforms}.
For this we compute the $SU(2)$ parts of the Levi-Civita connection first. Using \eqref{connection} it is found to be
\be
\begin{split}
p^+ =&\,\frac{\I}{4r}
\left[ \cR z^\Lambda\(\de\tzeta_\Lambda-F_{\Lambda\Sigma}\de\zeta^\Sigma\)
-\frac{\cR}{8\pi^2}\sum\limits_{\gamma} \hng{} \Zg{}\de \Igg{}
\right] ,
\\
p^3 =&\, \frac{1}{8r} \left[ \de\sigma
+\tzeta_\Lambda\de \zeta^\Lambda-\zeta^\Lambda\de \tzeta_\Lambda
+2\cR^2 K\cA_K
-\frac{\cR}{4\pi^2}\sum\limits_{\gamma} \hng{} \(\Igp \de\Zg{}-\Igm \de\bZg{}\) \right].
\end{split}
\label{SU(2)connect}
\ee
Then $\omega^3$ becomes
\bea
\omega^3 &= & \frac{1}{4 r^2}\, \de r\wedge \[ \de\sigma
+\tzeta_\Lambda\de \zeta^\Lambda-\zeta^\Lambda\de \tzeta_\Lambda
-\frac{\cR}{4\pi^2}\sum\limits_{\gamma} \hng{} \(\Igp \de\Zg{}-\Igm \de\bZg{}\) \]
\nn\\
&&
+\frac{\cR^2 K}{2 r}\, \de\log\frac{ r}{\cR^2}\wedge\cA_K
+\frac{1}{2r}\(\de\zeta^\Lambda\wedge \de\tzeta_\Lambda
-\I\cR^2 N_{\Lambda\Sigma}\de z^\Lambda \wedge\de\bz^\Sigma
+\frac{\I\cR^2}{2r}\,z^\Lambda\bz^\Sigma\cY_\Lambda\wedge\bar\cY_\Sigma\)
\nn\\
&& +\frac{1}{16\pi^2 r}\sum_\gamma\hng{}\Bigl(\de \Igp \wedge\de\(\cR\Zg{}\)-\de\Igm\wedge \de\(\cR \bZg{}\) \Bigr).
\label{om3}
\eea

We will not reproduce the full calculation here. We refer the reader to appendix B of \cite{Alexandrov:2014sya} for this.
We only give the final expression for $\omega^3$ here written in the form of \eqref{metom}

\bea
\omega^3&=&
\frac{\I \,\hat\Sigma\wedge \bar{\hat\Sigma}}{4 r^2\(1-\frac{2r}{\cR^2\Uin}\)}
+\frac{\I\cR^2}{4r^2}\,z^\Lambda\bz^\Sigma \cY_\Lambda\wedge\bar\cY_\Sigma
\nn\\
&&
-\frac{\I}{2 r}\,\Min^{\Lambda\Sigma}\(\cY_\Lambda
+\frac{\I\cR}{2\pi}\sum_\gamma \Om{\gamma} V_{\gamma\Lambda}\rIgamp{\gamma}\(\de Z_{\gamma}-\Uin^{-1}Z_\gamma\p K\)\)
\nn\\
&&\qquad
\wedge \(\bar\cY_\Sigma-\frac{\I\cR}{2\pi}\sum_{\gamma'} \Om{\gamma'} \bV_{\gamma'\Sigma}
\rIgamm{\gamma'}\(\de \bZ_{\gamma'}-\Uin^{-1}\bZ_{\gamma'}\bar\p K\)\)
\nn\\
&&
+\frac{\I}{2r\Uin}\(  \cY_\Lambda \Min^{\Lambda\Sigma}\bvl_\Sigma-\frac{\I\cR}{2\pi}\, \sum_\gamma \Om{\gamma}\cW_\gamma\de Z_\gamma\)
\wedge\(\vl_{\Lambda'} \Min^{\Lambda'\Sigma'}\bar\cY_{\Sigma'}+\frac{\I\cR}{2\pi}\, \sum_{\gamma'} \Om{\gamma'}\bar\cW_{\gamma'}\de \bZ_{\gamma'}\)
\nn\\
&&
+\frac{\I\cR^2 K}{2r}\Biggl\{ \cK_{ab}\de z^a\wedge \de\bz^b
-\frac{1}{K^2\Uin^2}\(\frac{1}{2\pi}\sum_\gamma \hng{}|\Zg{}|^2 \rIg- \vl_\Lambda \Min^{\Lambda\Sigma}\bvl_\Sigma\)^2
\p K\wedge \bar \p K
\Biggr.
\nn\\
&&\Biggl.
+\frac{1}{2\pi K}\sum_\gamma \hng{} \rIg\(\de Z_\gamma-\Uin^{-1}Z_\gamma\p K\)\wedge \(\de\bZ_\gamma-\Uin^{-1}\bZ_\gamma\bar\p K\)
 \Biggr\}.
\label{final-om3}
\eea
From this the metric can be easily read off through $\omega^3(X,Y) = g(X,J^3Y)$.
The one-form $\cY_\Lambda$ introduced in \eqref{holforms} is rewritten as
\be
\begin{split}
\cY_\Lambda
=&\,\de\tzeta_\Lambda-F_{\Lambda\Sigma}\de\zeta^\Sigma-\frac{\I}{4\pi}\sum_\gamma\hng{}V_{\gamma\Lambda}
\Biggl[ \rIg\de\Theta_\gamma +\cR\(\rIgp\de Z_\gamma+\rIgm\de\bZ_\gamma\)
\Biggr.
\\
&\, \Biggl.
\!\!\!\!+\frac{2\pi\cR}{\Uk}\,\vv_\gamma \(\frac{4\,\de r}{\cR^2}- \de K
-\sum_{\gamma'} \Om{\gamma'}\(\frac{\vv_{\gamma'}}{\cR}\, \de \Theta_{\gamma'}-
\frac{\rIgam{\gamma'}}{2\pi}\,\(\bZ_{\gamma'} \de Z_{\gamma'}+Z_{\gamma'}\de \bZ_{\gamma'}\)\)\)\Biggr].
\end{split}
\label{expand-Y}
\ee
An important feature of this result is that it shows that in the presence of instantons
$\cY_\Lambda$ has a non-vanishing projection along $\de r$.

The other one-form $\hat\Sigma$ becomes
\bea
\hat\Sigma= 2 \(1-\frac{2r}{\cR^2\Uin}\)\de r
+ \frac{\I}{4}\(\de \sigma+\tzeta_\Lambda\de \zeta^\Lambda-\zeta^\Lambda\de \tzeta_\Lambda+\cV\) ,
\label{Sig}
\eea

To be concise we introduced several notations once again:

\begin{itemize}

\item
The first one is an invertible matrix $\Min_{\Lambda\Sigma}$ which is an instanton corrected version of $N_{\Lambda\Sigma}$.
It is given by
\bea
\Min_{\Lambda\Sigma}&=&  N_{\Lambda\Sigma}-\frac{1}{2\pi}\sum_\gamma \Om{\gamma}\rIg \bV_{\gamma\Lambda}V_{\gamma\Sigma}.
\label{defmatM}
\eea

\item
We also used the following set of convenient notations :
\be
\begin{split}
\vv_\gamma=&\,\frac{1}{4\pi}\( Z_\gamma\rIgp+\bZ_\gamma\rIgm\),
\\
\vl_\Lambda=&\,\sum_\gamma \Om{\gamma} \vv_\gamma V_{\gamma\Lambda}.
\end{split}
\label{notvec}
\ee

\item
A potential $\Uin$ which is
\bea
\Uin =
K-\frac{1}{2\pi}\sum_\gamma \hng{}|Z_\gamma|^2 \rIg+ \vl_\Lambda \Min^{\Lambda\Sigma}\bvl_\Sigma,
\eea

and another potential labeled by charges is given by
\bea
\cW_\gamma = \bZ_\gamma \rIg- \rIgp \vl_\Lambda\Min^{\Lambda\Sigma} \bV_{\gamma\Sigma}.
\label{cWdef}
\eea

\item
We have already seen that the one-loop correction introduces a term proportional to the K\"ahler
connection to the $\sigma$ circle bundle. In presence of instanton corrections, this term is promoted to
\bea
\nn
\cV&=&
2\cR^2 K\(1-\frac{4r}{\cR^2\Uin}\) \cA_K
+\frac{8r}{\cR\Uin}\sum_\gamma\hng{} \( \vv_\gamma+\frac{1}{2\pi}\, \rIg V_{\gamma\Lambda}\Min^{\Lambda\Sigma}\bvl_\Sigma\)\cCf_\gamma
\\
&& +\frac{2r }{\pi\I\Uin}\sum_\gamma \Om{\gamma} \[\(\cW_\gamma+\frac{\cR\Uin}{8\pi\I r}\,\Igp\)\de Z_\gamma
\right.
\\ \nn && \left. \qquad \qquad\qquad\qquad
-\(\bar\cW_\gamma+\frac{\cR\Uin}{8\pi\I r}\,\Igm\)\de \bZ_\gamma  \],
\label{conn}
\eea
where we introduced the following linear combination of differentials of the RR-fields
\be
\cCf_\gamma= N^{\Lambda\Sigma}\(q_\Lambda-\Re F_{\Lambda\Xi}p^\Xi\)\(\de\tzeta_\Sigma-\Re F_{\Sigma\Theta}\de\zeta^\Theta\)
+\frac14\, N_{\Lambda\Sigma}\,p^\Lambda\,\de\zeta^\Sigma.
\label{connC}
\ee

\end{itemize}

Finally let us write down the expression for the metric in this case,
\bea
\de s^2&=&
\frac{2}{r^2} \(1-\frac{2r}{\cR^2\Uin}\)(\de r)^2
+\frac{1}{32r^2\(1-\frac{2r}{\cR^2\Uin}\)}\(\de \sigma +\tzeta_\Lambda \de \zeta^\Lambda-\zeta^\Lambda\de \tzeta_\Lambda+\cV \)^2
\nn\\
&&
+\frac{\cR^2}{2r^2}\,|z^\Lambda\cY_\Lambda|^2
+\frac{1}{r\Uin}\left|  \cY_\Lambda \Min^{\Lambda\Sigma}\bvl_\Sigma-\frac{\I\cR}{2\pi}\, \sum_\gamma \Om{\gamma}\cW_\gamma\de Z_\gamma\right|^2
\nn\\
&&
-\frac{1}{r}\,\Min^{\Lambda\Sigma}\(\cY_\Lambda
+\frac{\I\cR}{2\pi}\sum_\gamma \Om{\gamma} V_{\gamma\Lambda}\rIgamp{\gamma}\(\de Z_{\gamma}-\Uin^{-1}Z_\gamma\p K\)\)
\nn\\
&&\qquad
\times\(\bar\cY_\Sigma-\frac{\I\cR}{2\pi}\sum_{\gamma'} \Om{\gamma'} \bV_{\gamma'\Sigma}
\rIgamm{\gamma'}\(\de \bZ_{\gamma'}-\Uin^{-1}\bZ_{\gamma'}\bar\p K\)\)
\nn\\
&&
+\frac{\cR^2 K}{r}\Biggl( \cK_{a\bar b}\de z^a \de\bz^b
-\frac{1}{(2\pi K\Uin)^2}\left|\sum_\gamma \hng{}Z_\gamma \cW_\gamma\right|^2 |\p K|^2
\Biggr.
\nn\\
&&\Biggl.\qquad
+\frac{1}{2\pi K}\sum_\gamma \hng{} \rIg\left|\de Z_\gamma-\Uin^{-1}Z_\gamma\p K\right|^2
\Biggr).
\label{mett1}
\eea

Actually, the result  in \eqref{mett1} is valid for two cases. One of them is the case when
we restrict to the set of mutually local charges, for which it is exact.
The other corresponds to the inclusion of all D-instanton
charges \eqref{mett1}, but restricted to the one-instanton approximation.
This effectively means to truncate the instanton series expansion at the linear order in Donaldson-Thomas (DT) invariants $\Om{\gamma}$.
 Further simplificatons occur in the latter case. One can explicitly construct
the inverse of the matrix $\Min_{\Lambda\Sigma}$ from \eqref{defmatM} and one can solve for $\cR$ in terms of other coordinates through
the expression for the contact potential \eqref{phiinstmany}.

One more complication arises due to the presence of instanton corrections. The nice two-staged fibration structure of the
one-loop corrected space \eqref{doublefib2} is now broken down. Firstly, the dilaton is not factorized now. Instead it enters
non-trivially along with $z^a$ and $(\zeta^\Lambda,\tzeta_\Lambda)$ in the expression for $\cY_\Lambda$ due to presence
of terms proportional to $\de r$. Secondly, the metric on the subspace parametrized by the complex structure moduli acquires
a complicated dependence on the other coordinates. The only fibration that survives is the circle bundle of the NS-axion $\sigma$ 
so that the total space has the following topological obstruction 
\be
\label{quant-fib}
\begin{array}{rl}
S^{1}_\sigma\ \longrightarrow\ &\cM_{\text{HM}}
 \\
&\  \downarrow
\\
& \cB_{r,z,\zeta,\tzeta}^{\rm inst}\, .
\end{array}
\ee
The connection on this circle bundle can be easily read off from \eqref{mett1} and is given by,
$\tzeta_\Lambda \de \zeta^\Lambda - \zeta^\Lambda\de \tzeta_\Lambda + \cV$,
where $\cV$ is given by \eqref{conn}.

Finally, since we working in type IIA framework the metric must be symplectic invariant.
However, it is hard to see it in the expression \eqref{mett1}.
That it is indeed present has been verified explicitly in appendix B.4 in
\cite{Alexandrov:2014sya}.

\section{Universal hypermultiplet and Toda potential}

One very interesting case in our context arises when the compactification is done on 
a ``rigid Calabi-Yau" threefold, which does not have any complex structure moduli
as for them $h^{2,1}(\CY) = 0$. The vanishing of the Hodge number for type IIA string theory
compactified on $\CY$ therefore implies that there is only one hypermultiplet which in literatures is known as
universal hypermultiplet \cite{Strominger:1997eb}. The hypermultiplet moduli space in this situation is then
a four dimensional QK manifold. In this special case the QK geometry allows for a more explicit description as
it is defined as an Einstein manifold with non-vanishing cosmological constant and self-dual Weyl curvature.

Ignoring NS5-brane instantons, as we have been doing throughout this chapter, implies that
there is at least one continuous isometry preserved, the translation along the NS-axion $\sigma$
for the HM moduli space. With a proper choice of coordinate metric on such four-dimensional QK manifolds can be written
in the form of Tod ansatz \cite{MR1423177} parametrized by one real function,
\be
\de s^2
=-\frac{3}{\Lambda}\[\frac{P}{\rho^2}\(\de \rho^2+4 e^T\de z\de \bz\)+\frac{1}{P\rho^2}\(\de\theta+\Theta\)^2\],
\label{met-Toda}
\ee
where $T$ is a function of $(\rho,z,\bz)$ and is independent of $\theta$ parametrizing the direction of the isometry.
Furthermore,
\bea
P &=& 1-\hf\, \rho\p_\rho T,
\\
\de \Theta &=& \I\(\p_z P\de z-\p_{\bz} P\de\bz\)\wedge \de \rho-2\I \p_\rho(P e^T)\de z\wedge \de \bz,
\label{deThet}
\eea
whereas the Einstein self-duality condition of the metric is encoded in the Toda differential equation
to be satisfied by the function $T$,
\be
\p_z\p_{\bz} T+\p_\rho^2 e^T=0.
\label{eq-Toda}
\ee

This description is more convenient because all instanton corrections are encoded in just one function,
the Toda potential $T$ which is constrained by the Toda equation \eqref{eq-Toda}. Hence to extract
them it suffices to find appropriate solution of the Toda equation. This strategy was very successful at the
one-instanton level \cite{Ketov:2001gq, Ketov:2001ky,Ketov:2002vr, Saueressig:2005es, Theis:2014bia}\footnote{A similar strategy can be applied to derive
NS5-brane instantons as well \cite{Alexandrov:2006hx}, because generic 4d QK manifolds
can be parametrized by solutions of another, more complicated non-linear differential equation, which replaces the Toda equation
in the absence of the isometry \cite{Przanowski:1984qq}.}. However, the results beyond this approximation
are often not reliable because, to fix the ambiguities of integration typically some ansatz for $T$ is made.
Due to such unjustified simplifications those results can 
not be trusted in full generality.

But now that we have the twistor construction, we do not need to find the Toda potential from the differential equation.
Instead it serves just as a consistency check, that is once a proper identification of coordinates and the Toda potential is made one has to verify if the Toda
equation should be satisfied automatically. Indeed, this is the case if one identifies \cite{Alexandrov:2009vj},
\be
\rho=e^\phi,
\qquad
z=\frac{\I}{2}\, \txii{+}_0,
\qquad
\theta= -\frac{1}{8}\, \sigma,
\qquad
T=2\log (\cR/2).
\label{dict-Toda}
\ee
One can now actually compute the metric $\cM_{\rm HM}$  by restricting \eqref{mett1} to four dimensions. One can also explicitly verify that $T$ satisfies Toda equation.
For details of these checks and result for the metric we refer the reader to 
\cite{Alexandrov:2014sya}. 

Since there is no complex structure moduli in this case the only possible value the indices $\Lambda,\Sigma$ can take here
is $0$. We therefore drop the index altogether. Moreover the prepotential for the rigid Calabi-Yaus is necessarily quadratic as it
is a degree two homogeneous function of only one argument \cite{Bao:2009fg},
\be
F(X) = \frac{\lambda}{2}\, X^2,
\qquad\quad
\lambda\equiv \lambda_1 -\I \lambda_2=\frac{\int_\cB\Omega}{\int_\cA\Omega}\, ,
\label{prepUHM}
\ee
where $\lambda$ is a fixed complex number, given by the ratio of periods of the holomorphic 3-form $\Omega\in H^{3,0}(\CY)$
over an integral symplectic basis $(\cA,\cB)$ of $H_3(\CY,\IZ)$, with $\lambda_2>0$.
Choosing electric frame for the mutual local charges we can explicitly construct the solutions for the Toda coordinates \cite{Alexandrov:2014sya}
given by a set of transcendental equations,
\be
\begin{split}
e^T= &\,\frac{1}{2\lambda_2}(\rho +c)
-\frac{ e^{T/2}}{4\pi^2\lambda_2}\sum\limits_{q>0} \bOm{(0,q)}q\,
\cos\bigl(2\pi  q\zeta\bigr)
 K_1 \bigl(8\pi q e^{T/2}\bigr),
\\
z =&\, \frac{\I}{2}\,(\tzeta-\lambda \zeta)
-\frac{1}{4\pi^2}\sum\limits_{q>0} \bOm{(0,q)} q\,
 \sin\bigl(2\pi q\zeta\bigr) K_0  \bigl(8\pi q e^{T/2}\bigr),
\end{split}
\label{dilUHM}
\ee
where we have introduced the rational DT invariants defined by
\bea
\bar{\Omega}(\gamma) = \sum_{d|\gamma} \frac{\Omega(\gamma)}{d^2}.
\eea

In the general case where no restriction on charges is imposed the above equations become more involved owing to the presence
of complicated twistor integrals which in general can not be reduced to a product of Bessel functions unlike the situation above.

\section{Curvature singularity in presence of D-instantons} 

Just by mere inspection or a more diligent computation of the quadratic curvature (Riemann tensor squared) of the metric 
\eqref{1lmetric} one can find that there is a curvature singularity for $\chi_{\CY} > 0 $ at $r=-2c$.
In this section we would like to address the question that what happens to this singularity when D-instantons are included.

Looking at the kinetic terms corresponding to the dilaton and the NS-axion $\sigma$ in \eqref{mett1} and comparing them with similar terms in 
the one-loop corrected metric \eqref{1lmetric} we find that the factor $r+2c$ is promoted in presence of the instanton corrections to 
\be
\hf\,\cR^2 \Uin-r.
\label{singfactor}
\ee
Therefore it is natural to expect that setting the above to zero will give rise to a singularity equation, that
is the equation describing the singular hypersurface on $\cM_{\text{HM}}$. Substituting expressions for $\Uin$ and $r$
one arrives at
\bea
&&
K+\frac{4c}{\cR^2}
+\frac{1}{\pi}\sum_\gamma \hng{}\(\frac{\I}{8\pi}\(\Zg{}\Igp+\bZg{}\Igm\)-|\Zg{}|^2\rIg\)
\label{singeqB}\\ \nn
&& +\frac{1}{8\pi^2} \(\sum_{\gamma}\Om{\gamma} \( \Zg{}\rIgp+\bZg{}\rIgm\) V_{\gamma\Lambda}\)
\Min^{\Lambda\Sigma}\sum_{\gamma'}\(\Om{\gamma'} \( Z_{\gamma'}\rIgamp{\gamma'}+
\right. \right.
\\ \nn && \left. \left.\qquad
\bZ_{\gamma'}\rIgamm{\gamma'}\)  \bV_{\gamma'\Sigma}\)
=0.
\nn
\eea 

There is actually another more geometric way to extract the equation for singularity. Let us illustrate it in the case for one-loop
metric first. From \eqref{1lDarboux} one can easily find the expressions for Darboux coordinates near the north pole.
Using the procedure to construct $(1,0)$-forms in \ref{compmet} we obtain the following, 
\bea
\begin{split}
&\Pi^a = \de z^a,
\\ &
\cY_\Lambda = \de \tzeta_\Lambda - F_{\Lambda\Sigma}\de \zeta^\Sigma,
\\ &
\Sigma = \de r + 2c\de \log \cR + \frac{\I}{4}\(\de\sigma+\tzeta_\Lambda\de\zeta^\Lambda - \zeta_\Lambda\de\tzeta_\Lambda\).
\label{1lbasis}
\end{split}
\eea  
The curvature singularity occurs if the basis forms become degenerate. The only possibility for this to happen in \eqref{1lbasis} is 
\bea
\Re \Sigma = \de e^\phi + 2c \de \log \cR = 0,
\eea
which gives the corresponding equation for singularity.
Using the one loop relation between $\cR$ and the dilaton one obtains that it reduces to
\bea
r + 2c = 0,
\eea
which thus represents the equation for singularity for the one loop corrected HM moduli space. 
In the more complicated case when instanton corrections are present one can 
do the similar exercise as above with the $(1,0)$-forms defined in \eqref{holforms} to recover
\eqref{singeqB}.

We write the equation \eqref{singeqB} using $\cR$ instead of the dilaton because in a sense it is a more fundamental quantity,
as it is related to the 10-dimensional string coupling. In the weak coupling regime $\cR \to \infty$ the first term in this equation dominates 
to all others. Let us also note that this term is positive. It means that \eqref{singeqB} can only have solutions if the left hand side becomes negative somewhere in the moduli
space. For understanding whether this can happen, we need to analyze the equation in strong coupling regime $\cR \to 0$. The only way to do this is to exploit S-duality symmetry.
But to do this, it is required to make this symmetry manifest. 

To this end, we first need to go to the type IIB formulation. The electrically charged D-instantons on the type IIA side are then mapped
to contributions coming from the D(-1) and D1-brane instantons. The D(-1)-instantons are point like objects and are only labeled
by a single charge $q_0$. The latter, D1-instantons can wrap two dimensional cycles inside the Calabi-Yau and are labeled by 
charges $q_a$. Apart from the exponential corrections coming from $g_s$, there are perturbative and non-perturbative
$\alpha'$-corrections. The non-perturbative corrections in $\alpha'$ come from the worldsheet instantons. The resulting moduli space then
admits an isometric action of $SL(2,\IZ)$ group which is the S-duality symmetry on the type IIB side. Under this symmetry the perturbative $\alpha'$-corrections
mix with D(-1)-instantons and the worldsheet instanton corrections mix with D1-instantons. 

At this stage we use the relevant mirror maps to rewrite \eqref{singeqB} in manifestly $SL(2,\IZ)$-invariant form. 
We will use them in the following to rewrite \eqref{singeqB} in manifestly S-duality invariant form by performing Poisson resummation over $q_0$. 
\bea
\cR = \frac{\tau_2}{2},\qquad
z^a = b^a + \I t^a,\qquad
\zeta^0 = \tau_1, \qquad
\zeta^a = -(c^a - \tau_1 b^a).
\label{mirmap}
\eea
We remind the reader from chapter 2 that action of the $SL(2,\IZ)$ transformation represented by the matrix $\(\begin{array}{cc} a & b \\ c & d \end{array}\)$
 is realized on the type IIB coordinates in the following way,
\be
\tau\to\frac{a\tau+b}{c\tau+d},
\qquad
t^a\to |c\tau+d| \, t^a,
\qquad
\(\begin{array}{c} c^a \\ b^a \end{array}\)\to \(\begin{array}{cc} a & b \\ c & d \end{array}\)\(\begin{array}{c} c^a \\ b^a \end{array}\),
\label{Sdualtr}
\ee
where we combined the inverse 10-dimensional string coupling $\tau_2=1/g_s$ with the RR-field $\tau_1$ into
an axio-dilaton $\tau=\tau_1 + \I \tau_2$.

We will restrict to the case with D(-1)-instantons only. Since these D(-1)-instantons mix with one-loop corrections in the same $SL(2,\IZ)$
multiplet, working in this approximation is justified. One could already hope for the possibility of removing the one-loop singularity  
by the presence of D(-1)-instantons only. 

Since we drop the worldsheet instantons the prepotential in this case is given by 
\be
F(X)= -\kappa_{abc}\,\frac{X^a X^b X^c }{6X^0} +\frac{\I\zeta(3)\chi_\CYm}{16\pi^3}\, (X^0)^2,
\label{prep}
\ee
where the first term is the usual cubic term and the second one encodes the perturbative $\alpha'$-correction. Since we consider only $D(-1)$-instantons 
the D-brane charges are labeled by only one number $q_0$. In this case the DT invariant is independent of the charge and coincides with
the Euler characteristic of the Calabi-Yau, $\Om{q_0} = \chi_{\CY} = - \chi_{\CYm}$, where $\CYm$ is the mirror Calabi-Yau. 
The prepotential \eqref{prep} leads to the following expression for the exponential of K\"ahler potential,
\be
K = 8V-\frac{\zeta(3)\chi_\CYm}{4\pi^3},
\ee
where $V=\frac{1}{6}\, \kappa_{abc} t^a t^b t^c$ is the Calabi-Yau volume and we used the relation $z^a=b^a+\I t^a$
from the mirror map \eqref{mirmap}.

We will not give details of the resummation procedure here. It can be found in \cite{Alexandrov:2014sya}. We write the 
result only and analyze its consequences. The equation for singularity in manifestly S-duality invariant form turns out to be
\bea
\begin{split}
&1+\frac{\chi_\CYm\, E_{3/2}(\tau)}{64\pi^3 V\tau_2^{3/2}}
-\frac{2\chi_\CYm^2}{\bigl(64\pi^3 V\tau_2^{3/2}\bigr)^2}\[E_{3/2}^2(\tau)
\right.
\\ & \left. \qquad \qquad \qquad
-\frac{9}{4} \mathop{{\sum}'}\limits_{m,n \in \IZ}\mathop{{\sum}'}\limits_{m',n' \in \IZ}\frac{\tau_2^5(mn'-nm')^2}{|m\tau+n|^5|m'\tau+n'|^5}\]=0,
\label{finsingeq}
\end{split}
\eea
where the $E_{3/2}$ is the non-holomorphic Eisenstein series and is modular invariant,
\be
E_{3/2}(\tau)=\mathop{{\sum}'}\limits_{m,n \in \IZ}\frac{\tau_2^{3/2}}{|m\tau+n|^3}.
\ee

Now that we have obtained \eqref{finsingeq}, we can exploit the power of S-duality for going to strong coupling regime to weak coupling. 
To do this, let us first put $\tau_1 = 0$ for convenience. Then S-duality transformation acts by
\bea
\tau_2 \to \tau_2^{-1},\qquad
V \to V\tau_2^3.
\label{strongweak}
\eea
with this we can extract information about the strong coupling limit. Since in hindsight we know 
that the equation \eqref{singeqB} is S-duality invariant, we can work directly with it. After we perform the S-duality
transformation \eqref{strongweak} as above we can taking the limit
the limit $\tau_2 \to \infty$, corresponding to the strong coupling, as it was the usual weak coupling limit. 
The instanton contributions become exponentially suppressed and the left hand side of the singularity equation has the expansion,
\be
-\frac{\zeta(3)^2\chi_\CYm^2}{2(16\pi^3 V)^2}\,\tau_2^{-6}
+\frac{\zeta(3)\chi_\CYm^2}{6(16\pi^2 V)^2}\, \tau_2^{-4}
+\frac{\zeta(3)\chi_\CYm}{32\pi^3 V}\,\tau_2^{-3}+  \frac{\chi_\CYm}{96\pi V}\,\tau_2^{-1}
+1+O\bigl(e^{-2\pi\tau_2^{-1}}\bigr).
\ee
Unlike the case before the transformation, where the left hand side was always positive,
the dominant contribution now comes from the first term going $-\tau_2^{-6}$. Thus we conclude that the 
singularity remains even if one incorporates D(-1)-instantons. The situation actually becomes worse as compared to 
the one-loop corrected case. There the singularity existed for only $\chi_{\CY} > 0$, whereas now it exists 
for any sign of the Euler characteristic.

Finally to remark, we do not expect that any of the D-instantons can cure this situation. The only
possibility to get rid of the singularity appears to require the inclusion of NS5-instantons. Their exponentially suppressed contribution
at the weak coupling regime can drastically alter the situation at the strong coupling though S-duality. It will be very interesting to 
investigate their effect  and is one of the works for future.

\newpage

\chapter{Mirror symmetry at the quantum level}

\label{chapter5}

In the previous chapter, we worked with the hypermultiplet moduli space in the type IIA picture. As was mentioned,
mirror symmetry relates the HM moduli space of type IIA string theory compactified on a Calabi-Yau $\CY$ and that of type IIB string 
theory compactified on a mirror Calabi-Yau $\CYm$. Let us now turn to the latter formulation. Even though these two descriptions should be identical due
to mirror symmetry, it is convenient to have both of them at our disposal, because there are several situations which simplify in either of these 
frameworks. In particular, for type IIB theory there is a very powerful  S-duality. Many of the results presented in the thesis depend 
crucially on clever use of this symmetry, for example, finding fivebrane transition functions on the type IIB side, which we will present in chapter \ref{ch7}.
Finally, the mirror symmetry at the quantum level is still a conjecture, although in the light of the recent results, it does not seem to be 
an unattainable goal to establish it in the presence of all possible quantum corrections.  

Let us briefly summarize the differences between the HM moduli spaces in the two mirror formulations. First, instead of complex structure deformations of the Calabi-Yau $z^a$,
one considers the complexfied K\"ahler moduli $v^a$, where we are using the same index because for mirror pair of Calabi-Yaus $h^{2,1}(\CY) = h^{1,1}(\CYm)$.
The periods of RR 3-form are replaced by the periods of the even dimensional RR forms ($c^0,c^a,c_a,c_0$). Furthermore, one can combine $\tau_1 = c^0$ with
the ten dimensional string coupling constant $\tau_2 = 1/g_s$ to form an axio-dilatonic field $\tau = \tau_1 + \I \tau_2$. 

Next, we consider the type of symmetries that characterize each of the particular formulations. The type IIA theory is manifestly symplectically covariant, whereas
the pertinent symmetry for type IIB is $SL(2,\IZ)$ duality. The physical fields in type IIB form a representation under the action of this symmetry group. In particular, this implies that,
at the quantum level, our construction in type IIA framework must have a hidden $SL(2,\IZ)$ symmetry, which can be made manifest by passing to the type IIB side. 

In this chapter we are going to provide a description of the twistor space of a class of QK manifolds admitting an isometric action of $SL(2,\IZ)$. We also show
how it can be significantly simplified in terms of the contact hamiltonians. This simplification will be crucial in later chapters where we are going to present our results for fivebrane 
transition functions.

\section{D-branes and charge quantization}

In this section we address the problem of classifying D-instantons on the type IIB side. It turns out that the
simplistic description of instantons as some $n$-dimensional objects wrapping $n$-dimensional cycles 
and classified by integers enumerating these cycles is not enough. The correct way to describe them is through the derived category of coherent sheaves
\cite{Douglas:2000gi}. In particular, for a non-vanishing D5-brane charge $p^0$, a bound state of D5-D3-D1-D(-1) branes can be represented as 
a coherent sheaf $E$ on $\CYm$, of rank ${\rm rk}(E) = p^0$. The charge vector is given by the so called generalized Mukai vector in the even-dimensional
cohomology basis \cite{Minasian:1997mm},
\bea
\gamma' = {\rm ch}(E) \sqrt{{\rm Td}(\CYm)} = p^0 + p^a \omega_a - q'_a \omega^a + q'_0 \omega_{\CYm},
\label{IIBch}
\eea
where ${\rm ch}(E)$ and ${\rm Td}(\CYm)$ are the Chern character of the sheaf and the Todd class of the Calabi-Yau $\CYm$ respectively. From this one can obtain
expressions for charges explicitly,
\bea
\begin{split}
p^0 &= {\rm rk} (E), \\
p^a & = \int_{\gamma^a} c_1(E), \\
q'_a & = -\(\int_{\gamma_a} {\rm ch}_2(E) + \frac{p^0}{24} c_{2,a} \), \\
q'_0 & = \int_{\CYm} \({\rm ch}_3(E) + \frac{1}{24} c_1(E) c_2(\CYm)\),
\end{split}
\eea
showing that charges $q'_\Lambda$ are not integers. There seems to be an apparent clash with
the mirror symmetry because the charge vector in the type IIA formulation is identified with the 
integer homology class $H_3(\CY,\IZ)$. 

This puzzle can be resolved as follows \cite{Alexandrov:2010ca}. 
Let us compare the central charge associated to D-instantons in type IIB 
\bea
Z_{\gamma'} = \int_{\CYm} e^{-v^a\omega_a} \gamma',
\eea
and the central charge in type IIA given by the prepotential restricted to the large volume limit
\bea
F(X) = -\frac{\kappa_{abc}}{6}\,\frac{X^aX^bX^c}{X^0} + \frac12\, A_{\Lambda\Sigma} X^\lambda X^\Sigma.
\eea
It is important to keep the quadratic piece here, even though it does not affect K\"ahler potential.
Then identifying $z^a$ with $v^a$, the two expressions coincide if
\bea
q'_\Lambda = q_\Lambda - A_{\Lambda\Sigma} p^\Sigma.
\label{ratch}
\eea
For this identification, one should take into account the conditions satisfied by 
the matrix $A_{\Lambda\Sigma}$ is constrained by certain conditions,
\bea
\begin{split}
& ({\rm a})\,\, A_{00} \in \IZ,
\\ &
({\rm b})\,\, A_{0a} \in \frac{c_{2,a}}{24} + \IZ,
\\ &
({\rm c})\,\, A_{ab}p^b - \frac12\,\kappa_{abc} \,p^b p^c \in \IZ,
\\ &
({\rm d})\,\, \frac16\, \kappa_{abc} p^ap^bp^c + \frac{1}{12} c_{2,a} p^a \in \IZ,
\label{propA}
\end{split}
\eea
for any arbitrary $p^a \in \IZ$. 
The above conditions ensure that the symplectic transformation \eqref{ratch} maps the rational charges of the type IIB formulation \eqref{IIBch} to
the integer charges of the type IIA formulation.  
The quantization condition for the primed charges is given by the following,
\bea
q'_0 \in \IZ - \frac{p^0}{24} c_{2,a} - \frac12\,\kappa_{abc} p^b p^c, \qquad
q'_a \in \IZ - \frac{1}{24} p^a c_{2,a}.
\eea

It turns out that the above quantization conditions on D-brane charges gets reflected in the mirror symmetry and S-duality transformations as well.
To investigate this, first let us see how mirror symmetry relates the type IIA and type IIB RR fields.
They appear in the imaginary part of the D-brane action, whereas the real part is determined
by the central charge which gives the strength of the D-instantons. We have already discussed the D-brane action in the 
type IIA formulation in chapter \ref{chapter2}. In type IIB, instead it is given by  \cite{Aspinwall:2004jr,Marino:1999af}
\bea
S_{\rm D-IIB} = 2\pi g_s^{-1} |Z_\gamma| + 2\pi\I \int_{\CYm} \gamma' \wedge {\hat{A}}^{\rm even} e^{-\hat{B}_2},
\eea
where the RR potential $ {\hat{A}}^{\rm even}$ is
\bea
{\hat{A}}^{\rm even}\,e^{-\hat{B}_2} =  \zeta^0 - \zeta^a \omega_a - \tzeta'_a \omega^a - \tzeta'_0 \omega_{\CYm},
\label{RRscalar}
\eea 
and the shifted RR fields are related to the unshifted ones by
\bea
\tzeta'_\Lambda = \tzeta_\Lambda - A_{\Lambda\Sigma} \zeta^\Sigma.
\label{ratRR}
\eea
This shift is the same as in \eqref{ratch} and absorbs the quadratic term in the prepotential.

\section{S-duality in the HM moduli space}
In chapter \ref{chapter2} we have already discussed about the action of S-duality on the classical HM moduli space.
In this approximation, the action of the $SL(2,\IR)$ group on the type IIB fields is given by \eqref{SL2Z} and the classical mirror map is given by 
\eqref{symptobd}. 

In this chapter our goal is to give a description of a generic QK manifold admitting only an isometric action of $SL(2,\IZ)$. We
are going to provide explicit construction of the mirror map in presence of all possible quantum corrections. Although, we will
not be providing any concrete transition functions here, we will do the necessary groundwork to establish that it is indeed possible 
to preserve quaternion-K\"ahler structure while breaking all the continuous isometries.  

\vspace{2mm}

But before proceeding further there are a few comments to make. 

\begin{itemize}

\item The continuous $SL(2,\IR)$ symmetry is present only at the classical level.
Already, $\alpha'$ corrections break it. 
 However it was shown in \cite{RoblesLlana:2006is},\cite{Alexandrov:2009qq} that if  D(-1) and D1 instanton corrections are incorporated,
one can restore a subgroup of it, namely $SL(2,\IZ)$. Furthermore, isometric action of the latter is also expected to hold even when all quantum corrections
are incorporated. 

\item The RR fields in terms of period integrals of 10-dimensional gauge potentials are \cite{Louis:2002ny, Alexandrov:2008gh},
\bea
\begin{split}
& c^0 = \hat{A}^0,\qquad c^a = \int_{\gamma^a} \hat{A}_2,\qquad c_a = -\int_{\gamma_a} \(\hat{A}_4 - \frac12\, \hat{B}_2 \wedge \hat{A}_2\),
\\ &
c_0 = -\int_{\CYm} \(\hat{A}_6 - \hat{B}_2\wedge \hat{A}_4 + \frac13\,\hat{B}_2 \wedge \hat{A}_2\wedge\hat{A}_2\).
\end{split}
\eea
The above expressions are exact only at the classical level. In general we define the type IIB fields 
as those which transform accrding to  \eqref{SL2Z}. Obviously, this means that the mirror map \eqref{symptobd}
receives quantum corrections. As has been mentioned, one of our goals in this chapter is to obtain explicit expressions for them. 

\item 
At the quantum level, $SL(2,\IZ)$ transformations are given by  \eqref{SL2Z} with $a,b,c,d \in \IZ$, except that the D3-axion
$c_a$ which was invariant, now acquires a constant shift $c_a \mapsto c_a - c_{2,a}  \varepsilon(\gl{})$ \cite{Alexandrov:2010ca}. Here, $\varepsilon(\gl{}) $ is
the multiplier system of the Dedekind function
$\eta(\tau)$ given by
\be
\label{multeta}
e^{2\pi\I\, \varepsilon(\gl{}) }=\frac{\eta\left(\frac{a\tau+b}{c\tau+d}\right)}{(c\tau+d)^{1/2}\,
\eta(\tau)}.
\ee  
In particular, $e^{2\pi\I\, \varepsilon(\gl{}) }$ is the $24^{\rm th}$ root of unity corresponding to the fractional terms appearing in the 
quantization condition of the charges $q'_\Lambda$. Its appearance can be traced back to the fact that the S-duality transformations
and the Heisenberg shifts \eqref{Heis0} are not completely independent. Actually one should identify the Heisenberg transformation 
with the parameter $\eta^0 = b$ with the $SL(2,\IZ)$ transformation induced by the matrix $\displaystyle{\begin{pmatrix} 1 & b \\ 0 & 1 \end{pmatrix}}$. 
This identification indeed works if one takes into account this constant shift of the axion $c_a$.
Even though this issue does not seem to be crucial, there is another reason which shows that this modification is really needed. 
Later, in chapter 7 we shall  see that this modification is important also for consistency of the actions of the isometry groups generated by $SL(2,\IZ)$, 
monodromy and Heisenberg symmetry, on the fivebrane transition functions.

\end{itemize}

\section{Monodromy invariance and spectral flow}

Another important symmetry that has to be taken into account at the non-perturbative level is the monodromy which
arises from the possibility of going around the singularities in the moduli space of complexified K\"ahler class deformations $\cM_{K}(\CYm)$. 
This should be a symmetry because the final metric must be smooth everywhere. Here we are interested in the monodromy around the large volume point. 
This symmetry is generated by
\bea
b^a \mapsto b^a + \epsilon^a, \qquad \epsilon^a \in \IZ.
\eea
Since in \eqref{RRscalar}, the $B$-field is present alongwith the RR potential, to realize such symmetry 
one should also perform a symplectic rotation of RR-fields \cite{Alexandrov:2010ca},
\bea
\begin{split}
\label{Mon1}
M_{\epsilon^a} :\,\,\,\,\,\,& \zeta^0 \mapsto \zeta^0, \qquad \zeta^a \mapsto \zeta^a + \epsilon^a,
\qquad \tzeta'_a \mapsto \tzeta'_a - \kappa_{abc} \zeta^b \epsilon^c - \frac12\, \kappa_{abc} \epsilon^b\epsilon^c \zeta^0,
\\ &
\tzeta'_0 \mapsto \tzeta'_0 - \tzeta'_a \epsilon^a + \frac12\,\kappa_{abc} \epsilon^a\epsilon^b\zeta^c +\frac16\, \kappa_{abc} \epsilon^a\epsilon^b\epsilon^c\zeta^0.
\end{split}
\eea 
The above transformation can be rewritten in terms of the unprimed fields. Then the matrix inducing the symplectic transformation becomes
\be
\rho(M_{\eps^a})
=\(\begin{array}{cccc}
1\ & 0 & 0 & 0
\\
\epsilon^a & {\delta^a}_b & 0 & 0
\\
-L_a(\epsilon)\ & -\kappa_{abc}\epsilon^c & 0 & {\delta_a}^b
\\
 L_0(\epsilon) & L_b(\epsilon)+2A_{bc}\epsilon^c & 1 & \ -\epsilon^b\
\end{array}\) ,
\label{monmat}
\ee
where we introduced two functions
\be
L_a(\eps)\equiv \frac12\, \kappa_{abc} \eps^b \eps^c-A_{ab}\eps^b,
\qquad
L_0(\eps)\equiv\frac16\, \kappa_{abc}\eps^a\eps^b\eps^c+\frac{1}{12}\, c_{2,a}\eps^a,
\ee
which are integer valued because of the properties of the matrix $A_{\Lambda\Sigma}$ \eqref{propA}.

Note that, the D-brane action $S_{\rm IIB}$ remains invariant if the monodromy transformation is associated with
a similar transformation on the charge lattice, the so called spectral flow defined by
\bea
\gamma \mapsto \gamma[\epsilon] = \rho(M_{\eps^a}) . \gamma.
\label{sflow}
\eea
The symplectic matrix is integer valued,
and hence it preserves the charge lattice and hence is an admissible transformation. 

Since the D-instanton action remains invariant under combined actions of monodromy transformation and the spectral flow
associated with it, the BPS indices $\Omega(\gamma;v)$ should also be invariant. Although the moduli dependence 
for them is mild, it is crucial for the invariance to take it into account.

\section{S-duality in the twistor space}

The most prominent feature of type IIB formulation is its S-duality symmetry. For our case it is 
generated by the action of $SL(2,Z)$ group on the type IIB fields according to \eqref{SL2Z}. 
We have already seen that the problem of incorporating quantum corrections can be most conveniently tackled at the level of the
twistor space. Therefore, it is important to understand how S-duality acts on the twistor space. More precisely,
we want to obtain the twistor description when there is only an isometric action $SL(2,\IZ)$ group present. Our goal is twofold here.
First, we want to understand what kind of constraints are imposed at the level of the twistor space in this case. Secondly, we 
want to obtain the mirror map relating the type IIA and type IIB fields. Of course, since we do not specify concrete transition
functions, the latter results will be only formal. This question of finding appropriate transition functions will be addressed in a later chapter.

\subsection{Lift of the $SL(2,\IR)$ for tree level HM moduli space }

We start from the classical twistor space of the HM moduli space which carries an isometric action
of the continuous group $SL(2,\IR)$. In the classical approximation, the last two terms in the 
prepotential in  \eqref{vectIIAprepot}  are absent. 
 Furthermore if we redefine the Darboux coordinates as
\bea
\txi'_\Lambda = \txi_\Lambda - A_{\Lambda\Sigma} \xi^\Lambda,
\qquad
\alpha' = \alpha + \frac12\, A_{\Lambda\Sigma} \xi^\Lambda\xi^\Sigma,
\eea
the quadratic piece in the prepotential is also removed and one remains just with the first cubic term.
The Darboux coordinates are \footnote{For clarity of notations,
we drop the primes in the rest of this chapter and by $\Fcl$, we denote the classical cubic part of the prepotential.}
\bea
\nn
\xii{0}^\Lambda &=& \zeta^\Lambda + \cR(t^{-1} z^\Lambda - t \bz^\Lambda),
\\
\txii{0}_\Lambda &=& \tzeta_\Lambda + \cR(t^{-1} \Fcl_\Lambda (z) - t \bar{\Fcl_\Lambda} (\bz)),
\\ \nn
\ai{0} &=& -\frac12\,(\sigma+\zeta^\Lambda\tzeta_\Lambda)  - \cR \zeta^\Lambda (t^{-1} F^{\rm cl}_\Lambda - t \bF^{\rm cl}_\Lambda)
-\cR^2 (t^{-2}\Fcl + t^2 \bar{\Fcl}) + \frac{\cR^2}{2} (z^\Lambda \bar{\Fcl_\Lambda} + \bz^\Lambda \Fcl_\Lambda).
\label{Darboux0}
\eea
In any case the isometric action of $SL(2,\IR)$ must have a holomorphic representation on the Darboux coordinates.
To achieve this one has to also prescribe how the fiber coordinate $t$ transforms. This was found in \cite{Alexandrov:2012bu}
and the result reads as
\bea
t \mapsto 
\frac{c \tau_2 + \varpi  (c \tau_1 + d) +
\varpi |c \tau + d| }{(c \tau_1 + d) + |c \tau + d| - \varpi c \tau_2}
=\frac{1+\varpi_-^{c,d}\varpi}{\varpi_-^{c,d}-\varpi}=
-\frac{\varpi_+^{c,d}-\varpi}{1+\varpi_+^{c,d}\varpi},
\label{clt}
\eea
where $\varpi_{\pm}^{c,d}$ are the two roots of the equation $c\xi^0 + d = 0 $ and are given by
\bea
\varpi_{\pm}^{c,d} = \frac{c\tau_1 + d \mp |c\tau + d|}{c\tau_2},
\qquad
\varpi_+^{c,d} \varpi_-^{c,d} = -1.
\eea
Then the transformation of the Darboux coordinates under the action of $SL(2,\IR)$ 
are found to be \cite{Alexandrov:2008gh},
\bea
\begin{split}
\label{SL2_dar}
& \xi^0 \mapsto \frac{a\xi^0 + b}{c\xi^0 + d}, \qquad \xi^a \mapsto \frac{\xi^a}{c\xi^0 + d},
\qquad \txi_a \mapsto \txi_a + \frac{c\kappa_{abc}}{2(c\xi^0 + d)} \xi^a \xi^b,
\\ &
\begin{pmatrix} \txi_0 \\ \alpha \end{pmatrix} \mapsto 
\begin{pmatrix} d & -c \\ -b & a \end{pmatrix} \, \begin{pmatrix} \txi_0 \\ \alpha \end{pmatrix}
+ \frac16\, \kappa_{abc} \xi^a \xi^b \xi^c \, 
\begin{pmatrix} c^2/(c\xi^0 + d) \\ -[c^2(a\xi^0 + b) + 2c]/(c\xi^0 + d)^2 \end{pmatrix}.
\end{split}
\eea 
Moreover, one obtains the following transformations for the contact potential,the  K\"ahler potential on the twistor space and the contact one-form,
\bea
e^{\Phi} \mapsto \frac{e^{\Phi}}{|c\tau+d|},\qquad 
K_{\cZ} \mapsto K_{\cZ} - \log(|c\xi^0 + d|), \qquad
\cX^{[i]} \mapsto \frac{\cX^{[i]}}{c\xi^0 + d}.
\label{Sconttrans}
\eea 
So, it turns out that the K\"ahler potential changes by a K\"ahler transformation consistently, leaving the metric on the twistor space
$\cZ$ or the metric on the HM moduli space $\cM_{\rm HM}$ invariant.
 Furthermore the contact structure remains invariant, as the contact one form
rescales by a holomorphic factor only, ensuring that the S-duality acts by inducing a contact transformation.

\subsection{Modular constraint on the transition functions}

Now we want to deform the QK space of the previous subsection in a modular invariant fashion. More precisely, we
will assume that all the continuous isometries are broken and find a description of the QK spaces admitting isometric action of $SL(2,\IZ)$.
In such situation the twistor space description is the most efficient one. The inclusion of instantons that are responsible
for breaking of the continuous $SL(2,\IR)$ symmetry to its discrete subgroup $SL(2,\IZ)$,
refines the covering on the twistor space and the same happens for the associated set of transition functions. The first issue that one needs to address 
here is how S-duality is realized in this context.

 For this purpose, to avoid unnecessary complications, we first provide the simplest construction in terms of open patches and closed contours
\cite{Alexandrov:2012bu, Alexandrov:2013mha} and later comment on generalizations. 

Let the twistor space $\cZ$ be defined by the covering
\be
\cZ=\cU_+\cup\cU_-\cup\(\cup_{m,n} \cU_{m,n}\)
\label{coverZ}
\ee
where $\cU_\pm$ cover the north and south poles of $\CP$ as above and $\cU_{m,n}$ are mapped to each other
under the antipodal map and $SL(2,\IZ)$ transformations
\be
\varsigma\[\cU_{m,n}\]= \cU_{-m,-n},
\label{tauU}
\ee
\be
\cU_{m,n}\mapsto \cU_{\hat m,\hat n},
\qquad
\( \hat m\atop \hat n\) =
\(
\begin{array}{cc}
d & -c
\\
-b & a
\end{array}
\)
\( m \atop n \) ,
\label{mappatches}
\ee
including an invariant patch  $\cU_0\equiv \cU_{0,0}$.
Furthermore, we assume in addition that $c\xii{m,n}^0+d$ is non-vanishing in $\cU_{m,n}$ for all $(c,d)\ne (\pm m,\pm n)$,
including $(m,n)=(0,0)$, and has a simple zero for $(c,d)=(m,n)$.
Denoting this zero by $\htp^{c,d}$, the reality condition for $\xi^0$ and \eqref{tauU} imply that
\be
\varsigma\bigl[\htp^{c,d}\bigr]=\htp^{-c,-d}\equiv\htm^{c,d}.
\label{realtpm}
\ee
With the covering \eqref{coverZ} we associate the following set of transition functions
\be
\label{transfun}
\Hij{+0}=\Fcl(\xii{+}) ,
\qquad
\Hij{-0}=\bar{\Fcl}(\xii{-}) ,
\qquad
\Hij{(m,n)0}= G_{m,n}(\xii{m,n},\txii{0},\ai{0}),
\ee
where $\Fcl(X)$ is the cubic prepotential.\footnote{Note that this assumption
implies that $\cU_{km,kn}$ for $k\ge 1$ are all identical, in particular $\cU_{0,\pm k} = \cU_{\pm}$.
However, we consider all such patches to be different. This sacrifice in rigor renders the presentation to be less cumbersome.
A more precise analysis is actually possible, but it does not change any result.}
The functions $G_{m,n}$ are not arbitrary. Besides the reality conditions
\be
\overline{\varsigma\bigl[G_{m,n}\bigr]}=G_{-m,-n},
\label{realG}
\ee
they must transform in such a way so that the modular invariance is ensured. The action of S-duality transformation
must preserve the contact structure. A simple way to realize this is through demanding that the Darboux coordinates transform
according to \eqref{SL2_dar}, where the only new feature is that the Darboux coordinates are labeled by patch indices $[m,n]$. 
Under S-duality, the patches that cover the twistor space should reshuffle to each other.
In particular it means that a different patch with indices $[m',n']$ should be related to the patch indexed by $[m,n]$ by S-duality transformation
as the following, 
\be
\( m'\atop n'\) =
\(
\begin{array}{cc}
a & c
\\
b & d
\end{array}
\)
\( m \atop n \).
\label{mpnp}
\ee

Performing $SL(2,\IZ)$ transformations on the gluing conditions \eqref{Gluingcond} one then finds that
the holomorphic transition functions are required to satisfy the condition,
\bea
\begin{split}
G_{m,n}\mapsto &\, \frac{G_{m',n'}}{c\xi^0_{[m',n']}+d}
- \frac{c}{6}\,\kappa_{abc}\, \frac{3 \xii{m',n'}^a - 2 T^a_{m',n'}}{(c\xii{m',n'}^0 +d)(c\xii{0}^0 +d)}\,T^b_{m',n'}T^c_{m',n'}
\\
+ &\,
\frac{c^2}{6}\,\kappa_{abc}\, \frac { 3 \xii{0}^a \xii{m',n'}^b +T^a_{m',n'}T^b_{m',n'}}
{(c\xii{m',n'}^0 +d)(c\xii{0}^0 +d)^2}\, T^c_{m',n'}T^0_{m',n'}
\\
- &\,\frac{c^3}{6}\, \frac{\kappa_{abc}\,\xii{m',n'}^a \xii{m',n'}^b\xii{m',n'}^c  }
{(c\xii{m',n'}^0 +d)^2 (c\xii{0}^0 +d)^2} \,\(T^0_{m',n'}\)^2,
\end{split}
\label{master}
\eea 
where
\be
T_{m,n}^\Lambda=\p_{\txii{0}_\Lambda}G_{m,n}-\xii{0}^\Lambda \p_{\ai{0}} G_{m,n}.
\label{defT}
\ee
This constraint is one of the main conditions for the $SL(2,\IZ)$ invariant construction.
As we will see below, even though this constraint looks complicated due to non-linearities, they can be removed if one chooses to work 
using the contact hamiltonians which were introduced in chapter \ref{chapter3}.

\subsection{Modular action on the twistor fiber}

Although it was not mentioned explicitly, in the previous subsection while describing the modular invariant construction of the twistor space,
we tacitly assumed that the $\IC P^1$ coordinate transforms appropriately. By this, we mean that $t$ transforms in such a way that the Darboux coordinates 
admit the action of $SL(2,\IZ)$ group as in \eqref{SL2_dar}. In our considerations, the main complication comes from the fact that $\xi^0$ is not globally defined anymore,
and as a result the transformation of the fiber coordinate also receives corrections. However, quite remarkably, it is still possible to find the ``appropriate" transformation
of $t$ by considering the transformation of the K\"ahler potential $K_{\cZ}$ on the twistor space and requiring that it transforms by K\"ahler transformation \eqref{Sconttrans} 
\cite{Alexandrov:2013mha}. In the process of finding how $t$ transforms, we have come across a truely beautiful mathematical construct which we are going to
call as invariant points. This objects are generalizations of $\mp \I$ in the classical case, which remains invariant under the action of $SL(2,\IR)$ \eqref{clt}.
We are going to explain what these objects are in more details in the following. 

\begin{itemize}
\item
The first and most important thing to consider is the transformation of $t$. It was found in \cite{Alexandrov:2013mha} that this transformation is rational and is given by,
\be
\gl{c,d}[t]=\Cfm\,\frac{t-\htp^{c,d}}{t-\htm^{c,d}},
\label{transftt}
\ee
where the constant prefactor should satisfy $|\Cfm|=|\htm^{c,d}|$ and we used the notation $\gl{c,d}$
for the $SL(2,\IZ)$ action.

Furthermore, the transformation \eqref{transftt} should satisfy a certain group law. In particular, this implies that
 $\gl{c,d}[\gl{-c,a}[t]] = t$ since the two $SL(2,\IZ)$ transformations are inverse of each other. This allows to fix the prefactor in \eqref{transftt} as
\be
\Cfpm=\gl{c,d}\[ \htpm^{-c,a}\],
\qquad
\htp^{c,d}\Cfm=\htm^{c,d}\Cfp.
\label{pref-trt}
\ee
In fact, from the reality properties of $\htpm^{c,d}$, one can easily see that the second of the above
conditions is the same as the condition on the modulus of $\Cfm$. It is easy to see that this transformation is a natural generalization of the 
transformation \eqref{clt} valid in the classical case. 

\item 
Now we turn to one of the other key features. 
It turns out that along with the roots 
of $c\xi^0 + d =0$, the critical points of $\xi^0$ also play an important role. They are defined by 
\bea
\p_t \xii{0}^0 (\tinvpm) = 0,
\eea 
and are related to each other by the antipodal map $\varsigma[\tinvp] = \tinvm$. They are instanton corrected versions of the points $\tinvpm = \mp \I$. 
These objects are called invariant points because they are fixed points of $SL(2,\IZ)$ transformation. This can be
seen from the fact that $\gl{c,d} [\p_t \xii{0}^0] \sim \p_t \xii{0}^0$ \footnote{From general expectations, we assume that
these points belong to the patch $\cU_0$. This makes the construction less cumbersome. Furthermore, in the classical case
or in the case when there are two continuous isometries, these points indeed are in $\cU_0$. The instantons corrections, when specific transition functions are considered, are supposed to
be not behaving too ``wildly" . More precisely,$\tinvpm$ should be perturbations around their classical values. 
Of course, we can relax this assumption, but we do not expect the results to change significantly by taking this into account.}
, which will become clear from the expressions of the Darboux coordinates 
that we will present in a short while. It is important to realize that $\tinvpm$ should be viewed not just as functions on the QK space $\cM$, but 
also as sections of the bundle $\cZ \rightarrow \cM$. Therefore, their invariance is properly interpreted as invariance of the sections. Thus they
satisfy the following property,
\be
\gl{c,d}[\tinvpm]=\gl{c,d}[t]_{t=\tinvpm}=\Cfm\, \frac{\tinvpm-\htp^{c,d}}{\tinvpm-\htm^{c,d}}.
\label{invpoints}
\ee

\item 
Having introduced the critical points, we are now in a position to introduce the invariant kernel which play a crucial 
role for making the modular symmetry explicit. We define it as a $t$-independent shift of the kernel appearing in \eqref{Darbouxinteq}
in the following way such that the integration measure remains S-duality invariant, 
\bea
K(\varpi,\varpi')=\frac{1}{2}\(\frac{\varpi'+\varpi}{\varpi'-\varpi}+ \kk(t')\),
\qquad \frac{\de t'}{t'}\, K(t,t') \mapsto \frac{\de t'}{t'}\, K(t,t') 
\eea
where
\bea
\kk(t)=\frac{\tinvp\tinvm -t^2}{(t-\tinvp)(t-\tinvm)}.
\eea
Hence we have the invariant kernel as
\be
\begin{split}
\frac{\de t'}{t'}\,K(t,t')=&\,
\frac{\de t'}{2}\, \frac{(t-\tinvp)(t'-\tinvm)+(t-\tinvm)(t'-\tinvp)}{(t'-t)(t'-\tinvp)(t'-\tinvm)} .
\end{split}
\label{fullinvkernel}
\ee

\end{itemize}
Now, armed with the S-duality invariant kernel, invariant points and the points $\htpm^{m,n}$, we can proceed to find 
expressions for Darboux coordinates satisfying  \eqref{SL2_dar}\footnote{For convenience, we drop the constant shift in $c_a$ which can be always restored}.

\subsubsection{The non-perturbative mirror map}

From equation \eqref{Darbouxinteq}, we see that the Darboux coordinates depend on both the coordinates on the base $A^\Lambda,B_\Lambda,B_\alpha,Y^\Lambda$
and the fiber coordinate $t$. Typically these coordinates on $\cM$ has very simple relations with type IIA coordinates and are fixed by symplectic invariance of the 
type IIA construction. However, our goal here is to rewrite the integral equations for the Darboux coordinates in such a way that the action of $SL(2,\IZ)$ on them
is realized as \eqref{SL2_dar},
 provided the right transformation of $t$ given by \eqref{transftt}. For this one has to express the coordinates  $A^\Lambda,B_\Lambda,B_\alpha,Y^\Lambda$
by type IIB physical fields $\tau,b^a,t^a,c^a,c_a,\psi$, which admit a simple and more natural action of the modular group \eqref{SL2Z}. Here, we call this relation as the 
``mirror map". Essentially, such coordinate transformations provide generalizations of the classical mirror map \eqref{symptobd}, because the relations of 
$A^\Lambda,B_\Lambda,B_\alpha,Y^\Lambda$ with type IIA coordinates are simple, as we have already mentioned. 

To this end, let us first give the mirror map relation. The constant terms in the integral equations
for the Darboux coordinates are given by \footnote{The index $j$ stands for all possible patches and contours.} 
\cite{Alexandrov:2013mha},
\bea
A^\Lambda &=& \zeta^\Lambda_{\rm cl}
-\hf\, (\tinvp+\tinvm)\(\frac{Y^\Lambda}{\tinvp\tinvm}-\bY^\Lambda\)
+\hf\sum_j \oint_{C_j}\frac{\de\varpi}{2\pi\I \varpi}\,
\kk(\varpi)\(\p_{\txii{0}_\Lambda }\Hij{j0}-\xii{0}^\Lambda \, \p_{\ai{0} }\Hij{j0}\) ,
\nn
\\
B_\Lambda &=& \tzeta_\Lambda^{\rm cl}
-\hf\sum_j \oint_{C_j}\frac{\de\varpi}{2\pi\I \varpi}\,
\kk(\varpi)\p_{\xii{j}^\Lambda}\Hij{j0},
\label{qmirror}
\\
B_\alpha &=& (\sigma+\zeta^\Lambda \tzeta_\Lambda)^{\rm cl}
-\hf\sum_j \oint_{C_j}\frac{\de\varpi}{2\pi\I \varpi}\,
\kk(\varpi)\( \Hij{j0}- \xii{j}^\Lambda \p_{\xii{j}^\Lambda}\Hij{j0}\),
\nn
\eea
where the index {\rm\small ``cl"} refers to the quantities given by the classical mirror map. Now with this and the following result for 
mirror map of $Y^\Lambda$, one can check that the Darboux coordinates indeed satisfy  \eqref{SL2Z}. 

 Let us define\footnote{The prime on the sum indicates that it runs over $(m,n)\in \IZ^2/{(0,0)}$.}
\be
Y^\Lambda=
 \I\tau_2\,\frac{\tinvp\tinvm}{\tinvm-\tinvp}\,\cY^\Lambda+
\frac{\tinvp\tinvm}{2}{\sum_{m,n}}' \oint_{C_{m,n}}\frac{\de \varpi}{2 \pi \I}\,
\frac{ \tinvp(\tinvm-t)+\tinvm(\tinvp-t)}{(t-\tinvp)^2(t-\tinvm)^2}\,T^\Lambda_{m,n},
\label{defYL}
\ee
where
\be
\cY^0=1,
\qquad
\cY^a =
\frac{b^a\(\tinvm e^{\I U}-\tinvp e^{-\I U}\)-2\sqrt{\tinvp \tinvm}\,t^a}{(\tinvm-\tinvp)\cos U}
+ \I\tau_2^{-1}\tan U \,\cI^a
\ee
and
\bea
\cI^a &=& \frac{(\tinvm-\tinvp)^2}{4\pi}{\sum_{m,n}}' \oint_{C_{m,n}}\de \varpi\,
\frac{\varpi \,T^a_{m,n}}{(\varpi-\tinvp)^2 (\varpi-\tinvm)^2},
\label{defI}
\\
U &=&\frac{\tinvm-\tinvp}{4\pi}{\sum_{m,n}}' \oint_{C_{m,n}}\de \varpi\,
\frac{ \log\(1- \p_{\ai{0} }G_{m,n}\)}{(\varpi -\tinvp) (\varpi- \tinvm)}.
\label{defU}
\eea
Then, provided at the critical points one has
\be
\begin{array}{rclcrcl}
\xii{0}^0(\tinvp)&=&\tau,
& \qquad &
\xii{0}^0(\tinvm)&=&\bar\tau,
\\
\xii{0}^a (\tinvp) &=& -c^a + \tau b^a,
& \qquad &
\xii{0}^a (\tinvm) &=& -c^a + \bar \tau b^a,
\end{array}
\label{cond-spm}
\ee
\be
\tinvpm\p_t\xii{0}^a (\tinvpm) = \pm\frac{2\I\tau_2 t^a}{\cos U}\, \frac{ \sqrt{\tinvp\tinvm}}{\tinvm-\tinvp}-\I e^{\pm \I U}L^a,
\label{cond-spmder}
\ee
where $L^a$ is an arbitrary real vector.
Then $Y^\Lambda$ should transform as \footnote{$\rhoi{m,n}$ are defined as $c\xii{m,n}^0 + d \equiv -t^{-1}(t-\htp^{m,n})(t-\htm^{m,n}) \rhoi{m,n}(t)$.}
\be
Y^0 \ \mapsto\
\frac{\Cfm\htp^{c,d}}{c(\htp^{c,d} - \htm^{c,d})^2 \rhoi{c,d}(\htp^{c,d})},
\qquad
Y^a \ \mapsto\
- \frac{\Cfm\htp^{c,d}\xii{c,d}^a(\htp^{c,d})}{(\htp^{c,d} - \htm^{c,d})^2 \rhoi{c,d}(\htp^{c,d})}.
\label{transfYL}
\ee
All these above results are not too difficult to prove and the proofs are given in appendix C of \cite{Alexandrov:2013mha}. 

Now, we can find an explicit equation determining the invariant points. Using \eqref{qmirror} and \eqref{defYL}, 
the expression for the Darboux coordinate $\xi^0$ in the patch $\cU_0$ becomes,
\bea
\begin{split}
\xii{0}^0(t)=&\,
\tau_1+\frac{\I\tau_2}{\tinvm-\tinvp}\(\frac{\tinvp\tinvm}{t}-t\)
\\
&\,
+ \frac{(t-\tinvp)(t-\tinvm)}{2t}\, {\sum_{m,n}}' \oint_{C_{m,n}}\frac{\de \varpi'}{2 \pi \I}\,t'\,
\frac{(t-\tinvp)(t'-\tinvm)+(t-\tinvm)(t'-\tinvp)}{(t'-t)(t'-\tinvp)^2(t'-\tinvm)^2}\,T^0_{m,n}.
\end{split}
\label{resxi0}
\eea
Thus, from $\p_t \xii{0}^0 (\tinvpm) = 0 $, one can obtain the equation
\be
\frac{\tinvp+\tinvm}{(\tinvp-\tinvm)^3}
=\frac{1}{4\pi\tau_2}\, {\sum_{m,n}}' \oint_{C_{m,n}}\de \varpi \,
\frac{\varpi\,T^0_{m,n}}{(\varpi -\tinvp)^2 (\varpi- \tinvm)^2} .
\label{eqspm}
\ee
Supplemented by the condition $\varsigma\[\tinvp\]=\tinvm$,
it can be solved by perturbative expansion in powers of (integrals of) the transition functions $T^0_{m,n}$ generating deformations.

Finally, let us mention that our construction is invariant under the action of a $U(1)$ symmetry that rotates the phase of the 
fiber coordinate as $t\mapsto e^{\I\phi} t$. As a particular case, this entails that the invariant points are defined upto
this rescaling $\tinvpm \mapsto e^{\I\phi} \tinvpm$. This induces an extra symmetry in the complex coordinates $Y^\Lambda \mapsto e^{\I\phi} Y^\Lambda$
and is responsible for the auxiliary coordinate discussed below  \eqref{Darbouxinteq}. A way to fix this is to demand $\tinvpm$ to be purely
imaginary \cite{Alexandrov:2013mha}. This truly imposes an additional condition, as the defining equation for $\tinvpm$ \eqref{eqspm} does not fix it. 
That the invariant points are imaginary can be viewed as ``gauge fixing" of the $U(1)$ symmetry and it is assumed to be a part of our construction.\footnote{Notice
that there exists other choices for fixing the gauge as well. One can consider $\Im Y^0 = 0$, for example. However, we choose $\Re \tinvpm = 0$, as this is the most convenient 
one.}

\subsubsection{Modular invariant potential}

As a byproduct in  \cite{Alexandrov:2013mha} a modular invariant potential 
was also constructed using the constant part of the contact potential. The existence of this function
is surprising as it encodes all deformations of the QK space respecting modular symmetry in a non-trivial way. It is given by
\bea
\label{constpartcpchk}
\begin{split}
\Fi =&\,\frac{|\tinvp-\tinvm|\, e^{-\frac12{\sum\limits_{m,n}}'\oint_{C_{m,n}}\frac{\de \varpi}{2 \pi \I \varpi}
\,\frac{\tinvp\tinvm -t^2}{(t-\tinvp)(t-\tinvm)} \log\(1- \p_{\ai{0} }G_{m,n}\)   }}
{\tau_2^{1/2}\cos\[ \frac{1}{4\pi} {\sum\limits_{m,n}}'\oint_{C_{m,n}}\frac{\de\varpi}{\varpi}\,\log\(1-\p_{\ai{0}}G_{m,n} \)\]}
\Biggl\{\hf\,\Im \( Y^\Lambda \bF_\Lambda(\bY)\)
\Biggr.
\\
& \left.
-\frac{1}{16\pi}{\sum_{m,n}}' \oint_{C_{m,n}}\frac{\de\varpi}{\varpi}\[
\(\varpi^{-1} Y^{\Lambda}-\varpi \bY^{\Lambda} \)\p_{\xii{m,n}^\Lambda}G_{m,n}
+\(\varpi^{-1} F_{\Lambda}(Y)-\varpi \bF_\Lambda(\bY) \)T_{m,n}^\Lambda\]\right\}\, .
\end{split}
\eea
In \cite{Pioline:2009qt},\cite{Bao:2009fg}, an attempt to construct such a function was made although 
its geometric origin was unclear. The work in  \cite{Alexandrov:2013mha} suggests that the modular invariant
function $\Fi$ has its origin in the constant part of the contact potential. From physics point of view,
perhaps, $\Fi$ can be interpreted as an S-duality invariant partition function.

\subsection{Virtue of contact hamiltonians : an example}

It is hard to overstate the simplicity of the description using contact hamiltonians.
In this section we provide another instance of usefulness of this. Of course, as had been 
mentioned this construction is completely equivalent to the one presented before using normal transition function depending on coordinates of two 
different patches. However, the advantage is that, the latter is more efficient than the former and contact hamiltonians contain sufficient information so that
the transition functions can be constructed from them valid to all orders. 

From \eqref{trans-contact}, $\lambda = c\xii{m,n}^0 + d$ for our case because of \eqref{Sconttrans}, which implies
that the constraint on the contact hamiltonian will be 
\bea
\hHij{i}_{m,n} \mapsto \frac{\hHij{i}_{m',n'}}{c\xii{m',n'}^0 + d}, 
\qquad
\begin{pmatrix} m' \\ n' \end{pmatrix}
= \begin{pmatrix} a & c \\ b & d \end{pmatrix}
\begin{pmatrix} m \\ n \end{pmatrix},
\label{masterch}
\eea
which is a simple modular constraint, compared to the very complicated non-linear one given in \eqref{master}.

Instead of using \eqref{trans-contact}, one can choose a more pedestrian way\footnote{Of course, both yield the same result.
We give this just to present yet another method to find the same result in this specific case.}
One can explicitly find from the transformation
of the Darboux coordinates \eqref{SL2_dar} that  
\be
\{ \gl{c,d}\cdot h, \gl{c,d}\cdot f\}=\frac{\gl{c,d}\cdot \{h,f\}}{c\xi^0+d}-c\,\gl{c,d}\cdot\(h\(\p_{\txi_0} f-\xi^0\p_\alpha f\)\),
\ee
which can also be immediately seen from the Leibnitz property of the contact brackets. 

This comes in handy because the action of S-duality can be conveniently encoded in terms of its action on
the contact bracket as,
\be
\gl{c,d}\cdot e^{\{\hHij{i},\, \cdot\,\}} \cdot\gl{c,d}^{-1}
=e^{\gl{c,d}\cdot\{\hHij{i},\, \cdot\,\}\cdot\gl{c,d}^{-1}}
=e^{\gl{c,d}\cdot\{\hHij{i},\, \gl{c,d}^{-1}\cdot\,\}}
=e^{\{(c\xi^0+d)\gl{c,d}\cdot \hHij{i},\, \cdot\,\}},
\ee
proving \eqref{masterch}.

In the next chapter, we will first discuss how to reformulate the type IIA construction for D-instantons in chapter \ref{chapter 4} in the type IIB side,
with explicit invariance under S-duality. Then we will proceed on to providing the fivebrane transition functions
valid to all orders in instanton expansion.

\newpage

\chapter{The symmetry group of $\cM_{\rm HM}$ }

\label{ch6}


In this chapter, we are going to discuss about the fate of the symmetry group of the HM moduli space $\cM_{\rm HM}$.
We have already learned that instanton corrections break the classical continuous isometries to their discrete subgroups
in chapter \ref{chapter2}. We have also seen in the previous chapter that, monodromy invariance around the large
volume point is generated due to the shift of the $B$-field \eqref{bjacr}. Furthermore, symmetries under the continuous transformations
of RR and NS scalars in chapter \ref{chapter2}, get broken down to discrete transformations as well. Finally, there is the $SL(2,\IZ)$ symmetry
of the HM moduli space. 

The goal of this chapter is to obtain closure of the group generated by monodromy around the large volume point, Heisenberg shifts
and $SL(2,\IZ)$ together. It turns out, demanding that the full quantum HM moduli space remains invariant under the action of these symmetry
groups is sufficient to achieve this task. In the process we will find that, a rather non-trivial modification of the integer monodromy transformations
given in \cite{Alexandrov:2010ca} is required. Such modifications are actually not {\it ad hoc}, but rather originates from subtleties in the
quadratic refinement \eqref{qrD}, introduced in chapter \ref{chapter 4}. Finally, we will also try to unravel the possible connections 
and thereby perhaps similar interpretation of this modification to the Freed-Witten anomaly \cite{Freed:1999vc}.

\section{Revisiting classical HM moduli space}

At the classical level, one has to consider the algebra of a semi-direct product $SL(2,\IR) \ltimes N$, where
$N=N^{(1)} \oplus N^{(2)} \oplus N^{(3)}$ is a nilpotent algebra of dimension $3b_2 + 2$, satisfying 
\footnote{We use the abbreviation $\[A,B\] = A^{-1}B^{-1} AB$ to write the commutators.}
\bea
\[N^{(1)},N^{(1)}\] \subset N^{(2)}, \qquad \[N^{(p)},N^{(q)}\] = 0, \text{if} \,\,p+q\ge 3,
\eea
where the generators $N^{(1)},N^{(2)},N^{(3)}$ transform as $b_2$ doublets, $b_2$ singlets and one
doublet under $SL(2,\IR)$ respectively \cite{Alexandrov:2012au}.  The relation of this algebra decomposition 
with the symmetries due to $SL(2,\IR)$, \eqref{bjacr} and \eqref{Heis0} are as the following.

\begin{itemize}

\item
The group elements $T^{(1)}_{(\eps^a,\eta^a)}$ are obtained by exponentiating $N^{(1)}$ (with suitable admixture 
of higher order generators) consisting of \eqref{bjacr} and \eqref{Heis0}, with non-vanishing $\eta^a$.

\item
The group elements $T^{(2)}_{\tleta_a}$ are obtained by exponentiating $N^{(2)}$ correspond to the Heisenberg shifts with non-vanishing
$\tleta_a$.

\item 
The group elements $\T3_{(\tleta_0,\kappa)}$ is obtained by exponentiating $N^{(3)}$ and coincide with Heisenberg shifts
with non-vanishing $\tleta_0,\kappa$. 

\item Unlike the Heisenberg shifts belonging to $N$, the Heisenberg shift $\zeta^0 \mapsto \zeta^0 + \eta^0$ belongs to 
to $SL(2,\IR)$ and in fact coincides with $SL(2,\IR)$ transformation $\tau \mapsto \tau + \eta^0$.
\footnote{In presence of quantum corrections how to make this identification work was already discussed in chapter \ref{chapter5}.} 

\end{itemize}

\section{Quantum HM moduli space and its symmetries}

In this section,  we are going to describe the effects of quantum corrections to the classical symmetries of $\cM_{\rm HM}$. 
We have already seen how to realize quantum S-duality symmetry for the HM moduli space in chapter \ref{chapter5}. 
Furthermore, in presence of instantons, we have also seen that to be able to identify $SL(2,\IZ)$ transformation
generated by $\begin{pmatrix} 1 & b \\ 0 & 1 \end{pmatrix}$ and Heisenberg shifts $\zeta^0 \mapsto \zeta^0 + \eta^0$,
we need to modify the transformation of the axion $c_a$ by a constant shift proportional to $\varepsilon(\gl{})$. Now what we have to do is to 
make the other pieces fit together. For what concerns us, we will have to consider the semidirect product group $SL(2,\IZ) \ltimes N(\IZ)$, when 
quantum corrections are present. It is natural to assume that under the group $N(\IZ)$ consisting of integer
monodromy transformation and large gauge transformation the HM moduli space remains invariant. However, to achieve that,
just a naive discretization of the \eqref{bjacr} and \eqref{Heis0} does not work. Here we are going to provide 
the consistent set of discrete transformations \cite{Alexandrov:2010ca}. 

Before going further, let us recall some facts about the quadratic refinements and the characteristics appear therein, as these
characteristics will be present in the transformations of the axions due to the action of large gauge transformations.

\subsubsection{Quadratic refinements and characteristics}

The quadratic refinement $\sigma_D(\gamma)$ typically appears in chiral bosonic partition function
\cite{AlvarezGaume:1986mi,AlvarezGaume:1987vm,Witten:1996hc,Freed:2000ta}. In our context, as we have already discussed it is required 
for the consistency with the wall-crossing phenomena. One can write a general solution to \eqref{qrrel} as
\be
\sigma_D(\gamma) = \exp\[2\pi\I\(-\frac12\,p^\Lambda \(q_\Lambda+A_{\Lambda\Sigma}p^\Sigma\)
+ \(q_\Lambda+A_{\Lambda\Sigma}p^\Sigma\) \theta_{\text{D}}^\Lambda
- p^\Lambda \phi_{{\text{D}},\Lambda}\) \],
\label{quadraticrefinementpq}
\ee
whre $ \theta_{\text{D}}^\Lambda$ and $\phi_{{\text{D}},\Lambda}$ are the characteristics which are also known as generalized spin
structures on $\CY$. These are defined modulo integers. The terms proportional to the matrix $A$ arise because of the fact that we have changed the 
basis and non-integrality of the charge $\gamma$ in the type IIB side 
\eqref{ratch}, as we have already seen in the previous chapter.

Although, one can think about these characteristics as just half-integer numbers, the symplectic invariance of the D-instanton transition function requires
the quadratic refinement $\sigma_D(\gamma)$ to transform under symplectic rotations as the following,
\be
\label{sympchar}
Sp(2h_{1,1}+2,\IZ)\ni\rho={\scriptsize \begin{pmatrix} \cD & \cC \\ \cB & \cA \end{pmatrix}}\ :\quad
\begin{pmatrix} \theta_{\text{D}}^\Lambda \\ \phi_{{\text{D}},\Lambda} \end{pmatrix}
\ \mapsto\
\rho
\cdot \[
\begin{pmatrix} \theta_{\text{D}}^\Lambda \\ \phi_{{\text{D}},\Lambda} \end{pmatrix}
-\frac12
\begin{pmatrix} (\cA^T\cC)_d \\ (\cD^T\cB)_d  \end{pmatrix}
\],
\ee
where $(A)_d$ denotes the diagonal of a matrix $A$.

In addition to the D-instanton characteristics, there exist a similar set of objects $\Theta=(\theta^\Lambda,\phi_\Lambda)$,
characterizing the fibration of the NS-axion circle bundle over the torus parametrized by the RR scalars. Under a symplectic transformation,
they behave in the same way as the D-instanton characteristics through \eqref{sympchar}.

Now we are in a position to give the quantum corrected transformations for Heisenberg shifts and monodromy.
\begin{itemize}

\item 
First, let us consider the Heisenberg shifts. They were given in \cite{Alexandrov:2010ca}, and are as the following,
\bea
T_{\eta^\Lambda,\tleta_\Lambda,\kappa} :
&& \zeta^\Lambda \mapsto \zeta^\Lambda + \eta^\Lambda,
\qquad \tzeta_\Lambda \mapsto \tzeta_\Lambda + \tleta_\Lambda,
\\ \nn 
&& \sigma \mapsto \sigma + 2\kappa - \tleta_\Lambda(\zeta^\Lambda - 2\theta^\Lambda) 
+\eta^\Lambda(\tzeta_\Lambda + A_{\Lambda\Sigma} - 2\phi_\Lambda) - \eta^\Lambda\tleta_\Lambda.
\eea 

Here $(\eta^\Lambda,\tleta_\Lambda,\kappa)\in \IZ^{2h_{1,1}+3}$.

\item
Next, as was discussed in \cite{Alexandrov:2010np,Alexandrov:2010ca}, the monodromy transformation in \eqref{Mon1}
should be appended by a transformation of the NS-axion as
\bea
\sigma \mapsto \sigma + 2\kappa(M_{\eps^a}),
\eea
 where $\kappa(M)$ is the character of the monodromy group, while considering all instanton corrections. Since the monodromy group is
abelian in nature, this character becomes $\kappa(M_{\eps^a}) = \kappa_a \eps^a$. The compensating spectral
flow transformation \eqref{sflow} takes the following form explicitly,
\be
\label{spectrflow}
\begin{array}{c}
\displaystyle{
p^a[\eps]= p^a + \epsilon^a p^0,
\qquad
q_a[\eps]= q_a -\kappa_{abc}p^b \epsilon^c
-\frac{ p^0}2\,\kappa_{abc} \epsilon^b \epsilon^c ,}
\\
\displaystyle{q_0[\eps]= q_0 -q_a \epsilon^a+\frac12\, \kappa_{abc}p^a \epsilon^b \epsilon^c
+\frac{ p^0}6\,\kappa_{abc} \epsilon^a \epsilon^b \epsilon^c}.
\end{array}
\ee

\end{itemize}

\section{Failure of the group law}

Even these above non-trivial adjustments turn out to be not satisfactory for obtaining a faithful representation of the duality groups. 
Let us first see where it fails precisely.
The relevant actions of various symmetries on the type IIB fields are given in the appendix A.2 
in a tabular form in \cite{Alexandrov:2014rca}. Computing the commutator between S-duality and Heisenberg transformation , we find that 
apart from monodromy we get some anomalous terms,
\footnote{$S= \begin{pmatrix} 0 & -1 \\ 1 & 0 \end{pmatrix}$ is the non-trivial generator of $SL(2,\IZ)$.}
\bea
S^{-1}\, T^{(1)}_{0,\eta^a}\, S\cdot
\(\begin{array}{c}
b^a \\ c^a \\ c_a \\ c_0 \\ \psi \end{array}\)
&=&
M_{\eta^a}\cdot
\(\begin{array}{c} b^a \\ c^a \\ c_a \\ c_0 \\ \psi \end{array}\)
+\(\begin{array}{c} 0 \\ 0 \\ A_{ab}\eta^b\\ \eta^a\(\phi_a + \frac{c_{2,a}}{8}-\hf\, A_{ab}\eta^b\)
\\ \eta^a\(\kappa_a+\frac{c_{2,a}}{24}\) \end{array}\),
\label{twistheisen}
\eea
where we remember to have set the NS5-instanton characteristics $\theta^\Lambda = 0$.
This requirement comes from the fact that for Heisenberg shift $\eta^\Lambda = 0$,
the group law (see below \eqref{grlaw}) should hold. Furthermore, if one considers the fivebrane transition functions
which we present in the next chapter, under Heisenberg shift with $\eta^\Lambda = 0$,  should be invariant \cite{Alexandrov:2010ca}.
This is ensured only if $\theta^\Lambda = 0$. 
This somewhat simplifies the transformation of the characteristics by eliminating some terms in \eqref{sympchar},
\be
\begin{split}
\phi_a \ \mapsto\ &\,\phi_a + \frac12\, \kappa_{aac} \epsilon^c,
\\
\phi_0 \ \mapsto\ &\,\phi_0 - \epsilon^a\phi_a
- \hf \(L_0(\epsilon) - \epsilon^a L_a(\epsilon) + \kappa_{aac} \epsilon^a  \epsilon^c\).
\end{split}
\label{transchar}
\ee
However, the situation is still not satisfactory, because it is clear from \eqref{twistheisen} a consistent action
of all symmetries is not generated ensuring closure of the group. 
Although in \eqref{twistheisen} the anomalous terms (i.e. the terms due to which \eqref{grlaw} fails to hold),
in the transformation of $c_0$ and $\psi$ can be removed 
by appropriately choosing $\phi_a$ and the character $\kappa_a$, the one in the transformation of $c_a$ can not
be canceled. Thus we have to conclude that the monodromy, Heisenberg and S-duality transformations fail to form
a consistent group representation.

\section{Constructing group representations}

To resolve the inconsistencies, first we notice that we can absorb the D-instanton characteristics to RR and NS scalars.
This is because these characteristics appear with the RR scalars in particular combinations, as in \eqref{charabs},
and subsequently, to ensure the invariance of the HM moduli space metric one has to absorb them in the NS axion as well. 
\be
\label{charabs}
\begin{split}
\zeta^\Lambda-\theta_{\rm D}^\Lambda\qquad\qquad \mapsto\ &\, \zeta^\Lambda,
\\
\tzeta_\Lambda-\phi_{{\text{D}},\Lambda}+A_{\Lambda\Sigma}\theta_{\rm D}^\Sigma\qquad \mapsto\ &\, \tzeta_\Lambda,
\\
\sigma +\phi_{{\text{D}},\Lambda}\zeta^\Lambda-\theta_{\rm D}^\Lambda\(\tzeta_\Lambda+A_{\Lambda\Sigma}\zeta^\Sigma\)\ \mapsto\ &\, \sigma.
\end{split}
\ee
This now amends the monodromy transformation,
\be
\label{bjacr-mod}
M_{\epsilon^a}\ :\quad  \begin{array}{l}
\displaystyle{b^a\ \mapsto\ b^a+\epsilon^a,
\qquad
\zeta^a\ \mapsto\ \zeta^a + \epsilon^a \zeta^0,}
\\
\displaystyle{\tzeta_a\ \mapsto\ \tzeta_a -\kappa_{abc}\zeta^b \epsilon^c
-\frac12\,\kappa_{abc} \epsilon^b \epsilon^c \zeta^0+A_{ab}\eps^b ,}
\\
\displaystyle{\tzeta_0\ \mapsto\ \tzeta_0 -\tzeta_a \epsilon^a+\frac12\, \kappa_{abc}\zeta^a \epsilon^b \epsilon^c
+\frac16\,\kappa_{abc} \epsilon^a \epsilon^b \epsilon^c \zeta^0
-\hf\, A_{ab}\epsilon^a\epsilon^b+\frac{c_{2,a}}{8}\, \epsilon^a,}
\\
\displaystyle{\,\sigma\ \mapsto\ \sigma -A_{ab}\eps^a\zeta^b-\hf\( A_{ab}\eps^a\eps^b+\frac14\, c_{2,a}\eps^a\)\zeta^0 +2\kappa_a \epsilon^a.}
\end{array}
\ee
The modification in monodromy transformation compared to \eqref{Mon1} is clearly due to the anomalous terms that appeared in \eqref{twistheisen}.
With this, we proceed to show how to construct a group representation involving the discrete symmetry groups of Heisenberg shift, Monodromy and S-duality.

Below in table \ref{tab-U} we display the action of the corrected symmetry transformations on the type IIB coordinates. Provided that
one fixes the character of the monodromy group as $\kappa(M_{\epsilon^a}) = -\frac{c_{2,a}}{24}$   \cite{Alexandrov:2014rca}
indeed we find the group representation. In particular, we obtain
\bea
S^{-1} T^{(1)}_{0,\eta^a} S = M_{\eta^a},
\label{grlaw}
\eea
where $S=\begin{pmatrix} 0 & -1 \\ 1 & 0 \end{pmatrix}$ is one of the generators of the $SL(2,\IZ)$ group, the so called S-generator. 
The action of the other generator $T= \begin{pmatrix} 1 & 1 \\ 1 & 0 \end{pmatrix}$ coincides with the Heisenberg shift $\tau \mapsto \tau + 1$. 
Since $SL(2,\IZ)$ group is generated as a ``word" of these S and T-transformations $\gl{} = S^{m_1}T^{n_1}S^{m_2}T^{m_2}...$,
it suffices to verify the group law separately for them. As T-transformation is one of the Heiseberg shifts it commutes with $T^{(1)}_{0,\eta^a}$
and the only non-trivial commutator that we have to check is \eqref{grlaw}.

\!\!\!\!\!\!\!\!\!\!\!\!\!\!\!\!\!\!\!\!\!\!\begin{table}
\vspace{0.cm}\hspace{-2.1 cm}
\begin{tabular}{|c|c|c|c|c|c|}
\hline $\vphantom{\frac{A^{A^A}}{A_{A_A}}}$
& $b^a$ & $ c^a$ & $c_a$ &$c_0$ & $\psi$
\\
\hline $\vphantom{\frac{A^{A^A}}{A_{A_A}}}$
$S$ &  $c^a$     &  $-b^a$  & $c_a+\frac{c_{2,a}}{8}$
& $-\psi$ & $c_0$
\\
\hline $\vphantom{\frac{A^{A^A}}{A_{A_A}}}$
$T$ &  $b^a$     &  $c^a+b^a$  & $c_a-\frac{c_{2,a}}{24}$
& $c_0 $ & $\psi-c_0$
\\
\hline $\vphantom{\frac{A^{A^A}}{A_{A_A}}}$
$M_{\epsilon^a}$ &   $b^a+\epsilon^a$    &   $c^a$  & $c_a+\frac12\, \kappa_{abc} \epsilon^b c^c+ A_{ab} \epsilon^b$
& $\begin{array}{c}c_0-\epsilon^a c_a \\-\frac16\, \kappa_{abc} \epsilon^a (b^b+2\epsilon^b)c^c \\
- \frac12 A_{ab}\epsilon^a \epsilon^b +\frac{c_{2,a}}{8}\, \epsilon^a\end{array}$
& $\psi+\frac16\, \kappa_{abc}\epsilon^a c^b c^c- \kappa_a \epsilon^a$
\\
\hline $\vphantom{\frac{A^{A^A}}{A_{A_A}}}$
$T^{(1)}_{0,\eta^a}$&  $b^a$    &    $c^a+\eta^a$  & $c_a-\frac12\, \kappa_{abc} \eta^b b^c+A_{ab} \eta^b$
&  $c_0+\frac16\, \kappa_{abc} \eta^a b^b b^c + \frac{c_{2,a}}{24}\, \eta^a $
&  $\begin{array}{c}\psi+\eta^a c_a+\frac12\, A_{ab} \eta^a \eta^b\\
-\frac16\, \kappa_{abc}\eta^a b^b (c^c+2\eta^c)\end{array}$
\\
\hline $\vphantom{\frac{A^{A^A}}{A_{A_A}}}$
$T^{(2)}_{\tilde\eta_a}$
&  $b^a$  & $c^a$ & $c_a+\tilde\eta_a$
& $c_0$  & $ \psi$
\\
\hline $\vphantom{\frac{A^{A^A}}{A_{A_A}}}$
$T^{(3)}_{\tilde\eta_0,\kappa}$ &  $b^a$  & $c^a$ & $c_a$
&  $c_0+\tilde\eta_0$ & $\psi+\kappa$
\\
\hline
\end{tabular}
\caption{The action of generators of the discrete symmetry transformations in the type IIB coordinate basis.}
\label{tab-U}
\end{table}
\vspace{0.cm}

\vspace{2 cm}

\subsection{Commutators of different elements}

Here we present the results of different commutators among various elements generating the group 
using the transformation rules from table \ref{tab-U} ($T^{(1)}_{\eps^a,0} \equiv M_{\eps^a}$). 

\begin{itemize}

\item Commutators with $S$ :

\be
\begin{split}
S^{-1} \,M_{\epsilon^a}\, S=&\,\,
T^{(1)}_{0,-\epsilon^a}\,T^{(2)}_{2 A_{ab} \epsilon^b}\, T^{(3)}_{0,\,3L_0(\epsilon) - \epsilon^a L_a(-\epsilon)},
\\
S^{-1}\, T^{(1)}_{0,\eta^a}\, S=&\,\,
M_{\eta^a},
\\
S^{-1}\, T^{(2)}_{\tleta_a} \,S =&\,\,
T^{(2)}_{\tleta_a},
\\
S^{-1}\, T^{(3)}_{\tleta_0,\kappa}\, S=&\,\,
T^{(3)}_{-\kappa,\tleta_0};
\end{split}
\label{monS}
\ee

\vspace{1 cm}

\item  Commutators with $T$ :

\be
\begin{split}
[M_{\epsilon^a},T] =&\,\,
T^{(1)}_{0,\epsilon^a}T^{(2)}_{L_a(\epsilon)}
T^{(3)}_{-L_0(\epsilon^a),\,  \epsilon^a L_a(\epsilon) - L_0(\epsilon)},
\\
[ T^{(1)}_{0,\eta^a},T]=&\,\,
[T^{(2)}_{\tilde\eta^a},T]=\unit,
\\
[T^{(3)}_{\tleta_0,\kappa},T] =&\,\,
T^{(3)}_{0,-\tleta_0 };
\end{split}
\label{commT}
\ee

\item The nilpotent subgroup generated by Heisenberg and monodromy transformations :

\be
\begin{split}
\[M_{\epsilon^a}, T^{(1)}_{0,\eta^a} \]
= &\,\,
T^{(2)}_{-\kappa_{abc}\epsilon^b\eta^c} \, T^{(3)}_{\eta^a L_a(-\epsilon),-\epsilon^a L_a(\eta)},
\\
\[M_{\epsilon^a}, T^{(2)}_{\tilde\eta_a} \]
=&\,\,
T^{(3)}_{\epsilon^a \tilde\eta_a,0}.
\\
\[T^{(1)}_{0,\eta^a}, T^{(2)}_{\tilde\eta_a} \]
=&\,\,
T^{(2)}_{0,-\eta^a\tilde\eta_a },
\\
\[T^{(1)}_{0,\eta^a}, T^{(3)}_{\tilde\eta_0,\kappa} \]=&\,
\[M_{\epsilon^a},T^{(3)}_{\tilde\eta_0,\kappa} \]=
\[T^{(2)}_{\tilde\eta_a},T^{(3)}_{\tilde\eta_0,\kappa} \]=\unit.
\end{split}
\ee

\end{itemize}

From the above we see that the modifications proposed in  \cite{Alexandrov:2014rca} indeed gives rise to a nice group representation.

\section{Some remarks on modifications of the monodromy transformation}

The monodromy transformation of the axions ensuring consistent consistent group representation of the discrete symmetries
was given in \eqref{bjacr-mod}. We have seen how the amendments came into being compared to \eqref{Mon1} \cite{Alexandrov:2010ca}.
The crucial feature of the new terms appearing in \eqref{bjacr-mod} is that they have origins in the quadratic refinement for D-instantons, or more 
precisely the D-instanton characteristics appearing in \eqref{quadraticrefinementpq}. Additionally, these modifications in monodromy transformation
ensure that the NS5-brane characteristics transform homogeneously, due to which they can now be dropped. Since the other characteristics corresponding
to D-instantons are already absorbed in the axions, in our results they do not appear (see for example \eqref{tranNS5all} in the next chapter). 

As was mentioned in chapter \ref{chapter 4}, the quadratic refinement of the intersection form is defined by 
the homomorphism $\sigma : \Gamma \mapsto U(1)$. For the self-dual three form flux $H=\de B$, the solution for them is given by \cite{Alexandrov:2010ca},
\bea
\sigma_\Theta (H) = e^{2\pi\I(-\frac12 m_\Lambda n^\Lambda + m^\Lambda \theta_\Lambda - n^\Lambda \phi_\Lambda)},
\eea
where $\Theta=(\theta^\Lambda,\phi_\Lambda)$ are NS5-brane characteristics, as introduced before and $(m_\Lambda,n^\Lambda)$ are integers.
\footnote{It might seem to be a bit of overkilling to discuss these characteristics again, as we have already seen in case of amended monodromy transformation
all of them drop. But we discuss them here as we want to contrast how the subtleties related to characteristics are traded against the subtleties in the 
monodromy transformation \eqref{bjacr-mod}. We reiterate again that doing so was crucial \cite{Alexandrov:2014rca} as it made sure the 
discrete symmetries form a group representation.} 

After absorbing the D-instanton characteristics to the RR and NS-axions, their transformation is now modified by half-integer shifts.
As we are going to see below, similar half-integer shifts arising from ``monodromy-like" transformation of certain $U(1)$ gauge fields 
appear in the context of the so called Freed-Witten anomaly \cite{Freed:1999vc}. Even though, the anomalies considered by Freed and Witten are on the worldsheet in presence of D-branes,
it seems that they are closely analogous to the anomalous terms appearing in the monodromy transformation \eqref{bjacr-mod}. 
Before proceeding, let us describe Freed-Witten anomaly briefly (and rather crudely). Then we will finish by exploring possible (and conjectural) connections of
\eqref{bjacr-mod} with them.

\subsection{Freed-Witten anomaly}
Here we recall some facts about global anomalies in the worldsheet path integral of type II string theories in presence of D-branes \cite{Freed:1999vc}.
The worldsheets that we consider are oriented Riemann surfaces, mapped to spacetime $Y$ which admits spin structure as our model contains fermions.
For the simplest case of closed worldsheets without boundary, the worldsheet measure is well defined, provided $Y$ is spin.

The situation gets complicated if one starts considering D-branes. They are supported on an oriented submanifold $Q$  on which strings can end.
The boundary of the worldsheet $\p\Sigma$ maps to this $Q$. On $Q$ there is a field $A$ which is considered to be a $U(1)$ gauge field conventionally, turns
out to have correct interpretation as a Spin$^{\rm c}$ connection \cite{Freed:1999vc}. Furthermore,
the $B$ field considered in this discussion is topologically trivial and for convenience, additionally we set it to zero.  

The relevant contribution to the worldsheet path integral that one has to consider is then 
\bea
{\rm pfaff}\,(D) \times e^{2\pi\I \oint_{\p\Sigma} A},
\label{FWpint}
\eea 
where ${\rm pfaff}\,(D)$ is the pfaffian of the worldsheet Dirac operator and the second factor is the holonomy
of $A$ around the boundary of the boundary of $\Sigma$. 

The main result of  \cite{Freed:1999vc} concerns the computation of the anomaly in ${\rm pfaff}\,(D)$, which is a section of a line bundle
over the space of parameters mapping the worldsheets $\Sigma$ to the spacetime $Y$. This bundle carries a natural metric and connection.
The anomaly can then be understood as an obstruction to the existence of a global flat section having unit norm. It was shown by Freed and Witten that
this connection is flat, but there is a holonomy $\pm 1$ determined by a certain mod 2 class (which in this case is the second Steifel-Whitney class) of the 
normal bundle $N$ to the D-brane worldvolume $Q$.  

For the path integral to be well-defined, there must be a compensating anomaly coming from the $A$ field. For the case when $B=0$,
this can be understood as a consequence of $A$ being a Spin$^{\rm c}$ connection, from which it can be deduced that the $W_3(N) = 0$.
\footnote{$W_3$ is the canonical integral lift of the third Stiefel-Whitney class $w_3$ and is the obstruction to ${\rm Spin}^c$ structure.
It is related by the connecting homomorphism, the so called Bockstein map to $w_2$, as $W_3(N)=\beta(w_2(N))$.
From the long exact sequence in cohomology of $Q$ induced from the short exact coefficient sequence \cite{Freed:1999vc}, one can show that $2W_3(N) = 0$
and $W_3(N)=0$ precisely when $w_2(N)$ can be lifted to a class in $H^2(Q,\IZ)$. In fact, $W_3$ reduces to $\mod\, 2$ to the Stiefel-Whitney class $w_3$,
and this is actually true for all odd Stiefel-Whitney classes}

Let us analyze this case a bit further.
The anomaly cancellation for closed surfaces imply that the result for \eqref{FWpint} does not depend on spin structures and one can
set the number of right and left movers to be equal. This makes the Dirac operator real and hence ${\rm pfaff}\,(D)$ is real. This ${\rm pfaff}\,(D)$
is not a number and one can not choose a particular sign for it. Consider a family of $\Sigma$ parametrized by a circle $C$. Then we have a map
$\phi : \Sigma \times C \mapsto Y$ and $\phi(\p\Sigma\times C) \subset Q$. It was shown in \cite{Freed:1999vc} that going around the loop changes the 
sign as 
\bea
{\rm pfaff}\,(D) \mapsto (-1)^\alpha \, {\rm pfaff}\,(D),
\eea
where $\alpha = \(\p\Sigma \times C,\phi^{*}(w_2(Q))\)$. In particular, the ambiguity in definition exists if $w_2(Q)$ is non-zero (second Stiefel-Whitney class).

Given the normal bundle $N$ to $Q$ in $Y$ one has the following 
\bea
\!\!\!\!\!\!\!\!\!\!\!\!\!\!\!\!\!(1+w_1(Y) + w_2(Y) + ...) \!= \!(1+w_1(Q) + w_2(Q) + ...)(1+w_1(Q) + w_2(Q) + ...).
\eea
Since $Y$ is spin, $w_1(Y) = w_2(Y) = 0$ and hence 
\bea
\alpha = \(\p\Sigma \times C,\phi^{*}(w_2(N))\) \,\,\text{as} \,\, w_2(Q) = w_2(N),
\eea
which have a more natural formulation in K-theory \cite{Freed:1999vc}. 

Now for the path integral to be well-defined the second factor should have the exact same ambiguity, so as to cancel that of the first.
To understand its geometric origin, let us consider the Levi-Civita connection $\omega$ on $Q$. The structure group $SO(n)$ can be 
represented as a double cover Spin $(n)$. The trace of the holonomy connection is 
\bea
{\rm Tr}\,{\cal  P} \,\exp\(2\pi\oint_{\p\Sigma} \omega\).
\eea
The sign ambiguity here comes due to the fact that there are two ways to lift $SO(n)$ to $Spin(n)$. 
Then the product of the second factor and the above is well-defined if the worldsheet integration measure is well-defined as argued 
by Freed and Witten. The important point is that in the trace of the holonomy around $\p\Sigma$ the only spinors that appear have charge $1$ with respect to 
$A$. Such spinors are actually sections of $S(Q)\otimes \cL$, where $\cL$ is the line bundle whose connection is $A$ and $S$ is the spin representation of 
$SO(n)$. Such tensor product is called to be a Spin$^{\rm c}$ structure of $Q$. This shows why the geometric nature of the ``gauge field" $A$ should
be indeed that of a Spin$^{\rm c}$  connection. 

When we say ``Freed-Witten" anomaly in the following, we do not ascribe the same meaning to it as in the literature. Instead cavalierly we mean this particular 
sign ambiguity, the consideration of which as we have seen, is extremely crucial for the worldsheet path integral to be well-defined. 
In the next section we speculate if the half-integer terms in our modified monodromy transformation can be given a similar geometric interpretation
to that of the sign ambiguity considered by Freed and Witten.

\subsection{Connection of the monodromy transformation \eqref{bjacr-mod} with Freed-Witten anomaly}

One of the important contents of the global worldsheet anomalies described in the previous subsection 
is that it originates form monodromy around a closed loop of the ``$U(1)$ gauge field" $A$. As we have already found,
in our case the monodromy transformation acquires similar half-integer terms as in \eqref{bjacr-mod}. 

In \eqref{bjacr} monodromy transformation has a simple interpretation. 
As far as they are concerned, the transformations can be obtained by shifting the fields $b^a \mapsto b^a + \eps^a$,
in the following relation,
\bea
A^{\rm even} e^{-B} = \zeta^0 - \zeta^a \omega_a - \tzeta_a\omega^a - \tzeta_0\omega_{\CYm},
\label{RRdef}
\eea
where the RR-potential in $A^{\rm even} \in H^{\rm even}(\CYm,\IR)$.  While doing that the RR-potential has to be kept fixed. 
As we have already seen, for the classical case whereas the NS-axion remains invariant, for the quantum case it transformed
by the character of the monodromy group for \eqref{Mon1}. 

As an attempt to explain the anomalous terms in \eqref{bjacr-mod}, one can think of the following simple resolution.
If one defines the RR-potentials as
\bea
\(A^{\rm even}+\frac18 c_2(\CYm) + \cA\) e^{-B} = \zeta^0 - \zeta^a\omega_a - \tzeta_a\omega^a -\tzeta_0\omega_{\CYm},
\label{wrongRR}
\eea
shifting the $B$ field actually generates the anomalous terms in \eqref{bjacr-mod}, provided one does not transform the 
RR-potential and $\cA$ is a two-form such that $\cA\wedge B = A_{ab} \omega^a b^b$ and $\cA\wedge B\wedge B = A_{ab} b^a b^b\omega_{\CYm}$.
But then it requires us to have 
\bea
A_{ab} = \kappa_{abc} \cA^c.
\label{wrongA}
\eea
When such modified definition for the RR-potential is considered, the imaginary part of the D-instanton 
action is also affected,
\be
\label{Sbbrane}
S_{\gamma}= 8\pi e^{(\phi+\cK)/2} \left| \int_{\hat \cX}  e^{-\mathcal{J}} \gamma' \right|
+ 2\pi \I  \int_{\hat \cX}  \gamma' \wedge \(A^{\rm even}+\frac18\, c_2(\hat\cX)+\cA\) e^{-B} .
\ee

But the problem is that \eqref{wrongA} fails to hold. We actually have checked with various Calabi-Yau data.
Surprisingly for many of them \eqref{wrongA} actually works. However for a few it fails to hold and in the following we are 
going to provide an example of such.

Let us consider the Calabi-Yau embedded in the product of projective spaces $P^2\times P^3$,
\bea
  \begin{pmatrix}
    3| & 3 & 1 \\
    2| & 2 & 1  \\
  \end{pmatrix}.
\eea
The above is the so called configuration matrix defined as a pair $[{\bf n}| {\bf q}]$. 
It specifies the embedding space and degrees of the homogeneity of the defining polynomials.
For this example, we have 
\bea
f_{abc\alpha\beta} x^a x^b x^c y^\alpha y^\beta = 0,
\qquad
g_{a\alpha} x^a y^\alpha = 0.
\eea
However, the configuration matrix does not specify the coefficients $f_{abc\alpha\beta}$ and $g_{a\alpha}$ otherwise \cite{Hubsch:1992nu}. 
Then from the adjunction formula, one can find that 
$c_{2,1} = 44, c_{2,2}=36$. One can also compute the intersection numbers,
$\kappa_{111}=2,\kappa_{112}=5,\kappa_{122}=3,\kappa_{222}=0$. 
Furthermore, one can calculate the Euler characteristic and Hodge numbers following \cite{Hubsch:1992nu}. They come out to be $-150$ and
$h^{11}=2,\,h^{12} = 77$ respectively.  
Since we have $\frac12\,\left(\kappa_{aab}\eps^b-\kappa_{abc}\eps^b\eps^c\right) = 0\, {\rm mod}\, \IZ$,
one finds that
\bea
A_{ab} = \frac12 \kappa_{aab} \,\,{\rm mod}\,\, \IZ,
\eea
where we have used the fact that $L_a(\eps)$ is integral from \eqref{propA}. 
Then we obtain
\bea
A_{11} = 0, \qquad A_{22} = 0, \qquad A_{12}=\frac12.
\eea
The above values of for the elements of the matrix $A$ is inconsistent with \eqref{wrongA}.
From \eqref{wrongA}, we have
\bea
\begin{split}
A_{11} &=\, \kappa_{111}\cA^1 + \kappa_{112}\cA^2= 2\cA^1 + 5\cA^2 = 0 \,\,\, \mod\, \IZ,
\\
A_{12} &=\, \kappa_{121}\cA^1 + \kappa_{122}\cA^2 = 5\cA^1 + 3\cA^2 = \frac12\,\,\, \mod\,\IZ
\\
A_{22} &=\, \kappa_{221}\cA^1 + \kappa_{222}\cA^2 = 3\cA^1 = 0 \,\,\,\mod \,\IZ.
\end{split}
\eea
Then it means that, for some integer $m$, we have $\cA^1 = m/3$. Hence, from the first of the equations, for some integer $n$,
one obtains $\cA_2 = \frac15\left(n-\frac{2m}{3}\right)$. Plugging into the second equation, we get $A_{12} = \frac{4m+9n}{15} \, \mod\, \IZ$, which clearly is a contradiction.
Thus, we conclude that the way of modifying \eqref{wrongRR} is too naive. 

\vspace{2.5 mm}

The other natural possibility that one can consider is that the RR-potential transforms itself,
although, this proposal is still conjectural. If that is the case indeed,
one can interpret the anomalous terms in \eqref{bjacr-mod} have similar origins to that of Freed-Witten
anomaly, because of the following reasons.
There is a striking similarity between the origin of the anomalous terms acquired by the transformation
of the RR-potential $A^{\rm even}$ under integer monodromy transformation and the half-integer terms that
appeared from the $U(1)$ gauge field $A$ in the context of Freed-Witten anomaly. Since in our case
the RR-scalars appear as periods of the RR-potential, it is reasonable to expect that the anomalous terms in 
\eqref{bjacr-mod} have a similar geometric origin. Furthermore, one expects that that these half-integer terms appear 
from the subtleties of the one-loop determinant around the D-instanton background, in a similar vein with \cite{Freed:1999vc}.

\newpage

\chapter{Mirror symmetry in action : instantons in type IIB formulation}

\label{ch7}

In this chapter our goal is to describe $\cM_{\rm HM}$ including all possible instanton corrections. In contrast
to the situation a few years ago, now we have a better handle on the instanton corrections, thanks to the works in
\cite{Alexandrov:2008gh,Alexandrov:2009zh,Alexandrov:2009qq,Alexandrov:2010ca,Alexandrov:2012au,Alexandrov:2014mfa,
Alexandrov:2014rca}, which crucially depended on the advent of the twistor space formalism to describe quaternion K\"ahler manifolds,
which was developed in \cite{Alexandrov:2008nk}. The advantage of working at the level of the twistor space was mentioned in 
chapter \ref{chapter3}. The holomorphic structure of the twistor space $\cZ$ and various symmetries of the QK space lifted to the twistor space,
can be exploited to compute these instanton corrections.

In chapter \ref{chapter 4}, we have already seen the construction of D-instanton corrected HM moduli space on the type IIA side
to all orders in instanton expansion. The problem of incorporating NS5-instanton correction can be addressed most economically
on the type IIB side \cite{Alexandrov:2014mfa}. Since D5 and NS5-instantons transform as a doublet under $SL(2,\IZ)$, once we know 
the D5-instantons in type IIB, we can invoke the S-duality constraint \eqref{master}, or equivalently \eqref{masterch} to find fivebrane 
instantons. Starting from the D-instantons on the type IIA side, one just applies mirror map relations to find the bound state of
D5-D3-D1-D(-1) instantons on the type IIB formulation. Then one generates fivebrane instantons by application of S-duality constraint. 
What is still missing, is the modular invariant construction when one considers the bound state of D3-D1-D(-1) instantons. An attempt
to find it was made in \cite{Alexandrov:2012au} at the large volume limit and one-instanton approximation. Even at that level,
the picture was not satisfactory. The main obstacle in this regard stems from the fact that D3-brane instantons are modular invariant themselves. 
Unlike the case for fivebrane instantons which were {\it derived}, for D3-instantons the construction should be {\it consistent}
with modularity. To ensure this consistency is a rather non-trivial problem. However, the mathematical structure of D3-instantons
found in \cite{Alexandrov:2012au} is no less than fascinating. It discerns a rich interplay between mock modular functions and
contact geometry, {\it a priori} two completely different branches of mathematics,. It is one of our goals to investigate this problem further to 
have the final word for D3-instantons.

\section{D(-1)-D1 instantons}

Let us first recall the construction of  D(-1)-D1 instantons along the lines of  \cite{Alexandrov:2009qq}. They correspond to instantons 
with vanishing magnetic charge $p^\Lambda$. Through mirror symmetry, they are related to the A-D2 instantons in the type IIA side. 
The metric when these instantons are the only ones considered was given in chapter \ref{chapter 4} equation \eqref{mett1} where the expression 
is written in type IIA coordinates. In this case the Darboux coordinates $\xi^\Lambda$ are globally defined and the integrals appearing in the twistor lines
for the Darboux coordinates can be explicitly evaluated. Furthermore, since there is no $\txi_\Lambda$ and $\alpha$ dependence in this case,
the non-linearities in the S-duality constraint \eqref{master} disappear and it coincides with \eqref{masterch}, satisfying a simpler modular constraint 
\bea
G_{m,n} \mapsto \frac{G_{m,n}}{c\xi^0 + d} + \, {\rm reg.}
\eea
Furthermore, the transition functions originate from simple modular forms, due to which 
coincide with the contact hamiltonians, as was remarked in \ref{chapter3}. 

Let us briefly describe how $SL(2,\IZ)$ invariant construction of D(-1)-D1 instantons was obtained starting from electrically charged D2 instantons 
in the type IIA formulation as was done in \cite{Alexandrov:2009qq}. One starts with the expression for the D-instanton transition function \eqref{Dinsttrans}
where the charge lattice is restricted to $p^\Lambda = 0$. Then a Poisson resummation with respect to $q_0$ is performed. Thereby, one finds
a sum over two integers $(m,n)$ where the first one appears due to expansion of the dilogarithm counting multi-covering effects and the second one appears as a result of resummation,
i.e. $q_0$ is traded for $n$. Similar situation arises for the Darboux coordinates as well. In particular, it turns out that for each of the expression of 
the Darboux coordinates, there is a term that transforms as \eqref{SL2_dar} and another anomalous piece. However, such anomalous terms can be removed 
if one performs local contact transformations appropriately in the patches of definition of Darboux coordinates. As a result one indeed finds an $SL(2,\IZ)$ invariant
construction of the twistor space. 

The resulting twistor space can be described by \eqref{coverZ} where the patches $\cU_{m,n}$ surround the roots $t_+^{m,n}$ of the equation $m\xi^0+n = 0$.
The transition functions are given by
\be
G_{m,n}^{\rm D1}(\xi)
= -\frac{\I}{(2\pi)^3}\!\!\sum_{q_a\in H_2^+\cup\{0\}}\!\!\!\! n_{q_a}^{(0)}\,
\begin{cases}
\displaystyle\frac{e^{-2\pi \I m q_a\xi^a}}{m^2(m\xi^0+n)}, &
\quad  m\ne 0,
\\
\displaystyle (\xi^0)^2 \,\frac{e^{2\pi \I n q_a\xi^a/\xi^0}}{n^3}, &\quad  m=0.
\end{cases}
\label{defG1}
\ee
Here we set  $ n_{0}^{(0)}=-\chi_\CYm/2$ and used that
\be\label{instnr}
\begin{split}
\Omega(\gamma_{\rm D1}) =&\, n_{q_a}^{(0)} \ \quad\mbox{\rm for}\quad \gamma_{\rm D{1}}=(0,0,\pm q_a,q_0), \quad \{ q_a \} \ne 0,
\\
\Omega(\gamma_{\rm D(-1)})=&\, 2 n_{0}^{(0)}\quad\mbox{\rm for}\quad \gamma_{\rm D{(-1)}}=(0,0,0,q_0).
\end{split}
\ee
It is useful to note also that the contributions to \eqref{defG1} with $m=0$ are nothing
but the $\alpha'$-corrected part of the prepotential, $\sum_{n>0}G_{0,n}^{\rm D1}=F^{\alpha'\text{-loop}}+F^{\rm ws}$.
Since $\cU_{0,n} = \cU_+$, the transition functions to the poles are determined by adding the respective classical pieces. 

The resulting covering of the twistor space looks completely different from the melon-shaped covering on  the type IIA side from which we started.  
This fact can be reconciled by remembering that in order to find modular invariant description of the twistor space we had to perform 
certain gauge transformations. The generating functions can be singular outside their patch of definition. As a result, the gauge transformed 
Darboux coordinates can have different singularity structure than the ones we started with. In  \cite{Alexandrov:2009qq} through a 
careful analysis, it was shown that the constructions in the type IIA and type IIB formulations are indeed the same. 

Having found the $SL(2,\IZ)$ invariant description for D(1)-D1 instantons, we first proceed to obtain the fivebrane transition functions,
as was done in \cite{Alexandrov:2014rca,Alexandrov:2010ca}. We are going to assume that there is no bound state of the form D3-D1-D(-1)
present, which can be restated as setting $\Omega(\gamma) = 0$ for the charges $\gamma = (0,p^a,q_a,q_0),\, p^a \ne 0$.
We will finish this chapter by showing that the moduli space $\cM_{\rm HM}$ in presence of fivebrane instantons under the action of discrete
isometries discussed in chapter \ref{ch6}.

\section{Subtleties of NS5-brane contributions}

Before giving the procedure of finding the fivebrane contributions, let is make some remarks
on why NS5-brane contributions are rather difficult to find. Even though a result to all orders in the instanton 
expansion was found in \cite{Alexandrov:2014mfa,Alexandrov:2014rca}, there are several issues that 
require better understanding. 

\begin{itemize}

\item First of all the dynamics of NS5-branes is not actually well understood. The worldvolume theory, in particular 
for the case of multiple fivebranes is not known to the full extent. 

Let us consider type IIA formulation compactified on a Calabi-Yau $\CY$, with a stack of $k$ Euclidean  
NS5-branes whose worldvolume $W$ is $\CY$ itself. For $k=1$ the worldvolume supports five scalar fields (describing
transverse fluctuations of $W$ in $\IR^{10} \times S^1$, where $S^1$ is the M-theory circle), two symplectic Majorana-Weyl fermions 
and the two-form potential $B$, whose field strength $H$ is self-dual,
\bea
*_W H = \I H.
\label{SDH}
\eea
The worldvolume $W$ behaves as a magnetic source for the two form potential $B$ which propagates in 
the 10 dimensional bulk of spacetime. On the other hand, the field strength $H$ for the NS two form field B
acts a source of the RR 3-form potential $\hat{A}_3$. This flux behaves as if D-branes are bound to a single NS5-brane.

The problems are two fold. First, due to the self-duality condition \eqref{SDH} the 3-form flux can 
not be defined properly for the bound state of D-branes and NS5-branes if the intersection product 
of 3-cycles supprting D-branes is non-vanishing. This means that the partition function describing the 
dynamics of the NS5-brane should be interpreted as a section of some bundle.

One has to be rather careful while considering the situation $k>1$.
A way to do that is by following \cite{Belov:2006jd}, although a deeper understanding is required
According to it, $k$-fivebranes split into twisted sectors labeled by the pairs of integers $(p,q)$ such that
$pq = k$. It is interpreted as $k$ fivebranes wrapping $\CY$ recombine into $q$ fivebranes wrapping $p$
copies of $\CY$. The dynamics of a $(p,q)$ twisted sector is described by $q^3$ interacting self-dual
two forms $B_{ijk}$. $q$ among them $B_{iii}$ stay massless and the abelian two form $B = {\sum_{i=1}^q} B_{iii}$
describes center of mass degrees of freedom.

\item The NS5-instantons impart true stringy corrections to the HM moduli space $\cM_{\rm HM}$. In absence
of $\alpha$ dependence in the transition functions, one can exploit the QK/HK correspondence \cite{Alexandrov:2011ac}. It is a procedure 
to map the QK space having an isometry to an HK space with a hyperholomorphic line bundle (whose curvature is
a $(1,1)$-form with respect to all of three complex structures) and the transition functions on twistor space of the HK space does not have
explicit $t$ dependence. Although, while giving the twistor description of the D-instantons in chapter \ref{chapter 4}, we did not resort
to it, it was quite clear that we restricted to the symplectic subspace of $(\xi,\txi)$ on the twistor space. When NS5-instantons are 
present all continuous isometries are broken as we have seen in chapter \ref{chapter2} and one has to work using the
full-fledged contact geometry. 

\end{itemize}

\section{Fivebrane corrections}

Before proceeding, let us summarize the strategy we are going to take to derive fivebrane transition functions
on the type IIB side. We will include D(-1)-D1 instantons from the very beginning. The BPS rays for electric D-instantons with charge $\tilde{\gamma}=(0,0,0,\tilde{q}_0)$
are aligned along the imaginary axis and BPS rays corresponding to D-instantons with charge vector  $\tilde{\gamma}=(0,0,\tilde{q}_a,\tilde{q}_0)$ are colored by green in figure 7.1.
It was shown in \cite{Alexandrov:2009qq} how to rotate these BPS rays by first performing a Poisson resummation on 
the $\tilde{q}_0$ and then doing gauge transformation as shown in figure 7.1. It was found that, starting from the A-D2 instantons on the
type IIA side, one finds a modular invariant construction for D(-1)-D1 instantons on the type IIB side. However,
the D5-instanton transition functions (across the red BPS rays in figure 7.1) do not remain unaffected due to this procedure.
They become gauge transformed and the corresponding contact hamiltonians are shown in figure 7.2. 
We are almost done now.
The application of S-duality constraint 
\eqref{masterch} should give us fivebrane contact hamiltonian in presence of D(-1)-D1 instantons. In the following,
in a step by step fashion, we show how to implement this strategy concretely.

\subsection{Gauge transformation due to D(-1)-D1 instantons}

To display the effect of gauge transformation, we need to first introduce some conventions. We define an 
ordering of the charges as $\gamma > \gamma'$ iff $0<\arg(Z_\gamma Z_{\gamma'}^{-1}) <\pi$. 
Then for each charge $\gamma$ we define an associated set of D(-1)-brane charges
whose BPS rays lie in the same half-plane as $\ell_\gamma$
\be
\GamD{-1}_\gamma=\left\{\gamD{-1}=(0,0,0,\tilde q_0)\ :\
\tilde q_0\Re Z_\gamma>0
\right\},
\label{latDm1}
\ee
and another set of D1-brane charges
for which the BPS rays are between $\ell_\gamma$ and the imaginary axis
\be
\GamD{1}_\gamma=\left\{\gamD{1}=(0,0,\tilde q_a,\tilde q_0)\in H_2^+\cup H_2^-\ :\
\Nq(\gamD{1})=\Nq(\gamma) \ {\rm and}\
\begin{array}{c}
\gamD{1}> \gamma \quad \mbox{for}\ \Nq(\gamma)\ {\rm odd}
\\
\gamD{1}\le \gamma \quad \mbox{for}\ \Nq(\gamma)\ {\rm even}
\end{array}
\right\},
\label{latD1}
\ee
where $H_2^+$ is the set of charges corresponding to effective homology classes on $\CYm$,
$H_2^-$ is the set of opposite charges, and $\Nq(\gamma)$ denotes the quadrant which
$\ell_\gamma$ belongs to.\footnote{One can write
$\Nq(\gamma)=\left\lfloor\frac{2}{\pi}\, \arg \left(\I Z_\gamma\right)\right\rfloor$.}
Note that both the ordering and the two charge sets $\GamD{\pm 1}_\gamma$ may change after crossing a wall of marginal stability.
Given this, we take the BPS rays corresponding to gauge transformation generating transition function on $\cU_\gamma$ to lie on the anti-clockwise direction of the 
from the BPS ray $\ell_\gamma$ (see the figure 7.1).

\vspace{- 5 cm}

\begin{figure}
    \centering
    \includegraphics[scale=0.5]{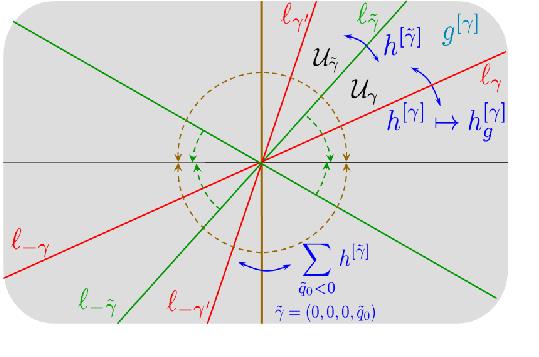}
    \caption{Example of the BPS rays corresponding to D5 (red), D1 (green) and D(-1) (brown) branes and the effect of the gauge transformation which rotates the two latter types of rays to the real line. $\cU_\gamma$ denotes the patch lying in the counterclockwise direction from the BPS ray $\ell_\gamma$.}
\end{figure}

\vspace{ 5.5cm}

The gauge transformation contact hamiltonian is then given by
\be
\gi{\gamma}=(-1)^{\Nq(\gamma)}\[\frac{1}{2}\sum_{\gamD{-1}\in\GamD{-1}_\gamma} \hHij{\gamD{-1}}+
\sum_{\gamD{1}\in\GamD{1}_\gamma}\hHij{\gamD{1}}\].
\label{fungengauge}
\ee
This has a very simple meaning. The D1-instanton BPS rays and correspondingly the discontinuities are rotated to either
of the positive or negative real axis, whichever is the closest. The D(-1)-instantons before gauge transformation were aligned 
along the imaginary axis. After the gauge transformation, they are split into two halves and each of them are also rotated to the real axes. 

As a result the contours of D(-1) and D1 instantons are now along the positive or negative real axes and the corresponding 
contact hamiltonians can be summed up as they depend only on $\xi^\Lambda$.  Poisson resumming over $\tilde{q}_0$ leads to the alternative twistor description,
complying with the S-duality constraints. Instead of BPS rays now we consider the usual S-duality invariant contours for
D(-1)-D1 case, which are circles surrounding the points $t_{\pm}^{m,n}$. The corresponding contact hamiltonians coincide with \eqref{defG1} as
$h_{m,n}^{\rm D1} = G_{m,n}$,. 

\subsection{D5-instantons after gauge transformation}

The gauge transformation imparts non-trivial effects on the D5-instanton transition functions. 
This is due to the fact that while rotating the BPS rays for D(-1) and D1-instantons, one necessarily crosses the 
BPS rays corresponding to D3 or D5 instantons for generic cases where the charges are mutually non-local. 
However, since the gauge transformation contact hamiltonians are dependent of $\xi$ only, the resulting shift in
the D5 contact hamiltonians are linear shifts \eqref{gaugehHxi},
\be
\hHij{\gamma}_g(\xi,\txi)=H_{\gamma}(\Xi^{(g)}_\gamma),
\qquad
\Xi^{(g)}_\gamma=  \Xi_\gamma+p^\Lambda\p_{\xi^\Lambda}\gi{\gamma}(\xi).
\label{Hgam-tr}
\ee
One can also compute the corresponding transition functions which take the following form,
\be
\Hij{\gamma}_g
= \hHij{\gamma}_g +2\pi^2 q_\Lambda p^\Lambda (\hHij{\gamma}_g)^2
-\frac{(-1)^{\Nq(\gamma)}}{4\pi^2}\sum_{\gamD{1}\in\GamD{-1}_\gamma\cup \GamD{1}_\gamma}
\bn_{\tilde q}\,e^{-2\pi\I\tilde q_\Lambda \xi^\Lambda}\cE\(4\pi^2 \tilde q_\Lambda p^\Lambda \hHij{\gamma}_g\),
\label{trHTijall}
\ee
where
\be
\cE(x)=1-(1+x)\, e^{-x}.
\label{fun}
\ee
The twistor space takes the form of figure 7.2. 

\begin{figure}
    \centering
    \includegraphics[scale=0.5]{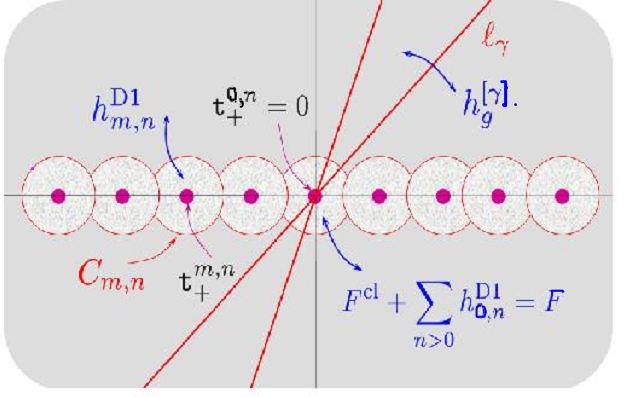}
    \caption{Schematic representation of the twistor data generating D(-1)-D1 and
                   D5-instantons in the type IIB picture}
\end{figure}

\subsection{Wallcrossing and gauge transformation}

Let us demonstrate that the gauge transformation \eqref{Hgam-tr} is consistent with wallcrossing phenomena.
For brevity, let us introduce the notation $\expe{x}=e^{2\pi\I x}$.

First, consider the original Kontsevich-Soibelman wallcrossing formula \eqref{ewall} when apart from D5-instantons,
there are only electric D-instantons having charge vector $\tilde{\gamma} = (0,0,0,.\tilde{q}_0)$ along with
the  bound states of D-instantons with charges $\gamma=(p^0,p^a,q_a,q_0)\,\, p^0\ne 0 $. The D-instantons with charge $\tilde{\gamma}$ are aligned along the 
imaginary axis. Then the KS formula is
\be
\prod_{\gamma_j\in \GamD{5}_1}  U_{\gamma_j}^{-\Omega^+ (\gamma_j)}
\prod_{\tilde q_0>0} U_{\gamD{-1}}^{-\chi} \prod_{\gamma_i\in \GamD{5}_2} U_{\gamma_i}^{-\Omega^+ (\gamma_i)}
\prod_{\gamma_i\in \GamD{5}_1}  U_{\gamma_i}^{\Omega^- (\gamma_i)}
\prod_{\tilde q_0>0} U_{\gamD{-1}}^{\chi} \prod_{\gamma_j\in \GamD{5}_2} U_{\gamma_j}^{\Omega^- (\gamma_j)}
=1,
\label{wcfD5D(-1)}
\ee
where we look at the upper half plane only in figure 7.1. + and - indices over the DT invariants denote chambers after and before wallcrossing respectively.
A similar analysis can of course be performed for the lower half plane. $\GamD{5}_1$ and $\GamD{5}_2$ denote BPS rays in the first and second quadrants respectively.

However, when one performs gauge transformation, the contours change. The BPS
ray corresponding to $\tilde\gamma$ now gets rotated to the real axis and the twistor space looks loke figure 7.2.
One can now raise the question : what does ensure consistency with wallcrossing, given such drastically different twistor space ?
The point is that the gauge transformation affects the transition function corresponding to the charge $\gamma$.
The corresponding KS operators are 
\be
\hUks_\gamma = W^{-1}_{\Nq(\gamma)} U_\gamma W_{\Nq(\gamma)},
\ee
where $U_\gamma$ {\it was} the KS operator for charge $\gamma$ before gauge transformation
and $W_n$ is the operator generating the gauge transformation in the $n$th quadrant.
In first and second quadrants they become
\bea
\begin{split}
W_1 =&\, \prod_{\tilde q_0>0} U^{-\frac{\chi}{2}}_{\gamD{-1}}
= \exp \[-\frac{\hat\chi}{8\pi^2} \sum_{\tilde q_0>0}\expe{-\tilde q_0\xi^0}\],
\\
W_2=&\, \prod_{\tilde q_0>0}  U^{\frac{\chi}{2}}_{\gamD{-1}}
= \exp \[\frac{\hat\chi}{8\pi^2} \sum_{\tilde q_0>0}\expe{-\tilde q_0\xi^0}\],
\end{split}
\eea
whereas for the other two quadrants the products run over negative charges \footnote{Here $\hat{\chi} = \chi \sum_{d|\tilde{q}_0} \frac{1}{d^2}$}.
Then it is easy to check that
\bea
\begin{split}
& \prod_{\gamma_j\in \GamD{5}_1}  \hUks_{\gamma_j}^{-\Omega^+ (\gamma_j)}  \prod_{\gamma_i\in \GamD{5}_2} \hUks_{\gamma_i}^{-\Omega^+ (\gamma_i)}
\prod_{\gamma_i\in \GamD{5}_1}  \hUks_{\gamma_i}^{\Omega^- (\gamma_i)} \prod_{\gamma_j\in \GamD{5}_2} \hUks_{\gamma_j}^{\Omega^- (\gamma_j)}
\\ 
=  & \prod_{\gamma_j\in \GamD{5}_1}\! \( W^{-1}_2 U_{\gamma_j}^{-\Omega^+ (\gamma_j)} W_2\)\!
\prod_{\gamma_i\in \GamD{5}_2} \! \(W^{-1}_1 U_{\gamma_i}^{-\Omega^+ (\gamma_i)}W_1\)\!
\\ \qquad\qquad &
\prod_{\gamma_i\in \GamD{5}_1} \!\( W^{-1}_1 U_{\gamma_i}^{\Omega^- (\gamma_i)} W_1\)\!
\prod_{\gamma_j\in \GamD{5}_2} \!\(W^{-1}_2 U_{\gamma_j}^{\Omega^- (\gamma_j)} W_2\)
\\ \nn
= &\, W^{-1}_2\[ \prod_{\gamma_j\in \GamD{5}_1}  U_{\gamma_j}^{-\Omega^+ (\gamma_j)}
\prod_{\tilde q_0>0} U_{\gamD{-1}}^{-\chi} \prod_{\gamma_i\in \GamD{5}_2} U_{\gamma_i}^{-\Omega^+ (\gamma_i)}
\right.
\\  & \left. \qquad \qquad  \qquad 	
\prod_{\gamma_i\in \GamD{5}_1}  U_{\gamma_i}^{\Omega^- (\gamma_i)}
\prod_{\tilde q_0>0} U_{\gamD{-1}}^{\chi} \prod_{\gamma_j\in \GamD{5}_2} U_{\gamma_j}^{\Omega^- (\gamma_j)} \]
W_2
\\
 =&\, 1,
\end{split}
\eea
where we have used \eqref{wcfD5D(-1)} and 
the contact structure represented by the gauge transformed operators is smooth across the wall.

Now one can consider charges corresponding to D1-D(-1) bound state having charge $\tilde{\gamma} = (0,0,\tilde{q}_a,\tilde{q}_0)$
and corresponding BPS rays in type IIA alongwith BPS rays corresponding to charges $\gamma$, as in figure 7.1.
\footnote{We assume that we are at the point in the moduli space
which does not belong to any line of marginal stability.}
For being able to do the analysis as above, one needs the gauge transformation operators
\be
W_\gamma=\prod_{\varepsilon_\gamma\tilde q_0>0} U_{\gamD{-1}}^{(-1)^{\Nq(\gamma)}\frac{\chi}{2}}
\prod_{\gamD{1}\in\GamD{1}_\gamma}U_{\gamD{1}}^{(-1)^{\Nq(\gamma)}n_{\tilde q_a}^{(0)}},
\ee
where $\varepsilon_{\gamma}=\sign \left(\Re Z_\gamma\right)$.
The transformed operators read as
\be
\hUks_\gamma = W^{-1}_{\gamma} U_\gamma W_{\gamma},
\ee
With this one should consider gauge transforming $U_\gamma$ as we did in the previous case. Doing exactly similar 
analysis one can explicitly prove that the contact structure and hence the twistor space does remain smooth after
the gauge transformation.

\subsection{Fivebrane instantons from S-duality}

Now we have all ingredients to reach our main goal --- the twistorial description of fivebrane instantons in the presence of
D1-D(-1)-instanton corrections.\footnote{We remind that our construction ignores the effect of D3-instantons.
Although such approximation is physically unjustified, at a formal level it can be achieved by setting to zero all DT-invariants
$\Omega(\gamma)$ for charges with $p^0=0,\ p^a\ne 0$. Note however that we do include the effect of D3-branes bound to D5-branes,
as required by invariance under monodromies.} To this end, we simply apply the modular constraint \eqref{masterch} to the gauge transformed contact 
hamiltonian in \eqref{Hgam-tr}. More precisely, we split the D5-D3-D1-D(-1) brane bound state charge $\gamma$ into two components,
$p^0$ corresponding to the D5-brane charge and $\hat{\gamma} = (p^a,q_a,q_0)$ as the rest. Then we identify $\hHij{\hgam}_{0,p^0}=\hHij{\gamma}_g$
and we apply the $SL(2,\IZ)$ constraint with the matrix
\be
\label{Sdualde}
\gl{}= \begin{pmatrix} a & b \\ k/p^0 & p/p^0 \end{pmatrix} \in SL(2,\IZ)\, ,
\ee
where the two integers $(p,k)\neq (0,0)$ have $p^0$ as the greatest common divisor, whereas $a$ and $b$ must satisfy $a p - b k = p^0$.
The integer $k$ will appear as NS5-brane charge. As for the other charges, it is convenient to pack them into
rational charges $n^a=p^a/k$, $n^0=p/k$ and the so-called invariant charges \cite{Alexandrov:2010ca}
\be
\begin{split}
\hat q_a = &\, {q_a  + \frac12 \,\kappa_{abc} \frac{p^b p^c}{p^0},}
\\
\hat q_0 =&\, { q_0  + \frac{p^a q_a}{p^0} + \frac13\, \kappa_{abc}\frac{p^a p^b p^c}{(p^0)^2}},
\end{split}
\ee
which are invariant under the spectral flow transformation.

Now a simple application of the S-duality constraint \eqref{masterch} gives the fivebrane contact hamiltonian
\be
\begin{split}
\hkp=&\,(p^0)^{-1}(k\xi^0+p)\,\, \gl{}\cdot \hHij{\gamma}_g(\xi,\txi)
\\
=&\, \frac{\bar\Omega_{k,p}(\hgam)}{4\pi^2} \frac{k}{p^0} (\xi^0+n^0)\sigma_D(\gamma)\, \expe{ S_{k,p;\hgam}},
\end{split}
\label{fivebraneh}
\ee
where we have used the following notations

(1) \be
\begin{split}
S_{k,p;\hgam}= &\, -k S_{n^\Lambda}+ \frac{p^0(p^0 \hat q_0-k \hat q_a (\xi^a + n^a))}{k^2(\xi^0 +n^0)}
- \frac{a}{k}\,p^0 q_0- c_{2,a} p^a \varepsilon(\gl{})
\\
&\,  -\frac{(-1)^{\Nq_{k,p}(\hgam)}}{2\pi\I}  \sum_{\gamD{1}\in\Gamma_{k,p;\hgam}}\bn_{\tilde q}\, p^\Lambda \tilde q_\Lambda \,
\expe{\tS_{k,p;\gamD{1}}}
\end{split}
\label{fivebraneaction}
\ee
with $S_{n^\Lambda} =\alpha - n^\Lambda \txi_\Lambda  + \Fcl (\xi + n)$.

\vspace{ 2 mm}

(2) S-duality transformed D1-brane twistorial action (note that both actions \eqref{fivebraneaction} and \eqref{Stildegam}
are regular at $k=0$ and reduce in this limit to the (gauge transformed) D-instanton twistorial actions $-\Xi^{(g)}_\gamma$ and $-\Xi_{\tilde\gamma}$,
respectively.)
\be
\tS_{k,p;\gamD{1}}= \frac{\tilde q_0 (p^0)^2}{k^2 (\xi^0 + n^0)}-\frac{p^0\tilde q_a \xi^a}{k(\xi^0+n^0)}-\frac{a}{k}\,p^0 \tilde q_0.
\label{Stildegam}
\ee

\vspace{2 mm}

(3) Rational Gopakumar-Vafa invariants $n^{(0)}_{q_a}$ constructed from standard invariants by using
\be
\bar{\Omega}(\gamma) = {\sum_{d|\gamma}} \frac{\Omega(\gamma)}{d^2}.
\ee

\vspace{2 mm}

(4) Transformed BPS indices that $\bar{\Omega}_{k,p}(\hgam) = \bar{\Omega}(\gamma,\gl{}\cdot z)$, that take into account that
the DT invariants are only piecewise constant and the moduli dependence of them is affected by S-duality transformation.

\vspace{ 2 mm}

(5) 
The transformed charge lattice 
\be
\Gamma_{k,p;\hgam}=
\GamD{1}_\gamma(\gl{}\cdot z)\cup \GamD{-1}_\gamma(\gl{}\cdot z).
\label{transGam}
\ee
The dependence on $z^a$ comes from the fact that the definitions \eqref{latDm1} and \eqref{latD1} depend on the central charge function. 

\vspace{2 mm}

(7) Target quadrant in the complex plane
$\Nq_{k,p}(\hgam)= \left\lfloor\frac{2}{\pi}\, \arg \left(\I \gl{}\cdot Z_\gamma\right)\right\rfloor$.

\vspace{ 5 mm}

The contours on the $\IC P^1$ associated with the contact hamiltonians \eqref{fivebraneh}
are given by the image under S-duality of the original BPS ray $\ell_\gamma$.
\be
\ell_{k,p;\hgam}=\{\varpi:\ Z_\gamma(\gl{}\cdot z)/(\gl{}\cdot t)\in\I\IR^-\}.
\label{BPSray-five}
\ee
It can be found that they join the points $\htpm^{k,p}$. ( as in figure 7.3).

\vspace{ -6 cm}

\begin{figure}[htbp]
\centering
\includegraphics[scale=0.5]{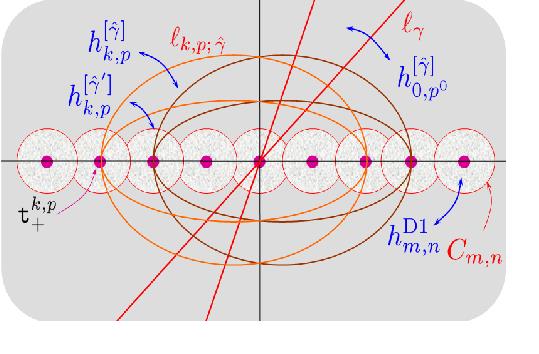}
\caption{Schematic representation of the twistor data generating D(-1)-D1 and all fivebrane instantons. BPS rays joining different points $\htpm^{k,p}$ correspond to different fivebrane charges $k$ and $p$. Different BPS rays joining the same points correspond to different reduced charges $\hgam$.}
\end{figure}

\vspace{7.5 cm} 

The expression for the transition function can be computed from the contact hamiltonian \eqref{fivebraneh}. The expression is valid
to {\it all orders} and is given by 
\be
\begin{split}
\Hij{\hgam}_{k,p}
=&\, \hkp+2\pi^2(\hkp)^2\left(\frac{\hat q_0 (p^0)^2}{k(\xi^0+n^0)}+\frac{2k^2\Fcl(\xi+n)}{(1+2\pi\I k\hkp)^2} \right)
\\
&\,
-(-1)^{\Nq_{k,p}(\hgam)}\,\frac{k(\xi^0+n^0)}{4\pi^2 p^0}\sum_{\gamD{1}\in\Gamma_{k,p;\hgam}}
\bn_{\tilde q}\,\expe{\tS_{k,p;\gamD{1}}} \cE\(\frac{4\pi^2 p^0 (\tilde{q}_\Lambda p^\Lambda)}{k(\xi^0+n^0)}\,\hkp\).
\end{split}
\label{tranNS5all}
\ee
It was checked in \cite{Alexandrov:2014rca} that the above transition function indeed satisfies \eqref{master}.
It exemplifies the power of the contact hamiltonian formulation.  Solving the constraint on the transition functions
is actually very cumbersome. Instead, the solution through the new parametrization is much more efficient.

\section{Monodromy and Heisenberg invariance of fivebrane instanton corrected HM moduli space}

In chapter \ref{ch6}, we have seen that requiring that the HM moduli space is invariant under the action of
the algebra generated by discrete symmetries, one can find a nice group representation for $SL(2,\IZ) \ltimes N(\IZ)$.
This led to the modification of the monodromy transformation \eqref{bjacr-mod} and as a consequence, unlike \eqref{twistheisen}
the anomalous terms disappear in \eqref{bjacr-mod}, so that one has $T^{(1)}_{0,\eta^a} S = S M_{\eta_a}$. The fivebrane transition
functions that we have derived in \eqref{tranNS5all}, is obtained by the application of the S-duality constraint \eqref{master}
on the D5-instanton transition function on the type IIB side. Thus with the amended transformations, it should be possible to 
check \eqref{grlaw} explicitly by considering the action of symmetries on the fivebrane transition function \eqref{tranNS5all}. 
This serves as a crosscheck that the action of the discrete symmetries  on the physical fields
given in chapter \ref{ch6} is indeed consistent.

In the following, we give the verification that the full contact structure of the twistor space remains invariant,
or equivalently $\cM_{\rm HM}$ remains invariant under the action of integer monodromy, Heisenberg and S-duality
symmetries. Before going further, let us recollect some facts about the characters $\varepsilon(\gl{})$ appearing in the 
S-duality transformation of the RR-axion $c_a$. 

\subsection{The character $\varepsilon(\gl{})$ and Dedekind sum}

The character $\varepsilon(\gl{})$, as we have already seen is defined by the following,
\be
\label{multeta}
e^{2\pi\I\, \varepsilon(\gl{}) }=\frac{\eta\left(\frac{a\tau+b}{c\tau+d}\right)}{(c\tau+d)^{1/2}\,
\eta(\tau)}.
\ee
In particular, $24\varepsilon(\gl{})$ is an integer and it has the explicit representation, 
\be
 \varepsilon(\text{g}) =
 \begin{cases}
      \frac{b}{24}\, \sign(d) & (c=0) \\
      \frac{a+d}{24c}-\hf s(d,c) -\frac18 & (c>0) \\
      \frac{a+d}{24c} + \hf s(d,c) + \frac18 & (c<0)
   \end{cases}
\label{Dedchar}
\ee
where $s(d,c)$ is the Dedekind sum. It can be written in terms of the following function
\be
\left(\left(x\right)\right)=\left\{\begin{array}{ll}
x-\lfloor x\rfloor -1/2, &\quad {\rm if \ }x\in \IR\setminus\IZ
\\
0, & \quad {\rm if \ } x\in \IZ
\end{array}\right.
\ee
as
\be
s(d,c) =  \sum_{r \text{ mod } |c|} \left(\left(\frac{r}{|c|}\right)\right) \left(\left(\frac{r d}{|c|}\right)\right).
\ee
It can be easily shown that 
\be
s(d,c)=  \sum^{c-1}_{r=1}  \frac{r}{c} \left(\left(\frac{rd}{c}\right)\right)
= (c-1)\left(\frac{d}{6c}\,(2c-1)-\frac14 \right)-\sum^{c-1}_{r=1}  \frac{r}{c} \left\lfloor\frac{rd}{c}\right\rfloor,
\label{Dedekindsum}
\ee
where we set $c>0$.
Thus, for the generators $S$ and $T$ of the $SL(2,\IZ)$ group, one obtains
\be
\varepsilon(S)=-\frac{1}{8},
\qquad
\varepsilon(T)=\frac{1}{24}.
\ee
It also satisfies a reciprocity relation which we will require for the proof,
\be
s(d,c)+s(c,d)=\frac{1}{12}\left(\frac{d}{c}+\frac{1}{cd}+\frac{c}{d}\right)-\frac14.
\label{reclaw}
\ee

\subsection{Invariance under monodromy}

We first discuss the invariance of the fivebrane instanton corrected contact structure under monodromy transformations.
The lift of \eqref{bjacr-mod} to the twistor space is given by the following action on the Darboux coordinates
\be
M_{\eps^a}\ :\quad
\begin{array}{rl}
\xi^0\ \mapsto\ & \displaystyle{\xi^0,
\qquad
\xi^a \ \mapsto\ \xi^a + \epsilon^a \xi^0,}
\\
\txi_a\ \mapsto\ & \displaystyle{\txi_a - \kappa_{abc} \epsilon^b \xi^c - \frac12\, \kappa_{abc} \epsilon^b \epsilon^c \xi^0
+A_{ab}\epsilon^b,}
\\
\txi_0\ \mapsto\ & \displaystyle{\txi_0 - \epsilon^a \txi_a + \frac12\, \kappa_{abc} \epsilon^a\epsilon^b \xi^c
+ \frac16\, \kappa_{abc} \epsilon^a\epsilon^b\epsilon^c \xi^0
-\hf\, A_{ab}\epsilon^a\epsilon^b+\frac{c_{2,a}}{8}\, \epsilon^a,}
\\
\alpha \ \mapsto\ &\displaystyle{ \alpha + \frac12 \left( \kappa_{abc} \epsilon^a \xi^b\xi^c +
\kappa_{abc} \epsilon^a\epsilon^b\xi^c \xi^0
+ \frac13\, \kappa_{abc} \epsilon^a \epsilon^b\epsilon^c (\xi^0)^2 \right) +\frac{c_{2,a}}{24}\, \eps^a},
\end{array}
\label{monDc}
\ee
and the fiber coordinate $t$ does not transform. Let us first find what constraint it entails on the transition function.
One can easily find that 
\bea
\begin{split}
\p_{\xii{i}^a} (M_{\epsilon^a}\cdot \Hij{ij})=&\,
\p_{\xii{i}^a} \Hij{ij} - \kappa_{abc} \epsilon^b T^c_{[ij]}
- \frac12\, \epsilon^b \epsilon^c T^0_{[ij]},
\\
\p_{\xii{i}^0} (M_{\epsilon^a}\cdot \Hij{ij})=&\,
\p_{\xii{i}^0} \Hij{ij} - \epsilon^a  \p_{\xii{i}^0} H
+ \frac12\, \kappa_{abc} \epsilon^a \epsilon^b  T^c_{[ij]}
+ \frac16\, \kappa_{abc} \epsilon^a \epsilon^b  \epsilon^c T^0_{[ij]},
\end{split}
\eea
where $T^\Lambda_{[ij]}$ is introduced in \eqref{master}. From the transformation of $\alpha$ one can
now obtain the constraint on the transition functions
\bea
\!\!\!\!\!\!\!\!\!\!M_{\epsilon^a} \cdot \Hij{ij} = \Hij{ij} + \frac12\, \kappa_{abc} \epsilon^a T^b_{[ij]} T^c_{[ij]} + \frac12\, \kappa_{abc} \epsilon^a \epsilon^b T^c_{[ij]} T^0_{[ij]}
+ \frac16\,\kappa_{abc} \epsilon^a \epsilon^b \epsilon^c (T^0_{[ij]})^2.
\label{constr-mon}
\eea
The D-instanton action remains invariant under monodromy transformation, if one supplements it
with the spectral flow transformation on the lattice of charges \eqref{spectrflow} with the parameter ${\epsilon'}^ a= \frac{p}{p^0}\, \eps^a$.  
To proceed we have to first
check the invariance of \eqref{fivebraneaction}.
For the last term, one should change the summation variable as
\bea
\tilde{q}_0 \mapsto \tilde{q}_a - \epsilon^{'a} \tilde{q}_a, \qquad \tilde{q}_a \mapsto \tilde{q}_a.
\eea
Then both the factors $p^\Lambda\tilde{q}_\Lambda$ and the exponential in $\tS_{k,p;\tilde{\gamma}}$ remain invariant. 
One could worry about having changed the summation range while doing this. But in fact it stays the same. Because
both $\gl{}\cdot Z_\gamma$ and $\gl{}\cdot Z_{\tilde{\gamma}}$  remain invariant if one performs the spectral flow transformation 
on the charge lattice. This proves that the last term is invariant. 

Let us calculate the phase $\nu(\eps)$ appearing in $M_{\eps^a} \hHij{\hgam}_{k,p} = \nu(\eps)\,\hHij{\hgam}_{k,p}$. Then
\bea
\begin{split}
\nu(\epsilon)=&\, \expe{-k\epsilon^a \kappa_a-p c_{2,a}\epsilon^a\[\frac18-\ep\(\frac{p}{p^0},\frac{k}{p^0}\)-\frac{p^0}{24}\]
-bq_a\epsilon^a-\frac{p(p-1)}{2}\, A_{ab}\epsilon^a\epsilon^b
\right.\\
&\,\left.
-\frac{p}{2p^0}\kappa_{abc}\epsilon^a p^b p^c+\frac{p}{2p^0}\, (p-b)\kappa_{abc}\epsilon^a\epsilon^b p^c
-\frac{p^2}{6p^0}\, (p-b)\kappa_{abc}\epsilon^a\epsilon^b \epsilon^c}
\\
= & \, {\bf E}\bigg(-c_{2,a}\epsilon^a \[\frac{b p^0}{12}\(1-p/p^0)^2\) - \frac{p}{2} s(k/p^0,p/p^0)
\right. 
\\ &
\left. \qquad
+\frac{p}{8}(p-p^0)\]
+ \(p/p^0 + 1\)\(1+b\) A_{ab} \epsilon^a p^b\bigg) ,
\end{split}
\eea
where we have used the expression for \eqref{Dedchar} and exchanged the arguments in the Dedekind sum by \eqref{reclaw}, and
\eqref{propA}. The last term in the above expression is clearly integer, because due to the $SL(2,\IZ)$ relation $a p/p^0 - bk/p^0 = 1$,
both of $b$ and $p/p^0$ can not be even. For the other term, we first use the expression for the Dedekind sum \eqref{Dedekindsum}.
We also remember that $c_{2,a}$ is even and hence the last term in \eqref{Dedekindsum} does not contribute. Then after a bit of algebra one
arrives at the following expression for the phase,
\bea
\nu(\epsilon) = \expe{-\frac{c_{2,a}\epsilon^a}{12}(d-1)\(b(d+1) - c(2d-1)\)},
\eea
where $c=p/p^0$ and $d=k/p^0$. It can be rewritten as
\bea
\begin{split}
\nu(\epsilon) = &\, \expe{-\frac{c_{2,a}\epsilon^a}{12}(d-1)\(b(ad-bc)(d+1) - c(2d-1)\)}
\\ 
= &\, \expe{-\frac{c_{2,a}\epsilon^a}{12} (d-1) \[-b^2 c (d+1) +c(d+1) - 3c d\]} 
\\ 
= & \, \expe{\frac{c_{2,a}\epsilon^a}{12} (d-1)(d+1)(b-1)(b+1)}.
\end{split}
\eea
In the second and third steps we have used the fact that $\frac{1}{6} d(d-1)(d+1)$ and $\frac12 d(d-1)$ are integers respectively. 
Now it is not difficult to see that $\nu(\epsilon) = 1$. First,$b$ and $d$ can not be simultaneously even. Secondly, $(b^2-1)(d^2-1)$ 
is not divisible by three only if $b=3k$ and $d=3l$, where $k,l \in \IZ$. But this is impossible, as it contradicts $ad - bc = 1 $. Hence, we have
proven that the contact structure remains invariant under monodromy transformation mapping contact hamiltonians to each other as
\be
M_{\eps^a}\cdot \hHij{\hgam}_{k,p}= \hHij{ \hgam[-p\eps^a/p^0]}_{k,p},
\qquad
M_{\eps^a}\cdot \ell_{k,p;\hgam}= \ell_{k,p;\hgam[-p\eps^a/p^0]},
\label{Moninv}
\ee
The transition functions \eqref{tranNS5all} are also mapped to each other as a consequence. However,
they satisfy the non-linear constraint \eqref{constr-mon} as the following \footnote{This is yet another instance
of the virtue of the contact bracket formulation. Instead of having to deal with non-linear \eqref{constr-mon}
one can work with simpler \eqref{Moninv}.}
\be
M_{\eps^a}\cdot \Hij{\hgam[p\eps^a/p^0]}_{k,p}= \Hij{ \hgam}_{k,p}
+ \frac12\, \kappa_{abc} \epsilon^a \Tmn{b} \Tmn{c} + \frac12\, \kappa_{abc} \epsilon^a \epsilon^b \Tmn{c} \Tmn{0}
+ \frac16\,\kappa_{abc} \epsilon^a \epsilon^b \epsilon^c (\Tmn{0})^2.
\label{Moninv-tr}
\ee
To end the proof that $\cM_{\rm HM}$ remains invariant under monodromy transformation \eqref{bjacr-mod} accompanied by
spectral flow transformation, we need to ensure that the rational DT invariants $\bar{\Omega}_{k,p}(\hgam)$ remains invariant. 
It was shown in \cite{Manschot:2009ia} that the BPS indices indeed remain invariant under combined action of monodromy transformations
and spectral flow transformation on the charge lattice with the same parameter. Hence the proof is complete. 

\subsection{Invariance under Heisenberg transformation}

For a direct check of the group law presented in chapter \ref{ch6}, one has to also verify that 
the action of Heisenberg shifts keep the contact structure invariant. From the commutators given in 
chapter \ref{ch6} clearly, the only Heisenberg transformation generator that can possibly
have non-trivial action on the fivebrane transition function \eqref{tranNS5all}or the corresponding
contact hamiltonian \eqref{fivebraneh} is $T^{(1)}_{0,\eta^a}$. Its action at the level of the twistor space is given by, 
\be
T^{(1)}_{0, \eta^a}\ : \quad  \begin{array}{c}
\displaystyle{\xi^a\ \mapsto\ \xi^a-\eta^a,
\qquad
\txi_a\ \mapsto\ \txi_a+A_{ab}\eta^b,
\qquad
\txi_0\ \mapsto\ \txi_0+\frac{c_{2,a}}{24}\,\eta^a,}
\\
\displaystyle{\alpha\ \mapsto\ \alpha+\eta^a\txi_a+\hf\, A_{ab}\eta^a\eta^b.}
\end{array}
\label{Heisact-Darboux}
\ee
As was the case with monodromy transformation, to ensure invariance of $\cM_{\rm HM}$ after fivebrane instanton corrections,
one has to perform a spectral flow transformation on the charge lattice $\hat{\gamma} \mapsto \hat{\gamma}[\eps]$, but now with the parameter 
$\epsilon^a = k\frac{\eta^a}{p^0}$. Thus, we have to establish that
\be
T_{0, \eta^a}\cdot \hHij{\hgam}_{k,p}= \hHij{ \hgam[-k\eta^a/p^0]}_{k,p},
\qquad
T_{0, \eta^a}\cdot \ell_{k,p;\hgam}= \ell_{k,p;\hgam[-k\eta^a/p^0]}.
\label{Heisinv}
\ee
For the transition functions, the invariance of the contact structure is encoded in the 
following transformation of transition functions into each other, 
\be
T_{0,\eta^a}\cdot \Hij{\hgam[k\eta^a/p^0]}_{k,p}= \Hij{ \hgam}_{k,p}.
\ee
While the second equation in \eqref{Heisinv} is a requirement for the contact structure to be invariant,
we have to check the first one that maps contact hamiltonians to each other for consistency. 
Let us compute the phase $\nu(\eta)$ in $T_{0,\eta^a} \hHij{\hgam}_{k,p} = \nu (\eta) \, \hHij{\hgam}_{k,p}$. 
Then a direct calculation leads to 
\bea
\begin{split}
 \nu (\eta)=&\,
\expe{\frac{k}{2p^0}\, \kappa_{abc} p^a p^b \eta^c +\frac{k}{2}\,c_{2,a}\eta^a \[s\(\frac{p}{p^0},\frac{k}{p^0}\)
-\frac{1}{4}\(p^0-1\)\]
\right.
\\
& \left.\qquad
-\frac{k(k-1)}{2}A_{ab}\eta^a\eta^b
+\eta^a A_{ab} p^b -\frac{ap^0}{k}\,p^\Lambda L_\Lambda(\epsilon)}.
\end{split}
\eea
Using \eqref{propA} the above can be rewritten as 
\be
\nu(\eta)
= \expe{  p^0 c_{2,a} \eta^a  \[ \frac{c}{2}\, s\(d,c \)
- \frac{c p^0}{8}\(1-c\) +\frac{a }{12} \(c^2 -1\) \]
-(a-1) \(c+ 1 \) A_{ab} \eta^a p^b },
\label{Hphase}
\ee
where we preferred to write the result in terms of $c=k/p^0$ and $d=p/p^0$.
Now note that the relation $ad - bc =1 $ ensures that $a$ and $c$ can not be simultaneously even.
Therefore, the last term in \eqref{Hphase} is an integer and thus disappears.
Furthermore, using the expression for the Dedekind sum \eqref{Dedekindsum} and taking into account that $c_{2,a} \eta^a$ is even,
one finds
\be
\nu(\eta) = \expe{ p^0 c_{2,a} \eta^a  \(c -1 \)\[
\frac{c }{8}\,(p^0-1)+\frac{d}{12}\(2c -1 \)+\frac{a }{12}\,(c+ 1 )\] }.
\label{Heisen1}
\ee
Now the situation is similar to the case of monodromy. The first term can be dropped because 
$\frac14 c(c-1)p^0(p^0-1) \in \IZ$. The rest can be rewritten as 
\be
\nu(\eta) = \expe{ \frac{d}{12}\,c_{2,a} \eta^a (c+1)(c-1)(a-1)(a+1)},
\ee
which is equal to one, as $6|(c+1)(c-1)(a-1)(a+1)$. This ends the proof of 
the invariance of the fivebrane transition function under Heisenberg transformations. 

Thus we have proven by direct action of the symmetries on the transition function
that the group law \eqref{grlaw} holds with the modified monodromy transformations,
which was one of the major results in \cite{Alexandrov:2014rca}.

Before we finish, we would like to point out some shortcomings of our constructions. 
The twistor space presented in figure 7.3 has a rather complicated structure. There are
open patches surrounding $\htpm^{m,n}$ which lie along the real line. On top of that,
there are fivebrane contours that divide the twistor space into several other patches as they connect
the points $\htpm^{m,n}$. The first problem comes from the fact that we have assumed the points
$\htpm^{m,n}$ do not have any accumulation point on the real line, that is they are separable.
Although, it seems to be a mild assumption, it is desirable to get rid of this. An even more serious problem 
arises with the fivebrane contours. They intersect in infinitely many points on the twistor space. Unfortunately,
the transition function that we have derived \eqref{tranNS5all} is not consistent with the cocycle condition
around them. The situation can get even worse if the points form a dense set on the twistor space. 
Unfortunately, the author does not know how to reconcile $SL(2,\IZ)$ invariance with the cocyclicity
at these points and this renders the problem for future investigations.
\newpage

\chapter{Discussion and conclusions}

\label{discconcl}

This thesis mainly pertains to understanding of the quaternion-K\"ahler manifolds that arise as
hypermultiplet moduli space of type II string theories compactified on Calabi-Yau threefolds. We have discussed how to 
incorporate various quantum corrections to this HM moduli space. The thesis follows the lead from \cite{Alexandrov:2011va}
and \cite{Alexandrov:2013yva}. We show how to include various quantum corrections step by step : starting from one-loop
to the tree-level metric, then discussing how to incorporate D-instantons and finally the NS5-instantons. They all are included coherently
with the quaternion-K\"ahler property of the HM moduli space metric. 

Our construction crucially relies on the twistorial techniques that had been introduced in chapter \ref{chapter3}. 
It generalizes the superconformal approach which is applicable in the presence of sufficient number of commuting isometries
on the HM moduli space. The advent of the twistor construction helps us to go beyond such restrictions and in particular,
this procedure is immensely important for incorporating various instantons as we have seen in the subsequent chapters. 
In this framework, various non-trivial and rather complicated physical effects got a simple and nice geometric interpretation. 

\begin{itemize}

\item It turns out that there is an even simpler description of the twistor spaces. Instead of working with 
the transition functions introduced in \cite{Alexandrov:2008nk}, one can introduce the so called contact hamiltonians \cite{Alexandrov:2014mfa}.
The advantage is that, it facilitates handling of various symmetries at the level of the twistor space. We have already seen the 
advantages in two instances : for realization of S-duality symmetry and monodromy invariance around the large volume point.
Having to deal with constraints imposed by symmetries on contact hamiltonians is much more easier compared to the  
constraints on the usual transition functions. 

The reason can be traced to the fact that unlike the parametrization with usual transition functions, the contact hamiltonian
depends on the Darboux coordinates on the same patch. A consistent realization of the symmetries necessarily requires that
the derivatives entering in the gluing conditions abide by certain constraints, as the action of a symmetry group at the level of the twistor space is defined by its action on the
Darboux coordinates. The complication with the usual transition functions comes from the fact that
there one has to care about the patches meticulously while taking derivatives and this gets reflected in the expressions for derivatives. 
The situation simplifies drastically with our new parametrization using contact hamiltonian. The constraints subsequently 
are significantly simpler on the contact hamiltonians. Our derivation of the fivebrane instanton corrections in chapter \ref{ch7}
crucially depended on this simplification. 

\item In \cite{Alexandrov:2014rca}, we found the fivebrane instanton correction to the HM moduli space. 
This is actually one of the main works that is  presented in this thesis.  
Our result is valid to all orders, extending it from the linear instanton analysis performed in \cite{Alexandrov:2010ca}. 
Achieving this result was possible owing to our new formulation using the contact hamiltonians. In chapter \ref{ch7},
as we have discussed, exploiting a rather intricate chain of duality arguments, especially exploiting S-duality
symmetry of the type IIB formulation, we derived the fivebrane transition functions on the type IIB side.

\item Our findings about the discrete isometry groups for the HM moduli space is very interesting. We have resolved the issue raised 
in \cite{Alexandrov:2010ca} regarding consistency of actions of the discrete symmetry groups on the HM moduli space.  In the process
we obtained a group representation and thereby we have shown that the fivebrane instanton corrected $\cM_{\rm HM}$ admits their action
in a consistent fashion. 

\item In \cite{Alexandrov:2013mha}, we have presented the results for the non-perturbative mirror maps, when the HM moduli space
admits no continous isometry. This is the situation when one considers the NS5-instantons. Our investigations for the NS5-instantons
to all orders in instanton expansion set off from that point. While accomplishing the task of describing such a QK manifold, we had come 
across several nice and at the same time intriguing geometrical objects such as the invariant points and a function encoding modular invariant deformations
of the twistor space to all orders. 

\item In \cite{Alexandrov:2014sya}, we have found an explicit expression for the HM moduli space metric valid to all orders when one 
restricts to the set of mutually local charges for the D-instantons. The same result is also valid for generic D-instantons to first order
in instanton expansion. We have also found that for the universal hypermultiplet case, one can recast the result in the form of the
Tod ansatz and the Toda potential constructed satisfies the Toda equation. This serves as a consistency check for our results and
as a byproduct, we have found the result for the Toda coordinates given by a set of transcendental equations.  

\end{itemize}

Although in light of all these results it seems that the final destination of describing the HM moduli space with all instanton corrections is not too far,
there are several problems that are still lurking around. We plan to investigate them in more details in future.
In the following, we present them in order.

\begin{itemize}

\item In chapter \ref{chapter 4} we came across the issue of curvature singularity in the HM moduli space. We have seen that 
the curvature singularity of the one-loop corrected metric at $r=-2c$ gets dressed up in presence of instantons. In particular,
D(-1)-instantons affect the position of the singularity on the moduli space and it turns out that the situation becomes even worse compared to the perturbative metric. 
Whereas, the singularity in the perturbative case was present only if the Euler characteristic of the Calabi-Yau was negative,
after considering D(-1)-instantons we find that it is present always. It is unlikely that other D-instantons can change this situation. 
The only possibility through which this singularity can be removed is perhaps the inclusion of NS5-instantons. They behave in a significantly
different fashion at the strong coupling regime and thus can drastically alter the behavior of the metric. We aim to study this 
situation in details in future.

\item Another important issue to be addressed is finding the NS5-instantons in the type IIA formulation.
Such a result must comply with the symplectic invariance of the type IIA side. Concerning the inclusion of NS5-instantons,
it will be also interesting to investigate if they regulate the exponentially divergent growth of BPS indices $\Omega(\gamma)$.
It is our hope that, the results presented in this thesis is a step forward to the understanding of these problems.

\item We need to understand the $SL(2,\IZ)$ invariant picture of D3-brane instantons better. 
An attempt towards it was made in \cite{Alexandrov:2012au} and from its findings, it is quite clear
that they indeed admit an $SL(2,\IZ)$ invariant description.
The rich interplay between mock modular forms and contact geometry as was found in \cite{Alexandrov:2012au} is probably only
the tip of the iceberg. A thorough understanding extending the construction to all orders in instanton corrections and at a generic point on the HM moduli space, will probably help us to unravel
much more deeper connections. Addressing this problem is one our goals in future. 

\item We hope to connect our results with other domains of string theory more intimately, for example with topological strings and BPS black holes. 
We cherish the idea of being able to extend at least some of our results to the context of vacua with $N=1$ supersymmetry and thus making
them useful for phenomenological model building.

\item An important step was taken by improving the action of the discrete symmetries at the quantum level on the HM moduli space $\cM_{\rm HM}$. 
Namely, we found that the closure of the duality group requires a modification of the integer monodromy transformation. As was said before, this adjustment had a 
double effect : on one hand it provided a consistent implementation of all discrete symmetry transformations, on the other hand it also resolved the 
apparent conflict of the fivebrane instantons, Heisenberg and monodromy symmetries \cite{Alexandrov:2010ca}. 

However, the proposed modification of the monodromy transformation on the RR-fields raised the following problem. We reiterate our observations
in the chapter \ref{ch6} here once again. Before the modification, monodromy transformation acted by shifting RR-fields according to \eqref{bjacr}.
This can be derived from the definition of RR-scalars in terms of the $B$-field and RR-potential $A^{\rm even} \in H^{\rm even}(\CY,\IR)$ 
\eqref{RRdef}, by just applying a shift on the $B$-field, while keeping the RR-potential fixed. With the modified monodromy transformation,
it would then be natural to ask if the anomalous terms in \eqref{bjacr-mod} can be generated in a similar way. As we have already explained 
in section 6.5, in the latter case the situation is not so simple. From what we have understood by now, the anomalous terms appearing in \eqref{bjacr-mod}
have rather non-trivial geometric origins, about which some conjectural remarks were made at the end of chapter \ref{ch6}. We hope to
have better justification for them in future, especially a better clarification from the standpoint of geometric understanding.

\end{itemize}

\providecommand{\href}[2]{#2}\begingroup\raggedright\endgroup

\end{document}